\documentclass[manuscript, nonacm]{acmart}

\AtBeginDocument{%
  \providecommand\BibTeX{{%
    Bib\TeX}}}

\usepackage{comment}
\usepackage{tabularx}
\usepackage{graphicx}
\usepackage{colortbl} 
\usepackage[normalem]{ulem}
\usepackage{lscape}
\usepackage{soul}
\usepackage{color}
	
\definecolor{blizzardblue}{rgb}{0.67, 0.9, 0.93}
\definecolor{celadon}{rgb}{0.67, 0.88, 0.69}
\definecolor{babypink}{rgb}{0.96, 0.76, 0.76}
\DeclareRobustCommand{\hlcyan}[1]{{\sethlcolor{blizzardblue}#1}}

\begin{document}

\title{A Longitudinal Study of Child Wellbeing Assessment via Online Interactions with a Social Robot}

\author{Nida Itrat Abbasi}
\email{nia22@cam.ac.uk}
\affiliation{%
  \institution{Department of Computer Science and Technology, University of Cambridge}
  \country{UK}
}
\author{Guy Laban}
\email{guy.laban@cl.cam.ac.uk}
\affiliation{%
  \institution{Department of Computer Science and Technology, University of Cambridge}
  \country{UK}
}
\author{Tamsin Ford}
\email{tjf52@cam.ac.uk}
\affiliation{%
  \institution{Department of Psychiatry, University of Cambridge}
  \country{UK}
}

\author{Peter B. Jones}
\email{pbj21@cam.ac.uk}
\affiliation{%
  \institution{Department of Psychiatry, University of Cambridge}
  \country{UK}
}
\author{Hatice Gunes}
\email{hg410@cam.ac.uk}
\affiliation{%
  \institution{Department of Computer Science and Technology, University of Cambridge}
  \country{UK}
}

\renewcommand{\shortauthors}{Abbasi et al.}

\begin{abstract}
Socially Assistive Robots are studied in different Child-Robot Interaction settings. However, logistical constraints limit accessibility, particularly affecting timely support for mental wellbeing. In this work, we have investigated whether online interactions with a robot can be used for the assessment of mental wellbeing in children. The children (N=40, 20 girls and 20 boys; 8-13 years) interacted with the Nao robot (30-45 mins) over three sessions, at least a week apart. Audio-visual recordings were collected throughout the sessions that concluded with the children answering user perception questionnaires pertaining to their anxiety towards the robot, and the robot's abilities. We divided the participants into three wellbeing clusters (low, med and high tertiles) using their responses to the Short Moods and Feelings Questionnaire (SMFQ) and further analysed how their wellbeing and their perceptions of the robot changed over the wellbeing tertiles, across sessions and across participants' gender. Our primary findings suggest that (I) online mediated-interactions with robots can be effective in assessing children's mental wellbeing over time, and (II) children's overall perception of the robot either improved or remained consistent across time. Supplementary exploratory analyses have also revealed that the gender of the children affected their wellbeing assessments with interactions effectively distinguishing between varying levels of wellbeing for both boys and girls for the first session and only for boys during the second session. The analyses have also revealed that girls have a higher opinion of the robot as a confidante as compared with boys. Findings from this work affirm the potential of using online mediated interactions with robots for the assessment of the mental wellbeing of children.

\end{abstract}



\maketitle




\section{Introduction}
Research into Socially Assistive Robots (SARs) in children's lives have become increasingly prevalent, addressing their potential roles in education \cite{tanaka2015pepper,Henschel2021}, autism interventions \cite{scassellati2018improving} and more recently in the field of mental health \cite{abbasi2022can,abbasi2024analysing}. Children have been shown to consider robots as companions and peers because of the non-intimidating, engaging nature of SARs that provide an interactive, responsive and non-judgemental platform for children to feel comfortable \cite{kory2019long, de2020effects}. However, interacting with SARs usually involves visiting a laboratory and thus providing logistical restrictions in availability and accessibility. In sensitive domains like mental health, emotional stigma, shame and attitudes towards seeking help (for example, from therapists \cite{thepthien2019self}, online support initiatives \cite{yang2019channel} and wellbeing applications \cite{punukollu2019use}) may affect self-disclosure in children \cite{smart_wegner2000, Corrigan2014}.In addition, children's inclinations to openly express their emotions are also heavily influenced by their gender as boys have been observed to show increased hesitation towards disclosing their genuine feelings compared to girls \cite{chandra2006stigma}.  

SARs have been shown to help with mood estimation \cite{gamborino2019mood} and improvement \cite{Laban_blt_2023}, alleviating anxiety during painful medical procedures \cite{rossi2022using} and identifying bullying instances among children \cite{bethel2016using}. However, most of these studies have involved visits to unfamiliar environments like university laboratories or hospitals that might affect children's interactions and comfort levels. Studies that have investigated the interaction of children with SARs in familiar settings have primarily been in the field of edutainment \cite{tanaka2015pepper} or autism research \cite{scassellati2018improving}. As discussions about the potential applications of social robots in well-being settings gain prominence, it is crucial to acknowledge that physical HRI research often relies on laboratory-constrained environments. This reliance limits the generalization of results, interactions, and insights across diverse populations and conditions. The COVID-19 pandemic, by accelerating the adoption of computer-mediated communication in households \cite{Choi2021}, has opened new avenues for HRI research. This shift towards virtual and computer-mediated communication (CMC) methods enables researchers to explore the interaction with robots in home environments via personal screens \cite{thri_covid_20}, allowing for rigorous and reproducible research that effectively supports individuals' well-being in their natural settings. For instance, CMC interactions with social robots have been leveraged to support informal caregivers \cite{coping_ijsr}, who often struggle to find time and energy for self-care, leading to overlooked health issues \cite{Revenson_book1_2016}. This approach offers them the opportunity to discuss their concerns with an online-mediated robot, ensuring continuous support in their natural environments. These strategies not only enhance the accessibility of emotional well-being interventions to a wider range of populations, thereby promoting inclusivity and diversity but also enable the generalization of HRI research findings to diverse groups that may be restricted from participating in in-person studies. Despite concerns about the loss of physical embodiment, recent studies have found no significant differences in participants' perceptions and interactions with robots in virtual versus in-person settings \cite{Laban2021,Honig2020,Gittens2021,Gittens2022}, providing strong evidence for the efficacy of CMC as an alternative means for HRI experimentation, especially in unique settings such as CRI with children from around the world. Thus, having a robotic platform that children can interact with in an environment familiar to them, and that can take into account any predispositions in their attitudes, especially in the context of mental wellbeing assessment, might help them to be more candid about their feelings and experiences.

In the context of long-term HRI, SARs have been shown to provide emotional support and companionship to the elderly who formed strong bonds with the robot over long-term interactions \cite{yamazaki2023long}. Even online longitudinal interaction with robots has been shown to enhance mood, mitigate feelings of loneliness and promote self-disclosure in adults \cite{Laban_blt_2023}. Thus, in this work, we present the \emph{first study} that has explored whether longitudinal online interactions with a robot can be used for the assessment of mental wellbeing in children. Further, we have explored whether co-location affects robotised wellbeing assessment by comparing our research findings with our prior investigation \cite{abbasi2022can, abbasi2024analysing} that has involved SARs for the assessment of mental wellbeing through an in-lab setting. We also provide an exploratory supplementary investigation into whether the gender of the participants affects the longitudinal assessment of mental wellbeing of children. For this purpose, we have gathered data from 40 children (20 girls and 20 boys, 8-13 years old) who interacted with a Nao robot (SoftBank Robotics), over the course of three sessions for about 30-45 mins in each session. Our primary findings suggest that (i) online interactions with the robot (using Zoom or Skype) can be effective in the assessment of mental wellbeing of children, and (ii) user perception of the robot largely improved or remained unchanged throughout the course of the three online sessions and was affected by the physical presence of the robot especially in the case of anxiety attributed towards the robot and the perceived intelligence of the robot. Our supplementary exploratory analyses have revealed that gender-based variability has been observed in terms of the longitudinal robot-assisted assessments as well as the children's perception of robot's behaviour and capabilities. We found that robot-assisted assessment was able to amplify the differences across varying levels of wellbeing for both boys and girls for the first session and only for boys for the second session. We have also observed that girls had a higher opinion of the robot as their confidante as compared to boys. In summary, this work builds upon our prior work where we observed children interacting with a robot in an in-lab setting \cite{abbasi2022can, abbasi2024analysing}. As mentioned above, our findings indicate that even in an online setting, robotised frameworks can be used to effectively differentiate between children belonging to varying levels of wellbeing. This consistency reinforces the adaptability and robustness of our approach paving the way for broader applications, across different settings, in assessing children's wellbeing. We hope that the findings from this work can provide a valuable stepping-stone for upcoming robot-led initiatives that are more accessible to children for the assessment of their mental wellbeing.

\section{Related Works}

\subsection{Wellbeing Assessments}

\begin{figure}[h!]
	\centering
	\includegraphics[width= 0.8\textwidth]{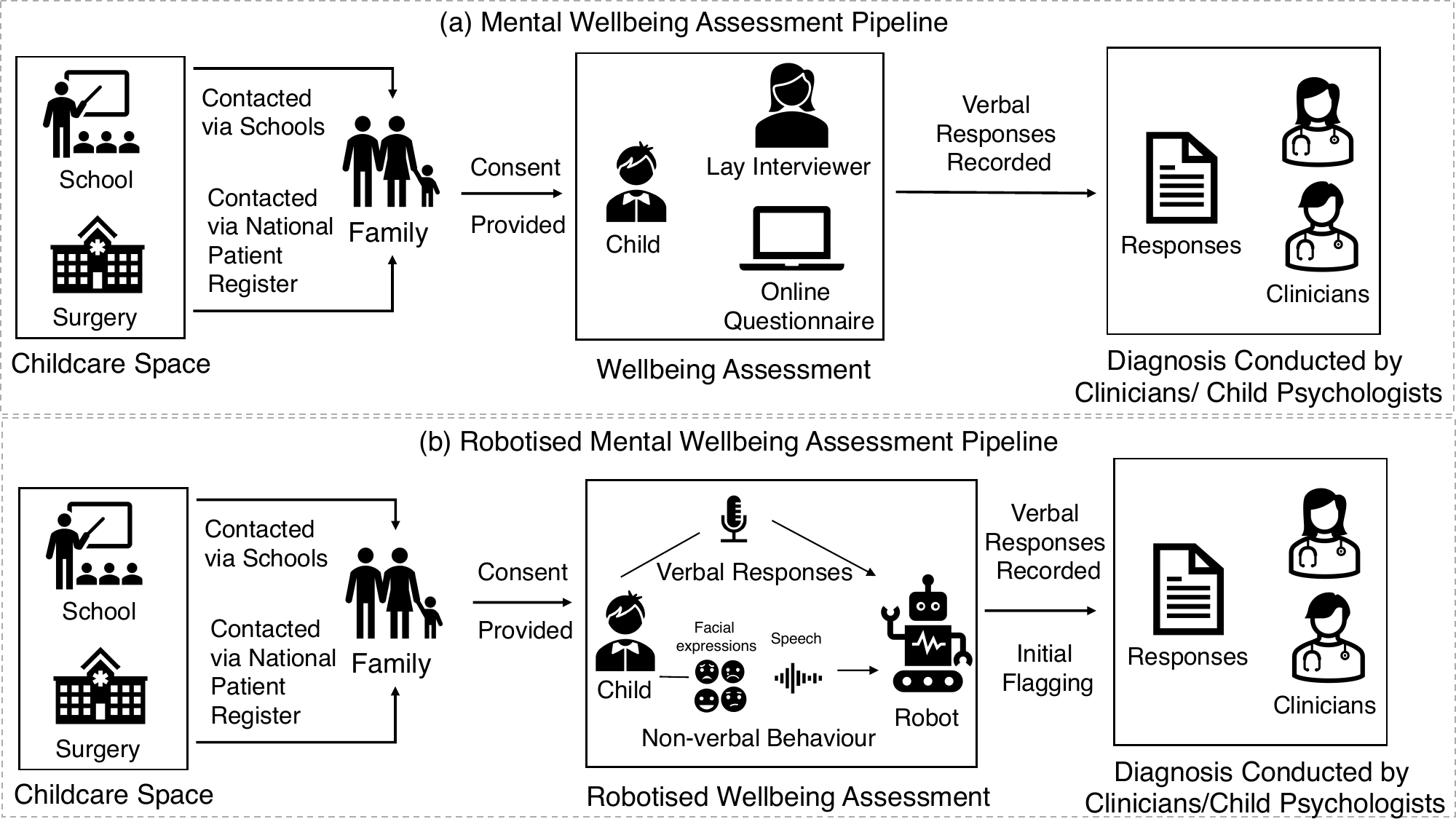}
	\caption{(a) Pipeline showing the procedure for mental wellbeing assessment of children in the UK \cite{ford2020data,mansfield2020oxwell,burn2022developing}. This pipeline is representative of certain assessment scenarios and is not standardised across all clinical and research assessments conducted in the UK. (b) Proposed robotised mental health assessment pipeline.}
	\label{pipeline}
\end{figure}

According to official government statistics published in 2023\footnote{https://www.england.nhs.uk/2023/11/one-in-five-children-and-young-people-had-a-probable-mental-disorder-in-2023/}, about 20.3 $\%$ of children (8-16 years) in the UK have probable mental health disorder. Early identification of mental wellbeing concerns is customary for the future welfare of children in order to mitigate any long-lasting consequences concerning their emotional wellness, social relationships and academic performance \cite{sorter2024addressing}. In the UK, several governmental \cite{campbell2021prevalence,ford2020data} and non-governmental \cite{mansfield2020oxwell,burn2022developing} initiatives have been conducted to understand the mental health of children and young people. Figure \ref{pipeline} (a) shows a depiction of the mental wellbeing assessment pipelines followed by some of these initiatives. The pipeline is not standardised across all clinical and research assessments concerning the mental health of children conducted in the UK. 
In children's mental health evaluations for research endeavours, once consent has been provided by the parents or legal guardians, mental health assessments are usually conducted by lay interviewers or through online forms. These assessments \cite[e.g.,][]{ford2020data,mansfield2020oxwell} usually follow structured or semi-structured interviews designed to measure several aspects of children's mental health including their emotional wellbeing. When conducted for research aims, this evaluation would be for screening and assessment and not for formal clinical diagnosis and follow-up treatment. 

However, these initiatives rely heavily on the assumption that the recorded responses of the children are authentic and representative of their true feelings \cite{tourangeau2000psychology}. Children's responses might be influenced by their reluctance to divulge sensitive information about themselves or being intimidated by the presence of other adults in the room whom they might consider authority figures \cite{OReilly2015}. As such, they may provide socially desirable answers that might not reflect their genuine thoughts and emotions \cite{tourangeau2000psychology,nock2001parent,godoi2020proteger}. Previous studies have shown that people gradually open up to robots over time \cite{Laban_blt_2023}, with those experiencing negative emotional states being more likely to share extensively with robots \cite{Laban2023}. Furthermore, SARs have been shown to be very effective in establishing trust particularly with children \cite{di2020shall}, who have often attributed robots as peers and friends \cite{fior2010children,kanda2004interactive}. As such, robots could potentially be a valuable addition to children's mental wellbeing assessment pipeline (Figure \ref{pipeline} (b)) possibly eliciting more candid responses from the children that accurately represent their feelings. In addition, robots' multimodal sensing capabilities can be leveraged to identify non-verbal cues that extends children's verbal responses, garnering insights from children's behaviour such as their facial expressions and vocal tone. Consequently, critical observations can be flagged for child mental health practitioners to facilitate a diagnosis so that necessary support, if needed, can be provided at an early stage \cite[see similar discussion in][]{robot_post}. 

\subsection{Digital Mental Health Support}

In recent years, mobile applications have emerged as a cost-effective and accessible means of delivering mental health interventions, offering a wide range of support to users on their personal devices. For instance, apps providing cognitive behavioral therapy (CBT), mood tracking, and emotional regulation strategies have gained popularity for their ability to scale and deliver interventions to underserved populations \cite{Guracho2023}. Mobile apps specifically designed for children, such as those targeting emotional self-awareness and mood monitoring, have shown potential to engage young users and offer additional support outside of traditional clinical settings \cite{Punukollu2019}. However, concerns regarding the lack of regulatory oversight, as well as limited evidence on the effectiveness and safety of many commercially available apps, present significant challenges in their widespread adoption and clinical recommendation \cite{Grist2017}. Socially Assistive Robots (SARs) present an alternative approach, offering a unique advantage through their physical cues of embodiment and interactive capabilities \cite{robot_post}. Unlike mobile apps, SARs engage users in multimodal interactions, incorporating voice, gestures, and facial expressions to foster rapport and elicit richer self-disclosures \cite{robot_post,spitale_acii22,Spitale2024,dISC2024}. The embodied nature of SARs allows for more naturalistic social interactions, which is particularly beneficial in mental health settings where trust and empathy are critical \cite{s21155166}. This interactive and dynamic form of engagement is especially important when working with children, who often require a playful and empathetic approach to feel comfortable discussing their mental state \cite{Rudenko2024}. Therefore, the use of SARs in this context can be a powerful tool for clinicians aiming to assess and support the emotional well-being of children in a non-threatening, child-friendly manner.

\subsection{Child robot interaction (CRI) in mental health.} 

Exploring health and wellbeing applications has been an emerging frontier in CRI literature. Studies have focussed on how SARs have been used to provide emotional support for a variety of health and wellness-related situations \cite{gamborino2019mood,rossi2022using,henkemans2017design,meghdari2018arash,bethel2016using,abbasi2022can}. For example, Gamborino et al \cite{gamborino2019mood} have used SARs for the assessment and enhancement of mood in children. Rossi et al \cite{rossi2022using} have investigated how robots have been used to alleviate stress and anxiety in children in hospital consultations.  Henkemans et al \cite{henkemans2017design} have used robots to provide motivation to children for self-management of diabetes. Meghdari et al \cite{meghdari2018arash} have observed that social robots provide comfort to pediatric cancer patients. 
Robots have also been used to investigate cases of bullying in children \cite{bethel2016using}. More recently, SARs have been successfully employed for the evaluation of mental wellbeing in children \cite{abbasi2022can}. However, all the aforementioned works have involved children's physical presence in unfamiliar settings like hospitals and research laboratories which might affect children's comfort levels and also contribute towards their hesitancy in opening up. 

\subsection{Remote access to robots.}
Having remote access to SARs might help in reducing the stress among participants that can be attributed to unfamiliar environments such as research labs and hospitals. For example, in adults, Laban et al  \cite{Laban_blt_2023} have observed that interactions with online mediated robots have promoted enhancement of mood and self-disclosure among their participants and that adults who are experiencing negative emotions tend to disclose more to a social robot despite being remote \cite{Laban2023}. In the case of children, speaking with a social robot online has encouraged diabetic children to maintain a diary \cite{van2014remote}. Lytridis et al \cite{lytridis2020distance} have used robots to provide personalised support and special education to children with ASD, without in-person interaction with the robot.
Pasupuleti et al \cite{pasupuleti2022exploring} have employed teleconferencing with a robot to promote good hand hygiene among children. Calvo et al \cite{calvo2022understanding} have demonstrated how repeated exposure to online social robots promotes trust among children in a storytelling task. Previous HRI studies have also reported that co-location with the robot does not necessarily improve the objective of the experimental tasks under consideration. For example, Gittens et al \cite{gittens2022zenbo} have shown that conducting HRI sessions online does not negatively affect the interaction experience. Kennedy et al \cite{kennedy2015comparing} have shown that the robot's physical presence did not improve children's learning. Hence, having a remote robot platform can mitigate the logistical constraints associated with CRI, enabling better accessibility for potential longitudinal interactions, and thus encouraging timely and consistent support in sensitive matters related to children's mental wellbeing.

\subsection{Longitudinal CRI.}
Long-term interactions facilitate an evaluation of the influence of novelty effect \cite{smedegaard2019reframing} on the child-robot interaction experience.  
For instance,  Kozima et al \cite{kozima2007longitudinal} have observed that long-term interactions with robots have elicited pro-social behaviour among preschoolers.
Abdelmohsen et al \cite{abdelmohsen2021virtual} have shown that longitudinal sessions with virtual robots have improved conversational skills in children with autism.
Russel et al \cite{russell2021use} have shown how robot-assisted intervention has helped in the reduction of stress and better coping among children with cystic fibrosis. 
Studies have also shown how the attention and engagement of children decrease over time upon prolonged interaction with robots. For example, Serholt et al \cite{serholt2016robots} have shown that the social responses from children had reduced over multiple interactions with their robot tutors. Salter et al \cite{salter2004robots} reported that children showed increased boredom after prolonged exposure to the robots in their schools. As such, understanding the long-term interactions between robots and children can provide compelling insights into developing personalised intervention strategies that account for individual variability, especially in the context of mental wellbeing assessments.

\subsection{Gender differences in mental health and HRI.} 
With regard to mental health, gender-based variability has been observed in children about their perceived wellbeing. For example, societal constructs concerning masculinity have been shown to dictate boys' attitudes towards mental health and seeking help when in need \cite{pearson2023masculinity}. Boys have also shown to have decreased awareness in relation to their mental health \cite{chandra2006stigma} and have increased stigma associated with mental health concerns like depression as compared to girls \cite{lindsey2010family}. Even in the field of HRI, the gender of the participants seemed to influence their perceived perception of the robot and their interaction experience. For example, Mutlu et al \cite{mutlu2006task} have shown that the male participants have reported higher self-perceived positive affect when interacting with a robot as compared to the female participants. Further, Strait et al \cite{strait2015gender} showed a difference in perceived ratings between males and females in relation to changes in verbal and non-verbal behaviours of the robot (i.e. women reacted favourably to increased politeness from the robot). Therefore, in order to develop a holistic understanding of the mental wellbeing of the current and future generations, it is important to account for gender-related differences when designing robot-assisted assessment tools for children.

\subsection{Research Questions and Contributions}

Children differ widely in their attitudes towards technology which might affect their interaction with a social robot \cite{jackson2008culture}. This is of specific importance when encouraging them to open up about their wellbeing. It has also been shown that wellbeing is influenced by multiple factors (family problems, financial issues, academic workload etc.) and is subject to change without any warning or discerning pattern \cite{gromada2020worlds}. Thus, we investigated whether and how online interaction (using Zoom or Skype) with a social robot can be used for the longitudinal assessment (three sessions over the course of 3-6 weeks) of mental wellbeing \textbf{(RQ1)}. Longitudinal CRI studies have also observed that habituation with the robot leads to children losing interest in the interaction, causing less engagement and attention to tasks \cite{salter2004robots}. Thus, we investigated how children's perception of the robot changed over the course of the longitudinal interaction \textbf{(RQ2)}. Specifically, we also investigated whether remote interactions with the robot affect the children's perception of the robot as compared to co-location  \textbf{(RQ2.1)}. Gender-based differences have also been observed with regard to participants' attitudes towards mental health \cite{chandra2006stigma} and their HRI experience \cite{strait2015gender}. Thus, we also conducted a supplementary exploratory investigation into whether and how gender affects the robot-assisted longitudinal assessment of the mental wellbeing of children and their perception of the robot's behaviour and capabilities \textbf{(RQ3)}. 

By undertaking a study to investigate these research questions, our work provides the following contributions to child robot interaction research field: Our work is the first study that (I) investigates whether remote and online interactions with a social robot (using Zoom or Skype) can be used for the evaluation of mental wellbeing in children; (II) compares how child mental wellbeing and user perceptions of the robot change longitudinally across time and sessions (3-6 weeks, 3 sessions at least a week apart); (III) explores whether and how child mental wellbeing and user perceptions of the robot vary with gender.

\begin{figure}
	\centering
	\includegraphics[width= 0.7\textwidth]{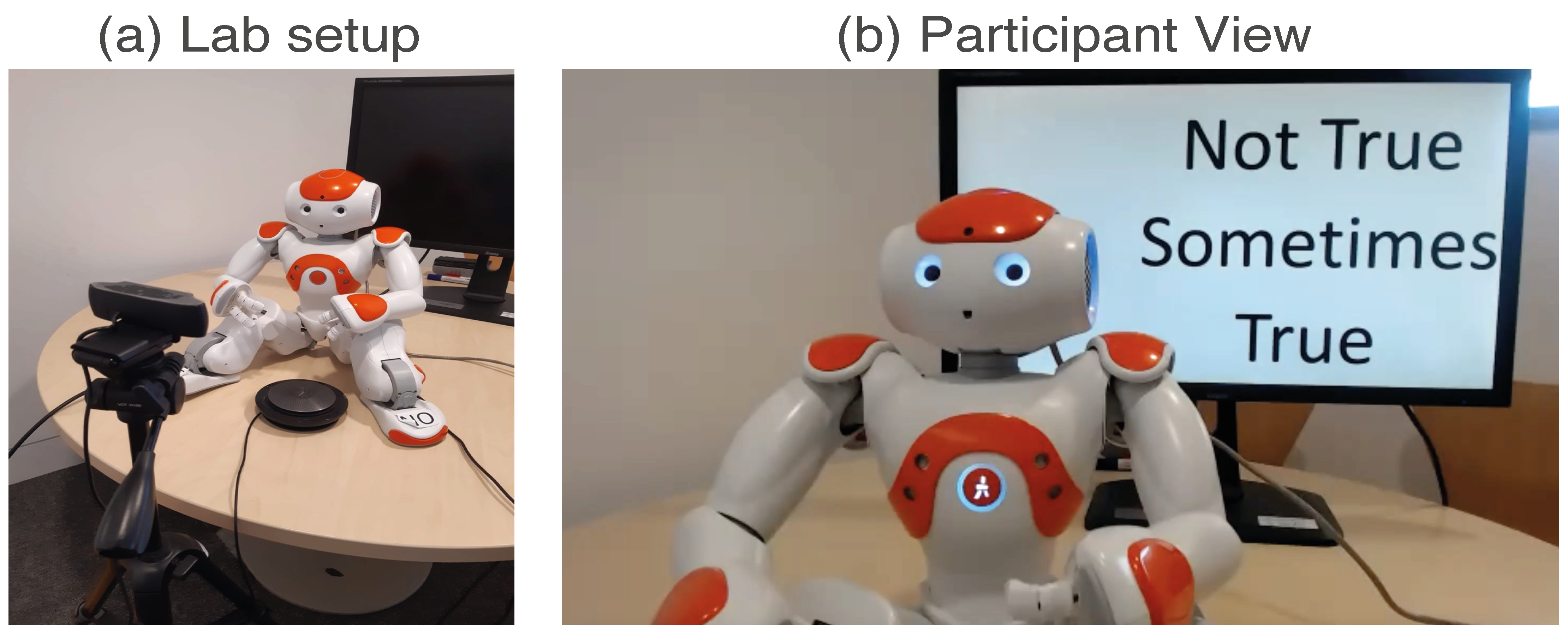}
	\caption{(a) Researcher view of the experimental setup, (b) Participant view of the interaction.}
	\label{setup}
\end{figure}
\section{Methods}
The following section describes the study design, participants, experimental protocol and data analyses conducted in this work.

\subsection{In-lab Study}
The experiment design for the present work was motivated by our prior investigation examining whether and how a humanoid robot can be used for the evaluation of mental wellbeing of children in a dyadic physical lab setting \cite{abbasi2022can,abbasi2023humanoid, abbasi2024analysing}. In our in-lab study, we collected data from 41 children\footnote{In order to balance out the gender of the participants, we collected data from about 13 more participants (as mentioned in \cite{abbasi2024analysing}) after the 28 participants mentioned in \cite{abbasi2022can}} (21 girls, 20 boys, mean age = $9.58 y.o.$, SD = $1.45 y.o.$). The participants were recruited from Cambridgeshire, United Kingdom. The experimental protocol consisted of the robot following a pre-scripted interaction and conducting the following five tasks sequentially: (i) Task 0: Icebreaking task; (ii) Task 1: Recall of a happy memory and a sad memory; (iii) Task 2: Short Moods and Feelings Questionnaire (SMFQ) \cite{angold1995development};  (iv) Task 3: Picture task inspired by the Child Apperception Test (CAT) \cite{bellak1949children};  and (v) Task 4: three subscales from the Revised Child Anxiety and Depression Scale (RCADS) \cite{chorpita2015revised}. Further details regarding the experimental setup and procedure can be found in \cite{abbasi2022can},\cite{abbasi2023humanoid} and \cite{abbasi2024analysing}.

In order to compare whether children's responses to a robot that is
remotely connected via Zoom or Skype differ from the children's responses to a robot that is co-located, in the context of mental wellbeing assessment,  a similar experimental protocol and tasks were adopted for the first session of this study. In this paper, we will refer to the study that used online interactions with the robot as the "remote study" while our previous study undertaken in the lab introduced in \cite{abbasi2022can,abbasi2023humanoid,abbasi2024analysing} will be referred to as the "in-lab study". It should be noted that new participants were recruited for the remote study in order to ensure that no individual participated in both the remote study and the in-lab study.

\subsection{Remote Study}
The experiment protocol followed during the  remote study has been mentioned below. It should be noted that in order to make valid comparison with our previous works \cite{abbasi2023humanoid,abbasi2022can,abbasi2024analysing}, we have followed the same experimental protocol in the remote study, only changing the mode of interaction from in-person to online.

\subsubsection{Participants} 
In planning for a repeated measures analysis to detect a medium effect size, a priori power analysis was conducted using G*power software \cite{RefWorks:410}, accounting for three groups and three repetitions per participant. Parameters included a medium effect size \textit{f}=0.25, an \(\alpha\) error probability of 0.05, desired power (1 - \(\beta\)) of 0.80, correlation among repeated measures of \textit{r} = 0.5, and a nonsphericity correction of \(\epsilon\)=0.75, indicating a moderate expectation of sphericity violation. The analysis recommended a total sample size of 36 participants to achieve an actual power (1 - \(\beta\)) of approximately 0.83, with a noncentrality parameter \(\lambda\)=10.13, critical F-value of 3.51, and degrees of freedom adjusted for the anticipated non-sphericity (numerator \textit{df}=1.5, denominator \textit{df}=49.5). This conservative approach ensures robustness against potential violations of sphericity and realistic assumptions about within-subject correlations, adequately powering the study to detect the specified effect size. To accommodate potential dropouts and ensure we have enough data for the supplementary analyses based on participants' gender, we have increased the sample size by 10\%, aiming for at least 20 units in each cell for drawing statistical inferences \cite{Simmons2011}. Further stratifications of the data beyond the three tertiles, such as those based on gender, are complementary to the main analysis presented in this paper. The final sample size was determined following previous CRI studies yielding similar results with long-term experimental designs \cite[e.g.,][]{Ahmad2017}, showing that a final sample size of 40 participants is appropriate.


42 individuals between the ages of 8 and 13 years old were recruited using one of the two recruitment approaches: (a) advertising using online flyers that were circulated across the schools, and (b) snowball sampling by leveraging contacts among the research team. Informed written consent was obtained from the parents/guardians prior to the participation of their children. The recruitment criteria involved exclusion on the basis of any existing neurological or psychological conditions declared by the parents/guardians. The recruitment criteria also included that the children were fluent in English which was the medium of communication of the study. This was confirmed based on the language of instruction in their respective schools to be English. The parents were asked to verify this on the consent form. Two participants were not included in the study mentioned here: one due to technical difficulties and another because of the inability to participate in the experiment sessions. Thus, the final sample consisted of 40 participants (20 girls and 20 boys, mean age = $10.04$, SD= $1.58$). One participant turned 14 during the course of the study sessions, however, was included in the study sample as they were 13 at the time of recruitment. The participants were from different parts of the world (Tajikistan: 13, UK: 8, Singapore: 7, Qatar: 6, India: 3, Other: 3) and interacted with the robot remotely using Zoom or Skype.  The study was approved by the Cambridge Psychology Research Ethics Committee at the University of Cambridge. 

\subsubsection{Setup}
The study was conducted across two locations: (1) a meeting room in the university (Fig. \ref{setup} (a)) where the robot was located and the researchers were present, and (2) remotely using either Zoom or Skype
where the participants were located (Fig. \ref{setup} (b)). The lab setup included the robot sitting on a table, in a crouching position with its hands on its legs and the Logitech camera placed about 1.5 m away from the robot. The camera was used to display the robot to the participant on the Zoom or Skype platform. In order to enable good audio quality during the experiment session, a Jabra disc microphone was placed in front of the robot. A screen for displaying visual aids was placed behind the robot. The child participant was requested to be seated in a quiet room and was asked to interact with the robot using smart devices like tablets, laptops etc. They were encouraged to use earphones for a good interaction experience, while the parents/guardians were requested to be within earshot, in case the children have any difficulties with regards to the technical aspects of the conference call like internet/wifi issues. They were asked not to help the children in relation to the experimental tasks and refrain from explaining the experiment activities.

\subsubsection{Robotic platform}
The study was conducted using the Nao humanoid platform from  Aldebaran Robotics\footnote{https://www.aldebaran.com/en/nao}. The Nao robot is equipped with human-like movement, visual sensors for object identification, and speech generation and is quite prevalent in CRI literature \cite{de2014child,arsic2022use}. The robot followed a `Wizard of Oz' paradigm in which the interaction dialogues were pre-programmed and did not change with the participants. However, the experiment flow, i.e. the continuation of the activities as to when the robot should respond or proceed with the subsequent activity, was determined by the researcher. This was done so that any interruptions to the experiment session due to the low quality of the conference call or internet issues do not affect the experiment tasks. Only one wizard who was a researcher conducted all the experiment sessions for all the participants. 

\subsubsection{Experiment protocol}
The researchers sent out the link for the video conference call prior to the experiment sessions. The sessions' dates and times were determined based on the children's availability. The children interacted with the robot for three sessions at least one week apart. For every interaction session, each child was requested to log in to their preferred video conferencing platform (Zoom or Skype), prior to the experiment session. The experimenter also joined the call using the camera and audio from their laptop and welcomed the child to the session. The experimenter described certain housekeeping rules in relation to the experiment session, including asking the child to speak clearly and sufficiently loudly and telling them about the regular breaks in between the activities. They were also instructed that they could pause the interaction, ask the robot to repeat what it was saying, and stop or skip parts of the activities, in accordance with how they felt. Once the child was comfortable and ready to begin the interaction, the experimenter switched the camera of Zoom or Skype to the Logitech camera placed in front of the robot. This was done so that the experimenters were no longer in the frame of view of the participants. In some cases, the experimenter might also switch the audio input from the laptop computer to the Jabra microphone placed in front of the robot, depending on what the participant felt comfortable with in terms of the audio quality of the conference call. The following are the steps that were taken to ensure that ethical practices during the child-robot interaction sessions:
\begin{enumerate}
\item Communication of study procedures: The experimenters made efforts to ensure that the study procedures were communicated with children in an age-appropriate manner. This was done verbally at the start of the sessions as well as throughout the session. The experimenters used simple and non-technical language to explain the experimental sessions and the role of the robot without delving into the study tasks in order to avoid priming and preparedness for answers. They ensured the children understood that they were not required to perform any task they did not want to. 
\item Study tasks and breaks: After each study task, the robot would ask the participants to see if they wanted to continue or take a break. If the children decided to take a break, the experimenter would come back into the field of view of the children and ensure that they felt comfortable resuming the study when they were ready to do so. The practice of allowing the children to take breaks and check in was designed in order to maintain a comfortable and supportive environment for the children.
\item Informed consent and assent: In addition to obtaining written consent from the parents, the study employed a child assent process to ensure that the children were comfortable with their participation. The experimenters verbally reassured that they could stop the session at any time. This opportunity to withdraw was clearly communicated to the children and they were reminded that they could take breaks and discontinue the interaction at any point they wished.
\item Ethical practices for child-robot interaction: Ethical considerations were at the forefront of the study design. Several measures were taken to minimise any risk of harm:
\begin{enumerate}
    \item Safety of the robot: The robot was tested and designed with the children’s safety in mind. It was ensured that the robot was safe to interact with and any potential for distress was minimised by pre-testing the interactions.
    \item Supervision and monitoring: All interactions were monitored by research staff to address any concerns immediately. The children were encouraged to express their feelings and any signs of discomfort were addressed promptly.
    \item Informed parents: Parents were informed about the nature of the study prior to the experiment sessions through the participant information sheet. They had the option to contact the research staff for any clarification/copy of the study tasks in order to ensure transparency and ongoing consent. They were also informed that they could contact the research staff if they would like to discuss the responses of their wards.
\end{enumerate}
\item Recording and privacy: The children were made aware that they were being recorded and they had the option to end the call as mentioned previously.

\end{enumerate}


The first interaction session started with the robot introducing itself, waving to the child, telling some jokes and playing tic tac toe with the child. The second interaction session began with the robot giving a brief introduction again about the session, waving to the child, and telling the children some jokes. Finally, the third interaction session started with the robot introducing the session, waving and playing a guessing game with the children. The decision to conduct three sessions was informed by prior literature \cite{nalin2012children, Ahmad2017} to capture initial reactions as well as allow for a more nuanced understanding of how children would interact with the robot differently over time. For all the sessions, the initial introduction task ended with the robot asking the children ice-breaking questions like "how was your day?", in order to get the children comfortable with the robot and its capabilities. After the introduction activity, the robot proceeded to perform the tasks as described in the next section (Sec. \ref{sec:studytask} Study tasks). The activity flow was controlled by the experimenter who was responsible for when the robot should proceed with the task. This was determined in accordance with when the child finished with their responses so that interruptions due to internet connection or the quality of the conference call did not affect the completion of the experiment tasks. 

\subsubsection{Study tasks}
\label{sec:studytask}
In order to investigate whether co-location with the robot affects wellbeing assessment in children, we followed the experiment tasks of our prior work \cite{abbasi2022can,abbasi2024analysing} which consisted of measures that are established standards in HRI and child psychology \cite{cervin2022multi,bremner2016personality, di2019assessment}. The first session consisted of the following tasks: \\
\textbf{Task1:}  The robot asked the children to recall a recent sad and happy memory. If the children responded with a recollection of either of the memories, the robot would then further ask the children to draw the memory on either the whiteboard of the Zoom platform or on a paper present in front of the child.\\
\textbf{Task2:} The robot administered the Short Moods and Feelings Questionnaire (SMFQ) \cite{angold1995development,sharp2006short}. The robot said aloud statements like "You felt lonely" from the questionnaire and the child could respond by verbalising their responses. The display screen was used to project the response ratings "Not true", "Sometimes" and "True" and the child was instructed by the robot to pick one of the response ratings in accordance with how they were feeling in the last two weeks.\\
\textbf{Task3}: The robot conducted a picture-based task which was inspired by the Child Apperception Test (CAT) \cite{bellak1949children}. The robot showed picture 7 from the CAT in the first session, using the computer screen and helped the children to form a narrative by asking questions like "What do you think is happening in this picture?", "What do you think happened before this picture?" and "What do you think happened after this picture?". The task was used to determine children's wellbeing based on the children's description of the pictures and the narrative they came up with. Similar tasks have also been used previously in HRI \cite{bremner2016personality}.\\
\textbf{Task 4:} The robot administered three subscales (generalised anxiety, panic disorder and low mood) from the Revised Child Anxiety and Depression Scale (RCADS) \cite{chorpita2015revised}. The robot made statements in accordance with the items of the questionnaire like "When you have a problem, your heart beats really fast". The children were instructed to respond using one of the response ratings "Never", "Sometimes", "Often", and "Always" that were displayed on the computer screen. 
\par For study sessions 2 and 3, the order of the tasks was changed in the following manner: (1) \textbf{Task 1}- recalling a recent happy and sad memory, (2) \textbf{Task 2}- the robot conducted the RCADS, and (3) \textbf{Task 3}- the robot administered the picture based task (picture 9 from the CAT in session 2 and picture 10 from the CAT in session 3 \cite{bellak1949children}). The SMFQ task was only present in the first session. 
The experiment sessions concluded with the robot either playing a game with the child (session 1 and session 2) or telling some jokes to the child (session 3). All three sessions concluded with the experimenter conducting the user perception questionnaires. Since one of the aims of this work is to understand whether co-location affects the perception of the robot for mental wellbeing assessment, we adopted the user perception questionnaires employed in our prior work mentioned in \cite{abbasi2023humanoid}. The first questionnaire was inspired by the questionnaire in \cite{murray2022learning} to measure the children's perception of the robot as a confidante. Open-ended questions and statement 10 ("Talking with the robot was similar to talking with a stranger") were excluded. The children could respond using Likert ratings: \emph{(1) Strongly dislike, (2) Somewhat dislike, (3) Neither like nor dislike, (4) Somewhat like, and (5) Strongly like}. Children's anxiety towards robots was measured using a questionnaire inspired by the RAS \cite{nomura2006measurement}. To enable clarity and better understanding for children, statement 3 of RAS was modified to "The robot may be unable
to understand my stories", statement 10 was changed to "Whether the robot understands what I am saying", and statement 11 was modified to "I may be unable
to understand what the robot is saying". Children could respond by verbalising one of the response ratings: \emph{(1) I do not feel anxiety at all,
(2) I hardly feel any anxiety, (3) I don’t feel much anxiety,
(4) I feel a little anxiety, (5) I feel much anxiety, and (6) I feel anxiety very strongly}. Finally, the third user perception questionnaire was inspired by the Godspeed questionnaire \cite{bartneck2009measurement}. We used the likeability and perceived intelligence subsections of the questionnaire and the children could respond using the questionnaire response ratings ranging from one to five (one corresponded with the negative characteristic like "ignorant" and five corresponded with the positive characteristic like "knowledgeable"). It should be noted that in order to maintain a uniform interaction experience for all participants, the robot behaviour was consistent. In other words, the robot did not build off the prior interactions with the participants.

\subsubsection{Measures}
\label{sec:measures}
The measures used in this work are as follows:\\
\textbf{Pre-study questionnaires:} Prior to the first experiment session, Qualtrics online links were sent to parents of the participating children which consisted of two questionnaires to be answered in relation to the children's wellbeing - (1) Self-report questionnaire (to be answered by the child) consisting of three sub-scales (generalised anxiety, panic disorder and low mood) of the RCADS (Cronbach's $\alpha$ = 0.91, mean score = 11.23 $\pm$ 9.19), and (2) Parent-report questionnaire (to be answered by the parent) consisting of three subscales from (generalised anxiety, panic disorder and low mood) of the RCADS (Cronbach's $\alpha$ = 0.89, mean score = 9.08 $\pm$ 7.54). The links to these questionnaires were attached in the email sent after receiving an expression of interest in participation from the parents. The children were not allowed to proceed with their first session until these questionnaires were completed. Total scores for self-report and parent-report questionnaires were computed separately using the rating scores of the RCADS (never = 0, sometimes = 1, often = 2, always = 3).\\
\textbf{In-study questionnaires:} Our aim in this work is to understand the feasibility of using remote interaction with the robot for wellbeing assessment of children. As such, we have only included children's responses to the SMFQ (task 2 in the first session; Cronbach's $\alpha$ = 0.85, mean score = 6.63 $\pm$ 5.44) and the RCADS (task 4 in the first session and task 2 in the second and third sessions; Cronbach's $\alpha$ = 0.92, mean score = 12.57 $\pm$ 10.17) for analyses in this work. We computed total scores from the children's responses to the SMFQ (not true= 0, sometimes= 1, and true = 2) and the RCADS (never = 0, sometimes = 1, often = 2, always = 3). \\
\textbf{Post-session questionnaires:} The user perception questionnaires were conducted by the experimenter after every experiment session. Total scores were computed for the first user perception questionnaire (Cronbach's $\alpha$ = 0.65, mean score = 34.42 $\pm$ 3.74), inspired by \cite{murray2022learning} to understand children's perception of the robot as a confidante, using the questionnaire ratings of \emph{1: Strongly dislike, 2: Somewhat dislike, 3: Neither like nor dislike, 4: Somewhat like, and 5: Strongly like}. For the second user perception questionnaire (Cronbach's $\alpha$ = 0.84, mean score = 18.95 $\pm$ 8.27), inspired by the RAS \cite{nomura2006measurement} to understand the anxiety attributed to the robot by the children, total scores (referred to as "robot anxiety") were computed using the response ratings of \emph{1: I do not feel anxiety at all,
2: I hardly feel any anxiety, 3: I don’t feel much anxiety,
4: I feel a little anxiety, 5: I feel much anxiety, and 6: I feel anxiety very strongly}. Finally, total scores from the third user perception questionnaire (Cronbach's $\alpha$ = 0.82, mean score = 47.59 $\pm$ 3.33), inspired by the Godspeed questionnaire \cite{bartneck2009measurement} to understand the user's likeability of the robot and the perceived intelligence of the robot, were computed using the questionnaire ratings of \emph{1- negative characteristic like "Unkind" to 5- positive characteristic like "Kind"}.

\subsection{Construct Validity Analyses}
\label{sec:validity}

In \cite{Abbasi2024}, we conducted validity analyses for our in-lab study described in \cite{abbasi2022can,abbasi2024analysing}. We have followed the same approach for the data collected from the online remote study described in this work. We conducted construct validity analysis through confirmatory principal component analysis (PCA) to validate established psychological assessment tools such as the SMFQ and RCADS in longitudinal CRI research presented here. This was done to ensure they accurately measure the intended constructs and to comprehend their limitations within CRI contexts. Similarly, we applied the same analysis to both self-reported and parent-reported RCADS to achieve a deeper understanding of model fit and to discern the differences more clearly.

\begin{table}[h!]

\begin{tabular}{ll}
\hline
\textbf{SMFQ Item} & \textbf{Loading on} \\
   & \textbf{Component}\\
\hline
1. You felt miserable or unhappy  & 0.73 \\
2. You didn’t enjoy anything at all   & 0.50 \\
3. You felt so tired I just sat around and did nothing & 0.60 \\
4. You were very restless  & - \\
5. You felt you were no good any more   & 0.83 \\
6. You cried a lot   & 0.50 \\
7. You found it hard to think properly or concentrate   & 0.47 \\
8. You hated yourself   & 0.82 \\
9. You were a bad person   & 0.62 \\
10. You felt lonely  & 0.58 \\
11. You thought nobody really loved you  & - \\
12. You thought you could never be as good as other kids & 0.69 \\
13. You did everything wrong & 0.77 \\
\hline
\end{tabular}
\caption{Loadings of SMFQ Items}
\label{tab:component_loadings_SMFQ}
\end{table}

\subsubsection{SMFQ}
A confirmatory PCA was conducted with the 13 SMFQ items to examine the underlying structure. After inspecting the correlation matrix, it was confirmed that there was no threat of multicollinearity and that discriminant validity of variables was maintained, with the highest correlation at .74. The KMO measure of .714 affirmed sampling adequacy for the analysis, with all individual KMO values above .5. Bartlett’s test of sphericity, 
\(\chi^2\)(78) = 224.40, \textit{p} \(<\) .001, indicated the data was suitable for PCA. The PCA revealed one component with an eigenvalue of 4.951, explaining 38.084\% of the variance. The scree plot justified retaining this single component. Table \ref{tab:component_loadings_SMFQ} shows the loadings of the different items of the SMFQ.
High loadings were observed for item 5 (.83) and item 8 (.82), while items with the lowest loadings were items 6 (.50) and 2 (.50). Items 4 and 11 were not loaded, suggesting their loadings were suppressed since they were below the .4 coefficient threshold considered significant for this analysis.

\begin{table}[h!]
\centering

\begin{tabularx}{\columnwidth}{Xccc}
\hline
\textbf{Item} & \textbf{Robot-Administered} & \textbf{Self-Reported} &  \textbf{Parent-Reported} \\
& \textbf{Loadings} & \textbf{Loadings} & \textbf{Loadings}\\
\hline							
1. You/I/My child worry(ies) about things  	&	0.62	&	-	&	0.49	\\
2. You/I/My child worry(ies) that something awful will happen to someone in your/my/the family  	&	0.67	&	0.47	&	0.50	\\
3. You/I/My child worry(ies) that bad things will happen to you/me/him or her  	&	0.64	&	0.67	&	0.57	\\
4. You/I/My child worry(ies) that something bad will happen to you/me/him or her   &	0.64	&	0.81	&	0.79	\\
5. You/I/My child worry(ies) about what is going to happen  	&	0.64	&	-	&	0.73	\\
6. You/I/My child think(s) about death  	&	0.73	&	0.63	&	0.74	\\
7. When you/I/my child have(has) a problem, you/I/he or she get(s) a funny feeling in your/my/his or her stomach  	&	0.51	&	0.56	&	0.59	\\
8. You/I/My child suddenly feel(s) as if you/I/he or she can't breathe when there is no reason for this  	&	0.53	&	-	&	-	\\
9. When you/I/my child have(has) a problem,  your/my/his or her heart beats really fast  	&	0.60	&	0.74	&	0.43	\\
10. You/I/My child suddenly start(s) to tremble or shake when there is no reason for this 	&	0.67	&	0.47	&	-	\\
11. When you/I/my child have(has) a problem, you/I/he or she feel(s) shaky 	&	0.64	&	0.70	&	-	\\
12. All of a sudden you/I/my child feel(s) scared for no reason at all 	&	0.68	&	0.50	&	-	\\
13. You/I/my child suddenly become dizzy or faint when there is no reason for this 	&	0.48	&	0.81	&	0.77	\\
14. Your/My/My child's heart suddenly starts to beat too quickly for no reason 	&	0.62	&	0.45	&	0.46	\\
15. You/I/My child worry(ies) that you/I/he or she would suddenly get a scared feeling when there is nothing to be afraid of 	&	0.57	&	-	&	0.41	\\
16. You/I/My child feel(s) sad or empty 	&	0.68	&	0.66	&	0.52	\\
17. Nothing is much fun anymore 	&	0.61	&	0.63	&	0.53	\\
18. You/I/ My child have(has) trouble sleeping 	&	0.54	&	0.63	&	0.44	\\
19. You/I/ My child have(has) problems with your/my/his or her appetite 	&	-	&	-	&	-	\\
20. You/I/My child have(has) no energy for things 	&	0.46	&	0.67	&	0.71	\\
21. You/I/My child are(am,is) tired a lot 	&	0.60	&	0.55	&	0.82	\\
22. You/I/My child cannot think clearly 	&	0.64	&	0.72	&	0.70	\\
23. You/I/My child feel(s) worthless 	&	0.61	&	0.81	&	0.85	\\
24. You/I/My child feel(s) like you/I/ he or she don't(doesn't) want to move 	&	0.60	&	0.82	&	0.75	\\
25. You/I/ My child feel(s) restless 	&	0.45	&	0.83	& -	\\
\hline							
\end{tabularx}
\caption{Component Loadings for Robot-Administered, Self-Reported and Parent-Reported RCADS. The item wordings were adapted to the respective mode of administration}
\label{table:rcads_loadings}
\end{table}

\subsubsection{RCADS - Robot administered}

A confirmatory PCA was conducted with the 25 robot-administered RCADS items to examine the underlying structure. After inspecting the correlation matrix, it was confirmed that there was no threat of multicollinearity and that discriminant validity of variables was maintained, with the highest correlation of .59. The KMO measure of .831 affirmed sampling adequacy for the analysis, with all individual KMO values above .5. Bartlett’s test of sphericity, 
\(\chi^2\) (300) = 1447.96, \textit{p} \(<\) .001, indicated the data was suitable for PCA. The PCA revealed one component with an eigenvalue of 8.79, explaining 35.16\% of the variance. The scree plot justified retaining this single component. Table \ref{table:rcads_loadings} shows the loadings of the different items of the robot-administered RCADS. High loadings were observed for several items, notably item 6 (.73) indicating strong relationships with the underlying factor, while the items with the lowest loadings was 25 (.45). Item 19 was not loaded, suggesting its loading was suppressed since it was below the .4 coefficient threshold for this analysis. 

\subsubsection{RCADS - Self reported}

A confirmatory PCA was conducted with the 25 self-reported RCADS items to examine the underlying structure. After inspecting the correlation matrix, it was confirmed that there was no threat of multicollinearity and that discriminant validity of variables was maintained, with the highest correlation of .83. The KMO measure of .65 affirmed sampling adequacy for the analysis, with all individual KMO values above .5. Bartlett’s test of sphericity, 
\(\chi^2\) (300) = 682.01, \textit{p} \(<\) .001, indicated the data was suitable for PCA. The PCA revealed one component with an eigenvalue of 9.05, explaining 36.21\% of the variance. The scree plot justified retaining this single component. Table \ref{table:rcads_loadings} shows the loadings of the different items of the self-reported RCADS. High loadings were observed for several items, notably item 25 (.83) indicating strong relationships with the underlying factor, while the items with the lowest loading was item 14 (.45). Items 1, 5,  8, 15, and 19 were not loaded, suggesting their loadings were suppressed since they were below the .4 coefficient threshold for this analysis.

\subsubsection{RCADS - Parents reported}

A confirmatory PCA was conducted with the 25 RCADS items to examine the underlying structure. After inspecting the correlation matrix, it was confirmed that there was no threat of multicollinearity and that discriminant validity of variables was maintained, with the highest correlation of .83. The KMO measure of .510 affirmed sampling adequacy for the analysis, with all individual KMO values above .5. Bartlett’s test of sphericity, \(\chi^2\) (300) = 811.35, \textit{p} \(<\) .001, indicated the data was suitable for PCA. The PCA revealed one component with an eigenvalue of 8.04, explaining 32.17\% of the variance. The scree plot justified retaining this single component. Table \ref{table:rcads_loadings} shows the loadings of the different items of the parent-reported RCADS.  High loadings were observed for several items, notably item 23 (.85) indicating strong relationships with the underlying factor, while the item with the lowest loadings was item 15 (.41). Items 8, 10, 11, 12, 19, and 25 were not loaded, suggesting their loadings were suppressed since they were below the .4 coefficient threshold for this analysis.

\subsubsection{Transferbility of the measurements}
The construct validity analysis revealed a dominant principal component underlying each of the scales, suggesting that each scale includes a consistent factor for measuring children's mental health in CRIs. However, not all items contributed equally to this factor. A comparison between robot-administered and self-reported RCADS highlighted the former's superior construct validity (i.e., demonstrating a better fit), suggesting that mediating scale items via robotic interaction enhances the validity of symptom disclosure. This is in line with previous transferability analysis conducted with data from CRIs of similar structure \cite{Abbasi2024}. The validity analysis not only elucidated the core factors measured by the scales in CRIs but also set the stage for the main analysis by establishing the superior construct validity of robot-administered assessments. This foundation enables a deeper exploration of how robotic mediation can improve psychological assessments. These results underpin the rationale for comparing wellbeing changes between tertiles for each session and across all sessions. The identification of a primary factor confirms that the scales can effectively distinguish between different levels of wellbeing measured using RCADS, making the Kruskal Wallis tests appropriate for identifying statistically significant differences between groups defined by tertiles of SMFQ scores.  The construct validity analysis ensures that the instruments used in the study are measuring what they are intended to measure with a high degree of accuracy, thereby lending credibility and support to the subsequent formal analyses of between-subject and within-subject differences.

\subsection{Data analyses}
\subsubsection{Data clustering.}
We used the responses from the SMFQ to cluster the participants into three tertiles (low, med and high) similar to the framework followed in \cite{abbasi2022can,abbasi2024analysing}. SMFQ thresholds were also taken into consideration in determining the children's categorisation into the wellbeing clusters. For example, if tertile categorisation resulted in a child with an SMFQ score of 9 being placed in med tertile, we included the child in the high tertile as the threshold for SMFQ is a total score of 8 \footnote{https://www.seattlechildrens.org/globalassets/documents/healthcare-professionals/pal/ratings/smfq-rating-scale.pdf}. Tertile edges were computed separately for girls and boys in order to ensure similar participant numbers across the three wellbeing clusters (low tertile, med tertile and high tertile). Since the SMFQ is used to monitor symptoms of mood disorders in children, children belonging to the low tertile and med tertile were very unlikely to obtain a diagnosis of mood disorders while children in the high tertile were more likely to obtain a diagnosis of mood disorders.

Following the tertile categorisation, these well-being clusters were used as independent variables to analyse their relationship with outcomes assessed by the RCADS. This approach, well-supported in mental health research, facilitates structured comparisons of how varying levels of mood, as indicated by SMFQ tertiles, influence specific outcomes such as anxiety and depression symptoms \cite{Shields2023,Johnson2021}. Given the distinct yet related constructs measured by SMFQ (general mood) and RCADS (specific anxiety and depression symptoms), this method allows us to assess how mood categories align with different levels of psychological symptoms. While a correlation between SMFQ and RCADS is typically expected \cite[e.g.,][]{lisa2022}, the CRI context, where the robot administered RCADS, required careful consideration, as we could not assume this correlation would hold in the same way as in traditional child mental health studies.

\subsubsection{Statistical analyses.}
Statistical analyses were performed using either the statistical toolbox in Matlab \footnote{https://uk.mathworks.com/products/statistics.html} or using IBM SPSS \footnote{https://www.ibm.com/spss}. Our primary analyses consisted of investigating wellbeing and user perception changes for the overall population (40 participants) across the varying levels of wellbeing (tertiles) as well as across the sessions. We also conducted a secondary exploratory analysis by further stratifying the data according to the gender of the participants. For this purpose, we employed non-parametric statistical analyses as our study data (RCADS, SMFQ and user perception responses) did not follow a normal distribution (Shapiro-Wilk test). We conducted Kruskal Wallis tests for between-subject analyses while investigating wellbeing or user perception changes between the tertiles for each session as well as across the tertiles for all the sessions combined. We employed Friedman's test to conduct within-subject analyses while comparing changes in wellbeing and user perception responses longitudinally across the experiment sessions. Between-subject pairwise comparisons while investigating change in responses across girls and boys, between the children's RCADS responses to the robot and the parent's RCADS responses to the pre-study questionnaire, and across the remote study data collected in this work and our in-lab study \cite{abbasi2022can,abbasi2023humanoid,abbasi2024analysing} were performed using Wilcoxon rank sum tests. Within-subject pairwise comparison between the children's responses to the children's RCADS responses to the robot and their self-report responses to the pre-study questionnaire was performed using the Wilcoxon Signed Rank test. 

To control for Type I errors due to multiple hypothesis testing, we applied the Holm-Bonferroni correction to p-values from the confirmatory hypothesis tests. This method sequentially adjusts significance levels for each hypothesis, prioritising smaller p-values to rigorously control Type I error while preserving sensitivity \cite{Aickin2011}. Both corrected and uncorrected p-values are reported in the results. 
Post-hoc tests for between-subject (Kruskal-Wallis test) and within-subject (Friedman's test) analyses and the pairwise comparisons (across each investigated feature) were corrected using Bonferroni corrections ($0.05/3$). 

Incorrect responses were replaced by the closest response rating by the researchers (e.g., a "no" response to the SMFQ statements was replaced with "not true"). Missed responses to the statements of the questionnaires during the experiment session were imputed with the mode of the responses for the participant under consideration, for that questionnaire. In the case of three participants, data (two participants for the self-report questionnaire and one participant for the user perception questionnaires for session 3) did not get recorded due to technical issues and was replaced with mean value across all participants\footnote{https://www.theanalysisfactor.com/seven-ways-to-make-up-data-common-methods-to-imputing-missing-data/}. However, the inputted data was not used to compute reliability scores (Cronbach's $\alpha$) and to evaluate the construct validity of the measurements mentioned in Sec. \ref{sec:measures} Measures and Sec. \ref{sec:validity} Construct Validity Analyses.

\begin{table}[h!]
\LARGE
 \centering
   \resizebox{0.8\textwidth}{!}
{\begin{tabular}{ |p{3cm}||p{3cm}|p{0.8cm}|p{3cm}| |p{0.8cm}||p{3cm}|p{0.8cm}| }
 \hline
 \multicolumn{7}{|c|}{(a) Remote study} \\ 
 \hline
	&	Overall	&	N	&	Girls 	&	N	&	Boys	&	N		\\	\hline
Low tertile	&	SMFQ <=3	&	15	&	SMFQ <=3	&	9	&	SMFQ <=4	&	8		\\	\hline
Med tertile	&	3<SMFQ <=7	&	11	&	3<SMFQ <=7	&	5	&	4<SMFQ <=7	&	4		\\	\hline
High tertile	&	SMFQ >7	&	14	&	SMFQ > 7	&	6	&	SMFQ > 7	&	8		\\	\hline
 \hline
 \multicolumn{7}{|c|}{(b) In-lab study \cite{abbasi2022can,abbasi2024analysing}} \\ 
 \hline
&	Overall	&	N	&	Girls 	&	N	&	Boys	&	N	\\	\hline
Low tertile	&	SMFQ <=2	&	16	&	SMFQ <=2	&	7	&	SMFQ <=2	&	9	\\	\hline
Med tertile	&	2<SMFQ <=4	&	12	&	2 < SMFQ <=5	&	9	&	2 < SMFQ <= 5	&	6	\\	\hline
High tertile	&	SMFQ >4	&	13	&	SMFQ score >5	&	5	&	SMFQ >5	&	5	\\	\hline
\end{tabular}}
\centering
\caption{Tertile edges and participant number (N) for the SMFQ scores for (a) remote study, and (b) In lab study \cite{abbasi2022can,abbasi2024analysing}.}
\label{SMFQ}
\end{table}

\section{Results}
\subsection{Tertile categorisation for identification of wellbeing clusters.}
The SMFQ data collected during the first session was used for tertile categorisation of the participants into three wellbeing clusters similar to \cite{abbasi2022can, jachens2019effort, munir2014occupational}. Table \ref{SMFQ} shows the tertile edges and number of participants for the different population groups (overall, girls and boys) for both the remote study conducted in this work and our in-lab study \cite{abbasi2022can,abbasi2024analysing}, across the varying wellbeing clusters (low tertile, med tertile and high tertile).

\begin{figure}[h!]
	\centering
	\includegraphics[width= 1\textwidth]{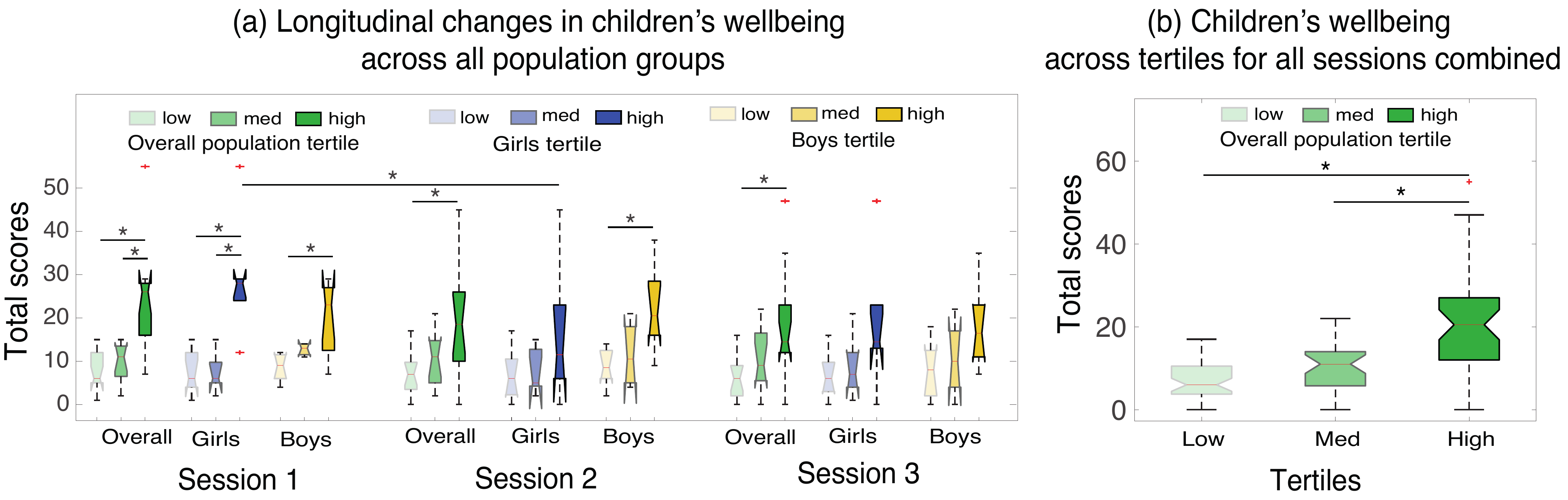}
	\caption{(a) Longitudinal changes in wellbeing (RCADS scores: dependent variables) across different population groups (overall, girls and boys) categorised into wellbeing tertiles (low, med and high; independent variables) using the SMFQ responses. (b) Children's wellbeing (RCADS scores) across tertiles for all sessions combined. $*p<0.05$ corrected.}
	\label{wellslong}
\end{figure}

\subsection{Effectiveness of online interactions with a social robot in wellbeing assessment (RQ1).}

Figure \ref{wellslong} (a) shows the longitudinal changes in the children's wellbeing across different population groups (overall shown in green, girls shown in blue and boys shown in yellow) and across different tertiles (low, med and high) for the three sessions of the study (independent variables: tertiles; dependent variables: RCADS responses).
For the overall population group, Kruskal Wallis H test has shown a statistically significant difference ($H(2) =18.98, p =	<.001$; $p=0.007$, Holm-Bonferroni corrected) in the RCADS responses for session 1 between the tertiles. Post-hoc pairwise comparisons using Dunn's tests, with a Bonferroni correction applied, have indicated statistically significant differences between low tertile and high tertile ($Mdn_{low}= 6, Mdn_{high}= 26, p = 0.00$) and between med tertile and high tertile ($Mdn_{med}= 11, Mdn_{high}= 26, p = 0.01$). 

In the case of the RCADS responses for session 2, Kruskal Wallis H test has shown a statistically significant difference ($H(2) =10.73, p =	0.005$; $p=0.010$, Holm-Bonferroni corrected) between the tertiles. Post-hoc pairwise comparisons using Dunn's tests, with a Bonferroni correction applied, have indicated statistically significant differences between low tertile and high tertile ($Mdn_{low}= 7, Mdn_{high}= 18.5, p = 0.004$). 

While considering the RCADS responses for session 3, Kruskal Wallis H test has also shown a statistically significant difference ($H(2) =12.02, p = 0.002$; $p=0.010$, Holm-Bonferroni corrected) between the tertiles. Post-hoc pairwise comparisons using Dunn's tests, with a Bonferroni correction applied, have indicated statistically significant differences between low tertile and high tertile ($Mdn_{low}= 6, Mdn_{high}= 18.5, p = 0.004$). 

We have also observed that while comparing across tertiles for all sessions combined for all the participants (Figure \ref{wellslong}(b)),  Kruskal Wallis H test has shown a statistically significant difference ($H(2) =41.29, p = <.001$; $p =0.007$, Holm-Bonferroni corrected) between the RCADS responses. Post-hoc pairwise comparisons using Dunn's tests, with a Bonferroni correction applied, have indicated statistically significant differences between low tertile and high tertile ($Mdn_{low}= 6, Mdn_{high}= 20.5, p = 0.00$) and between med tertile and high tertile ($Mdn_{med}= 11, Mdn_{high}= 20.5, p = 0.00$). This combined analysis has enabled us to identify overall trends and perform clearer comparisons between groups, thereby reinforcing the reliability of our findings. A similar pattern has emerged with the RCADS scores increasing for different levels of wellbeing (from low to high) across sessions (as seen in Fig. \ref{wellslong} (a)) or the overall population group as well as for all the sessions combined(Figure \ref{wellslong}(b)).

\begin{figure}[h!]
	\centering
	\includegraphics[width= 0.75\textwidth]{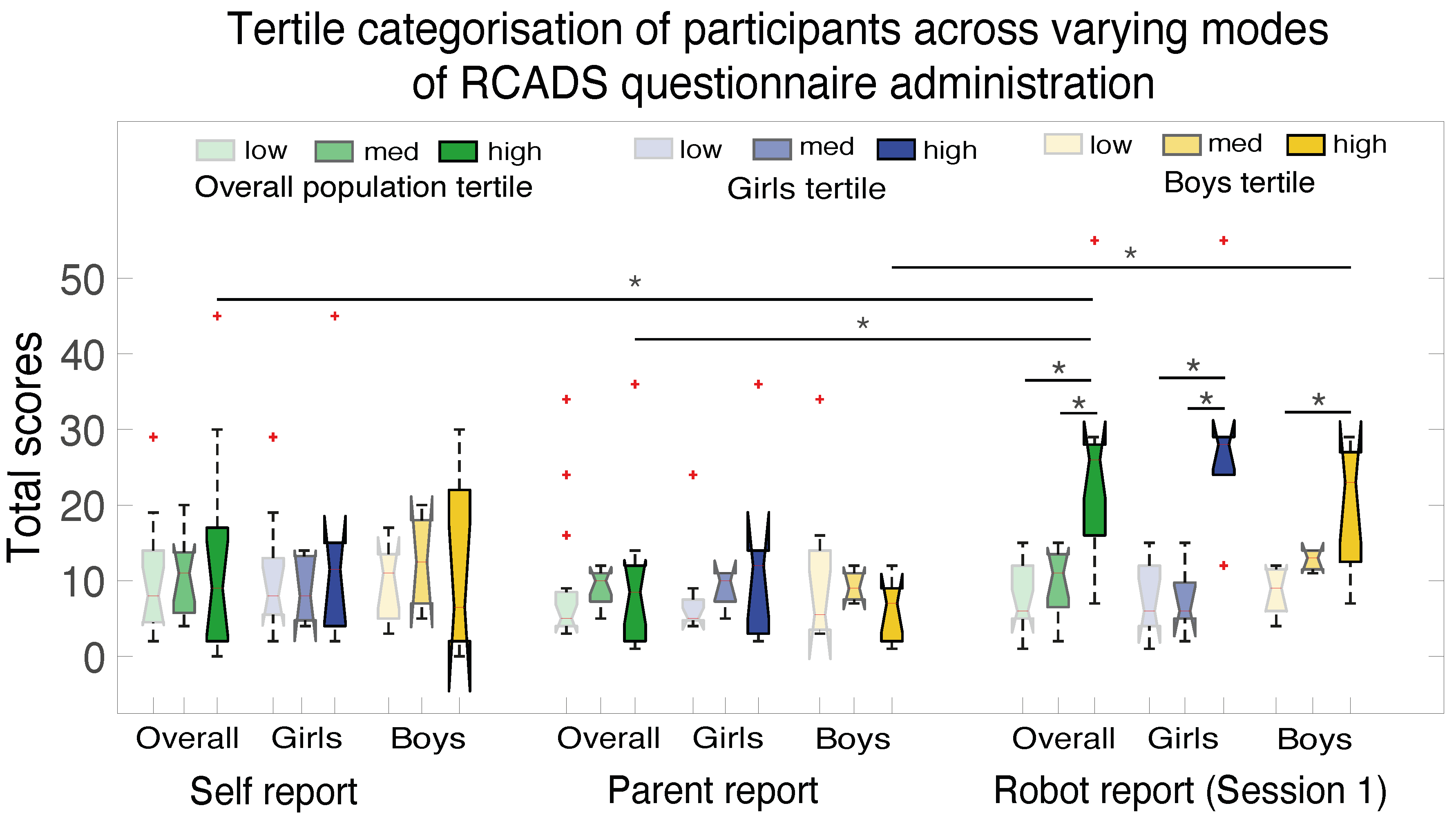}
	\caption{RCADS responses (dependent variables) across session 1 for the different modes of test administration from all population groups categorised into wellbeing tertiles (low, med and high; independent variables) using the SMFQ responses. $*p<0.05$ corrected.}
	\label{modesRR}
\end{figure}

Since self-report responses and parent-report responses for the RCADS were collected before the experiment sessions, we have compared the responses of the RCADS across the three modes of test administration (self-report, parent-report and robot-report for session 1). As seen from Figure \ref{modesRR}, Wilcoxon signed-rank test with a Bonferroni correction applied, has indicated a statistically significant difference between the self-report responses and the responses during session 1 with the robot for the high tertile ($Mdn_{self}= 9, Mdn_{robot}= 26, p = 0.04$). Further, Wilcoxon rank sum test with a Bonferroni correction applied has indicated a statistically significant difference between the parent-report responses and the responses during session 1 with the robot for the high tertile ($Mdn_{parent}= 8.5, Mdn_{robot}= 26, p = 0.04$). We have not found any statistically significant differences across the tertiles for the RCADS responses between session 1 for the remote study with the robot and the RCADS responses for our in-lab study \cite{abbasi2022can,abbasi2024analysing} for the overall population. 

In summary, our results show that across all three sessions, RCADS responses have been significantly higher in the high tertile than in the low tertile. RCADS responses have also been significantly higher for the high tertile when comparing between modes of questionnaire administration (robot-report, self-report and parent-report). 

\begin{figure}[h!]
	\centering
	\includegraphics[width= 0.65\textwidth]{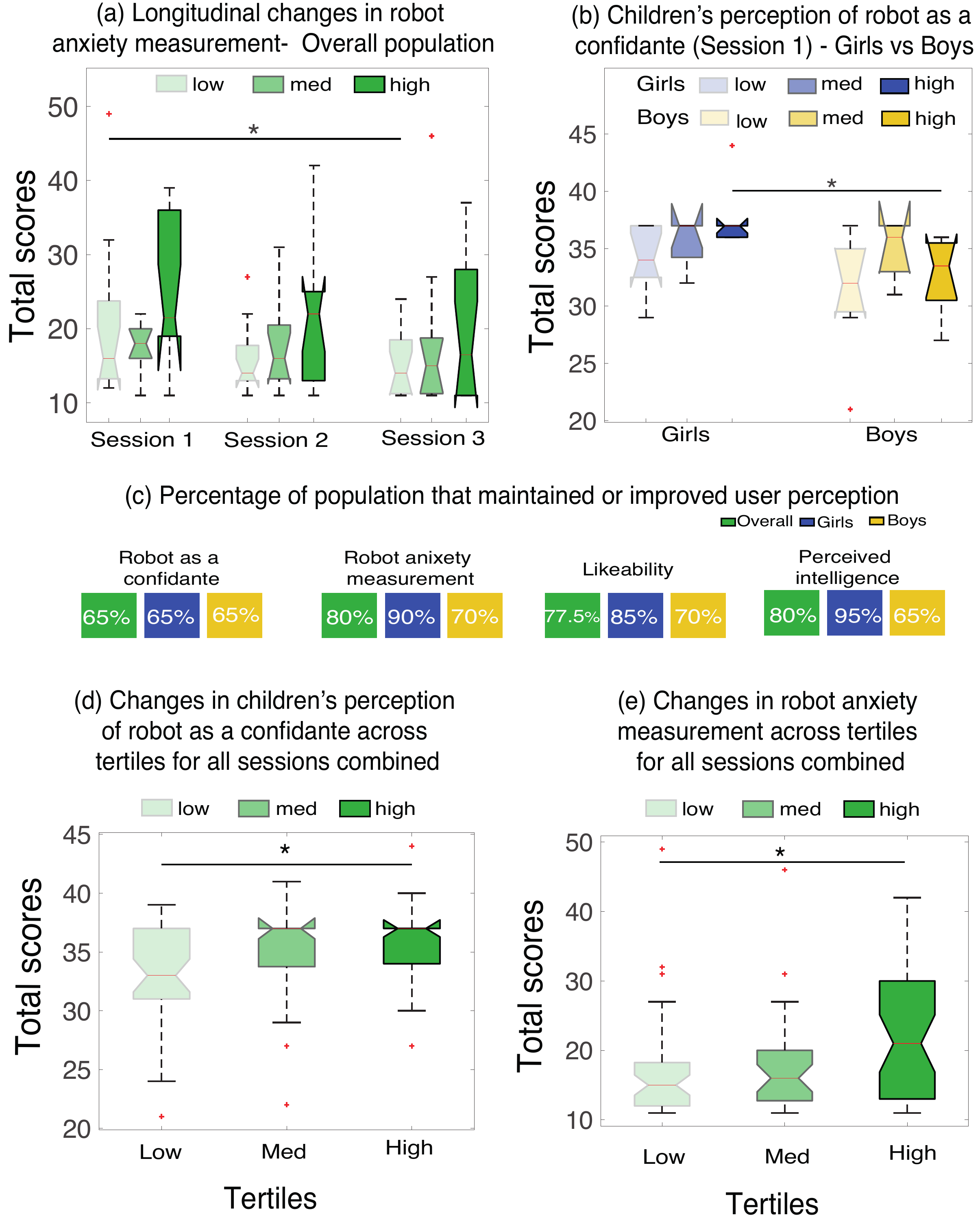}
	\caption{(a) Longitudinal changes in children's anxiety measurement attributed to the robot across all population groups (overall, girls and boys) categorised into wellbeing tertiles (low, med and high) using the SMFQ responses. (b) Gender-based variability in the role of the robot as a confidante in the remote study. (c) Total percentage of the population that had consistent or improved user perception. User perception changes across tertiles for all sessions combined for (d) perception of the robot as a confidante and (e) robot anxiety measurement. $*p<0.05$ corrected.}
	\label{userlong}
\end{figure}

\subsection{The effect of the longitudinal interactions and co-location on user's perception of the robot (RQ2).}
As seen in Figure \ref{userlong}, Friedman's test indicated a statistically significant difference in the robot anxiety measurement scores (dependent variable) for the overall population group belonging to the low tertile, across the three sessions (independent variable) $\chi^2 (2) =6.87, p = 0.032$. Post hoc analysis with Wilcoxon signed-rank tests was conducted with a Bonferroni correction applied, resulting in statistical significance between the robot anxiety measurement responses for session 1 and session 3 ($Mdn_{session1}= 16, Mdn_{session2}= 14, p = 0.041$). 
We have observed (Figure \ref{userlong} (c)) that out of 40 participants (20 girls and 20 boys), most participants have either maintained or improved their perception of the robot. We defined improvement as an average increase in the self-reported ratings of the robot's likeability, perceived intelligence, and role as a confidante, and an average decrease in the response ratings for the robot anxiety measurement i.e. the anxiety that the children attributed to the robot.

\par We did not find any statistically significant difference in the responses to the perception questionnaires (robot as confidante, robot anxiety measurement and likeability and perceived intelligence of the robot) across the tertiles throughout any of the sessions (session 1, session 2, and session 3). However, we have observed (Figure \ref{userlong} (d)) that the Kruskal Wallis H test has shown a statistically significant difference ($H(2) = 11.13, p =	0.004$; $p = 0.012$, Holm-Bonferroni corrected) in the children's perception of the robot as a confidante (dependent variable) across tertiles (independent variable) for all the sessions combined. Post-hoc pairwise comparisons using Dunn's tests, with a Bonferroni correction applied, have indicated statistically significant differences between low tertile and high tertile ($Mdn_{low}= 33, Mdn_{high}= 37, p = 0.004$). 

Kruskal Wallis H test has also shown a statistically significant difference ($H(2) = 8.82, p =	0.012$; 
 $p = 0.024$, Holm-Bonferroni corrected) in the robot anxiety measurement (dependent variable) across tertiles (independent variable) for all the sessions combined (Figure \ref{userlong} (e)). Post-hoc pairwise comparisons using Dunn's tests, with a Bonferroni correction applied, have indicated statistically significant differences between low tertile and high tertile ($Mdn_{low}= 15, Mdn_{high}= 21, p = 0.012$). We did not find any statistical significance between tertiles for all sessions combined in the case of children's likeability of the robot and their perceived intelligence of the robot.

\begin{figure}[h!]
	\centering
	\includegraphics[width= 0.75\textwidth]{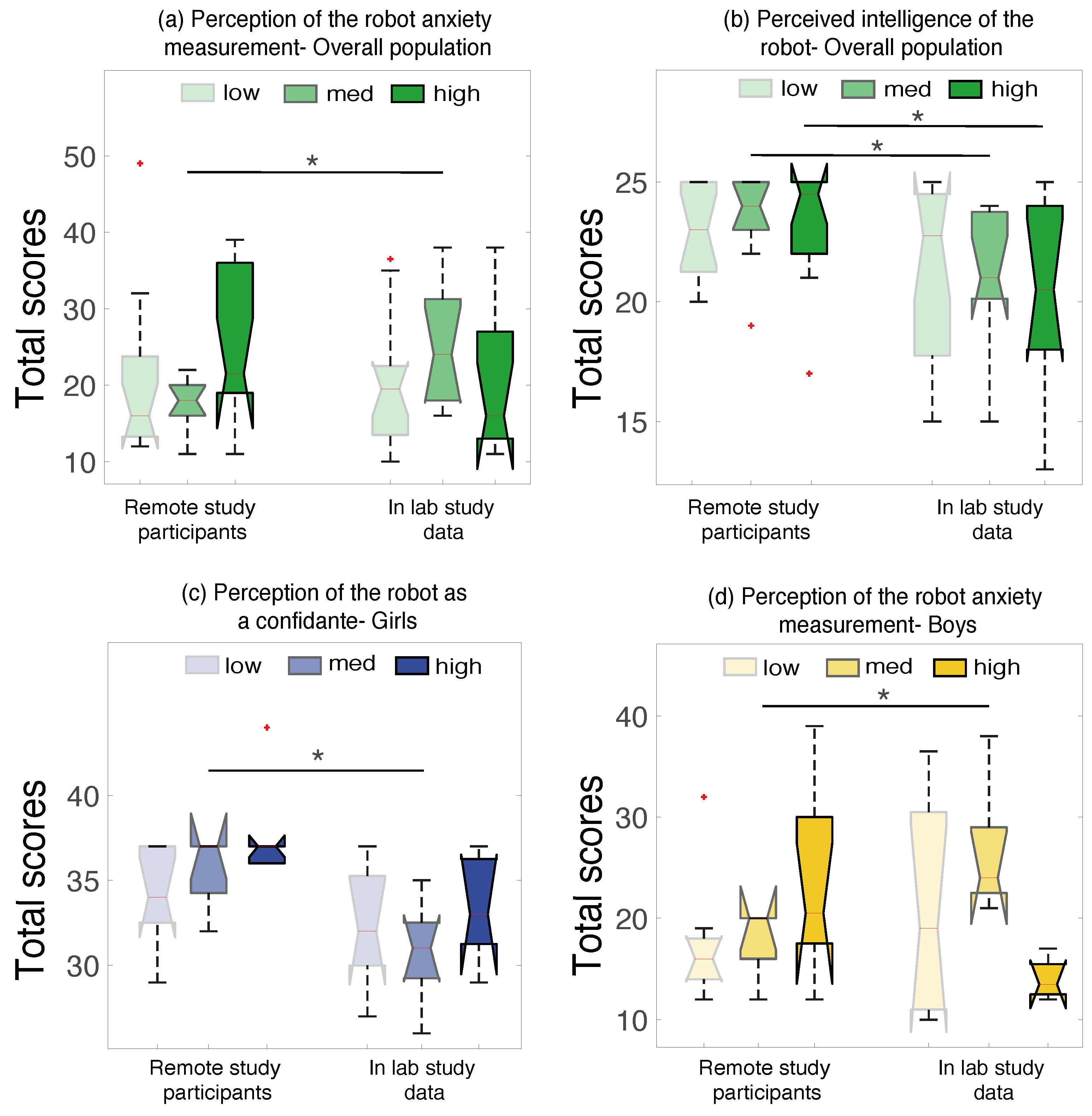}
	\caption{User perception changes across the remote study and our in-lab study \cite{abbasi2023humanoid} categorised into wellbeing tertiles (low, med and high) using the SMFQ responses. $*p<0.05$ corrected.}
	\label{userIRRR}
\end{figure}

As seen from Figure \ref{userIRRR} (a), Wilcoxon rank sum test with a Bonferroni correction applied has indicated a statistically significant difference for the robot anxiety measurement in session 1 of the remote study and our in-lab study \cite{abbasi2023humanoid} for the med tertile of the overall population group ($Mdn_{remote}= 18, Mdn_{in-lab}= 24, p = 0.038$). Further, Wilcoxon rank sum test (Figure \ref{userIRRR} (b)) with a Bonferroni correction applied has also indicated a statistically significant difference for the perceived intelligence of the robot in session 1 of the remote study and our in-lab study \cite{abbasi2023humanoid} for the medium tertile of the overall population group ($Mdn_{remote}= 24, Mdn_{in-lab}= 21, p = 0.045$). Wilcoxon rank sum test with a Bonferroni correction applied has also indicated a statistically significant difference between the perceived intelligence of the robot in session 1 of the remote study and our in-lab study \cite{abbasi2023humanoid} 
for the high tertile of the overall population group ($Mdn_{remote}= 24, Mdn_{in-lab}= 20.5, p = 0.045$). 

To sum up, our key findings indicate that longitudinally, robot anxiety measurement has been seen to decrease for the overall population. Further, the physical presence (remote vs. in-lab) of the robot does affect users' anxiety attributed to the robot (robot anxiety increases for the low and med tertile but decreases for the high tertile) and the perceived intelligence of the robot (decreases across all tertiles).

\subsection{Gender-based differences in the effectiveness of longitudinal wellbeing assessment and user perception (RQ3).}
From our exploratory supplementary analyses, we have observed that in the case of girls (Figure \ref{wellslong} (a)), Kruskal Wallis H test has shown a statistically significant difference ($H(2) =10.64, p =	0.005$; $p = 0.020$, Holm-Bonferroni corrected) in the RCADS responses (dependent variable) for session 1 between the tertiles (independent variable). Post-hoc pairwise comparisons using Dunn's tests, with a Bonferroni correction applied, have indicated statistically significant differences between low tertile and high tertile ($Mdn_{low}= 6, Mdn_{high}= 28, p = 0.006$) and between med tertile and high tertile ($Mdn_{med}= 6, Mdn_{high}= 28, p = 0.042$). 

Further, Friedman's test indicated a statistically significant difference (Figure \ref{wellslong}(a)) in the RCADS scores (dependent variable) for girls belonging to the high tertile, across the three sessions (independent variable) ($\chi^2 (2) =11.14, p =	0.004$; $p = 0.020$, Holm-Bonferroni corrected). Post hoc analysis with Wilcoxon signed-rank tests was conducted with a Bonferroni correction applied, resulting in statistical significance between the RCADS responses for session 1 and session 2 ($Mdn_{session1}= 28, Mdn_{session2}= 11.5, p = 0.007$).

In the case of boys, as seen in Figure \ref{wellslong} (a), Kruskal Wallis H test has shown a statistically significant difference ($H(2) =7.91$, $p = 0.019$, Holm-Bonferroni corrected) in the RCADS responses (dependent variable) for session 1 between the tertiles (independent variable). Post-hoc pairwise comparisons using Dunn's tests, with a Bonferroni correction applied, have indicated statistically significant differences between low tertile and high tertile ($Mdn_{low}= 9, Mdn_{high}= 23, p = 0.016$). 

Kruskal Wallis H test has also shown a statistically significant difference ($H(2) =9.22, p =	0.010$;  $p = 0.024$, Holm-Bonferroni corrected) in the RCADS responses (dependent variable) for session 2 between the tertiles (independent variable). Post-hoc pairwise comparisons using Dunn's tests, with a Bonferroni correction applied, have indicated statistically significant differences between low tertile and high tertile ($Mdn_{low}= 8.5, Mdn_{high}= 20.5, p = 0.009$). 

We did not find any statistically significant difference longitudinally (between session 1, session 2, and session 3) between the RCADS responses for boys across all tertiles. We have also not observed any statistically significant differences (Figure \ref{modesRR}) between the modes of RCADS administration (robot report, self-report, and parent-report) across all the tertiles for girls. In the case of boys (Figure \ref{modesRR})  Wilcoxon rank sum test with a Bonferroni correction applied has indicated a statistically significant difference between the parent-report responses and the responses during session 1 with the robot for the high tertile ($Mdn_{parent}= 7, Mdn_{robot}= 23, p = 0.013$).  Further, even in the case of girls and boys, we have not found any statistically significant differences across the tertiles for the RCADS responses between session 1 for the remote study with the robot and the RCADS responses for our in-lab study presented in \cite{abbasi2022can,abbasi2024analysing}.

We also did not find any statistical significance while comparing user perception responses to the questionnaires 
for girls or boys, across the tertiles, throughout any of the sessions. However, Wilcoxon rank sum test (Figure \ref{userlong}) has indicated a statistically significant difference between the children's perception of the robot as a confidante for the high tertile between girls and boys ($Mdn_{girls}= 37, Mdn_{boys}= 33.5, W= 67, p = 0.012$; $p = 0.030$, Holm-Bonferroni corrected). While considering user perception changes for girls (Figure \ref{userIRRR}(c)), Wilcoxon rank sum test with a Bonferroni correction applied has indicated a statistically significant difference for the perception of the robot as a confidante in session 1 of the remote study and our in-lab study \cite{abbasi2023humanoid} 
for the medium tertile of the overall population group ($Mdn_{remote}= 37, Mdn_{in-lab}= 31, p = 0.038$). Further, in the case of boys (Figure \ref{userIRRR}(d)), Wilcoxon rank sum test with a Bonferroni correction applied has indicated a statistically significant difference for robot anxiety measurement in session 1 of the remote study and our in-lab study \cite{abbasi2023humanoid} for the med tertile of the overall population group ($Mdn_{remote}= 20, Mdn_{in-lab}= 24, p = 0.048$).

In summary, across genders, we have observed that the RCADS responses were significantly higher for the high tertile for both boys and girls in session 1 and only for boys in session 2. We also observed that RCADS responses to the robot were significantly more negative for boys as compared to their parent's perception of their wellbeing. Finally, girls have a higher perception of the robot as their confidante as compared with boys and the physical presence of the robot does affect this perception.

\section{Discussion}
The study explored the use of online robotised platform for the assessment of mental wellbeing of children where 40 children (20 girls and 20 boys, 8-13 years old) interacted with the Nao robot over Zoom or Skype for three sessions. Our findings indicate that online interactions with the robot can be used to effectively assess children's wellbeing. Further, children's perception of the robot either improved or stayed consistent and gender differences were also observed with girls viewing the robot more positively as a confidante.

As seen from the construct validity analyses mentioned in Section \ref{sec:validity}, we employed a confirmatory PCA to explore the underlying structure of the SMFQ and the RCADS. We found that both scales predominantly revolve around a single principal component, suggesting a consistent underlying factor in each questionnaire explaining high variance of childrens' well being. It is important to note that not all items in these questionnaires contributed equally to this primary factor. Specifically, some items in both the SMFQ and RCADS did not significantly load onto the principal component, indicating that they might not align as closely with the core factor that these scales measure within CRI. It should be acknowledged that RCADS (as used in this study) is aimed at assessing the concepts of generalised anxiety, panic disorder, and low mood \cite[see][]{chorpita2005psychometric}, thereby potentially affecting factor loadings. 

Nevertheless, these results affirm the statistical effectiveness of these instruments in capturing the nuances of children's mental health in these interactive settings during longitudinal CRI, showcasing the potential for integrating social robots in psychological assessments. Similar to our previous results in single-session CRI \cite{Abbasi2024}, the comparison between robot-administered and self-reported RCADS reveals significant distinctions that emphasize the potential benefits of integrating robotic assistance into mental health assessments. This outcome represents a statistical validation rather than a behavioural indication. The analysis indicated that the robot-administered RCADS demonstrated a better fit in terms of construct validity, suggesting that interacting with a robot may facilitate disclosure of symptoms from children to assess their mental health in a statistically valid and reliable way. Moreover, in line with the literature, parents' reports of RCADS also showed worse fit compared to the robot-administered version of RCADS, highlighting potential bias in parents' reporting of their children's mental health symptoms \cite{Villatoro2018} compared to their children's reporting of it to robots. As such, we can infer that robot-administered questionnaires can be considered reliable (See Section \ref{sec:measures}) and valid (See Section \ref{sec:validity}) and as such can be utilised as credible ground truth for the evaluation of mental wellbeing of children. 

Our results (Figure \ref{wellslong}) also show that interacting with the robot online via Zoom or Skype has amplified the difference between the wellbeing clusters, especially the low tertile and high tertile. The effect is sustained over the course of three sessions as well as for all the sessions combined.  We also observed a decrease in the RCADS ratings for the overall population over the course of three sessions. This could be a result of the \emph{regression to mean} phenomenon which is common in longitudinal studies \cite{marsh2002multilevel} and indicates that \emph{extreme observations have a tendency to move closer to the average value upon subsequent measurements} \cite{barnett2005regression}. The decrease in the RCADS ratings might also be attributed to the improvement in wellbeing stemming from the familiarity with a robot.
Crossman et al \cite{crossman2018influence} have demonstrated that interaction with a robot has increased the positive mood in children. Robots have also been used to alleviate anxiety during medical procedures \cite{rossi2022using}. Even in longitudinal HRI studies focussing on adult participants, interaction with robots has led to a decrease in stress levels \cite{russell2021use,coping_ijsr} and enhancement of participants' mood \cite{Laban_blt_2023}.

As seen from our findings (Figure \ref{modesRR}), the robot-assisted mode of RCADS administration seems to be very effective in categorising the population across wellbeing clusters (especially low tertile and high tertile) as compared to traditional modes of self-reporting and parent-reporting of the children's wellbeing. This could potentially be due to what prior studies have been reporting \-- i.e., that children consider robots as peers \cite{kumar2020toy} and social companions \cite{nalin2012robotic} which could contribute to them being more candid about their emotions in front of a robot. Our finding is in line with our previous work \cite{abbasi2022can,abbasi2024analysing} that also measured how wellbeing assessment of children differed across different modes of questionnaire administration, where the robot mode consisted of the children being physically co-located with the robot (i.e., a co-present robot). 

With regard to whether co-location affects wellbeing assessment, our findings suggest that there were no statistical differences found across the tertiles while comparing our remote study with our in-lab study presented in \cite{abbasi2022can,abbasi2024analysing}. This is in line with previous research which suggests that co-location does not necessarily improve the objective of the CRI experience. Robinette et al \cite{robinette2016assessment} have demonstrated that there was minimal difference observed between the instruction-giving capabilities of a co-present robot, a remote robot and a virtual avatar robot. This is supported by Kennedy et al \cite{kennedy2015comparing} who have shown that learning was not improved among children while interacting with the robot in person. Our study findings indicate that forthcoming robot-assisted initiatives for the assessment of wellbeing can be effectively conducted online, thus providing a logistical advantage to children to open up about their feelings from a comfortable environment \textbf{(RQ1)}.

With regard to the longitudinal changes in children's perception of the robot \textbf{(RQ2)}, we have observed (Figure \ref{userlong}) that most participants have maintained or improved their perception of the robot. This is in line with the previous literature \cite{conti2019kindergarten} where prolonged exposure to robots has improved the opinions of Kindergarten children about the robot. Moreover, as seen from Figure \ref{userIRRR}, some children (low and med tertile) were more anxious in front of a physical robot while some children (high tertile) were more anxious in front of the remote robot. This individual variability in attitudes could be due to their past experiences with robots and/or technology in general. Prior research has indicated that people's attitudes towards robots were governed by their prior knowledge about robots or even technology \cite{sandoval2014human}, with negative representation causing negative outlooks about robots \cite{horstmann2019great}. We can also observe that children's overall longitudinal perception (i.e. combined for all sessions) of the robot as a confidante as well as their anxiety towards the robot for all sessions combined does differ across varying levels of wellbeing (Figure \ref{userlong}). However, from our previous work \cite{abbasi2023humanoid}, we have observed that the wellbeing of the children is independent of their anxiety towards the robot. This difference may arise from children's varying perceptions of the robot that is physically present as compared to one that is remote, as observed from our findings (Figures \ref{userIRRR}). Additionally, prolonged exposure to robots might also alter children's perception towards robots, potentially establishing a connection between their wellbeing and their attitude towards robots. Prior research \cite{rikkers2016internet,rapinda2021examining,stenseng2020time} has indicated possible links between emotional and behavioural concerns and prolonged use of digital technology such as electronic gaming and the internet. Therefore, future online robot-assisted initiatives should consider participants' outlooks, sentiments and existing/past emotional and behavioural concerns so that the technology can be utilised most effectively in the assessment of mental wellbeing of children.

From our exploratory supplementary analyses regarding the effect of gender \textbf{(RQ3)} on the longitudinal robot-assisted evaluation of mental wellbeing of children, we have observed that the interaction with the online robot was able to effectively differentiate between the high tertiles and low tertiles across boys and girls for session 1 and only for boys for session 2. Strait et al \cite{strait2015gender} have observed that females respond favourably to changes in robot behaviour while males tend to prefer consistency. In our study, the experimental tasks used to evaluate wellbeing remained consistent across sessions, potentially leading to the statistically significant differences in wellbeing clusters for boys and not girls in session 2. Further, the effectiveness of the online robot mode of test administration (compared with co-present robot, self-report and parent-report responses) appears to be consistent across genders. Therefore, our study shows that the benefits of using online robot-assisted methods for assessing mental wellbeing hold true regardless of the participant's gender. In addition, while the longitudinal perception of the online robot largely improved or remained consistent across genders, there was gender-related variability in how the robot was perceived as compared to our in-lab study presented in \cite{abbasi2023humanoid}. For instance, while comparing the responses of girls in the remote study and responses of girls in our in-lab study \cite{abbasi2023humanoid}, we observed a decrease in the robot's perception as a confidante (Figure \ref{userIRRR}(c)). We also observed a 
predominant increase in robot anxiety in boys (Figure \ref{userIRRR}(d)) between the remote study and our in-lab study presented in \cite{abbasi2023humanoid}. 

In addition, when comparing responses to the perception of the role of the robot as a confidante of boys and girls in the remote study (Figure \ref{userlong} (b)), we observed that girls had a higher opinion of the robot as compared to boys. The gender-based differences in user perception between the remote study and our in-lab study \cite{abbasi2023humanoid} could be attributed to the general difference in the attitudes across children with regard to technology \cite{jackson2008culture}. Further, girls' higher willingness to confide and share their emotions with others as compared to boys \cite{chandra2006stigma,pearson2023masculinity}) could also contribute to their higher perception of the robot as a confidante in the remote study as compared to boys.  Thus, in the design of future robot-led technologies for the assessment of mental wellbeing, it is crucial to take into account the differences in children's perceptions of the robot and their willingness to open up, to enable an environment where they feel comfortable expressing their feelings truthfully. 

Moreover, in order to effectively aid the psychological support teams, it is important to consider how frequently should robotised assessments be conducted to gain insights into the children's mental wellbeing. As we observed from our results in the remote study mentioned here and our in-lab study \cite{abbasi2022can,abbasi2024analysing}, one-off sessions, as well as longitudinal interactions with the robots, can both be effectively useful in understanding the mental wellbeing of children. From an HRI perspective, users' responses, perceptions and engagement with regard to longitudinal interactions can be varied and must be taken into account while designing interactions. For example, studies have shown that prolonged exposure to robots helps participants promote more self-disclosure \cite{Laban_blt_2023} and enables them to develop more positive opinions of the robot\cite{conti2019kindergarten}. Studies have also shown that interaction with robots has resulted in improvement of mood as well as wellbeing of both children \cite{crossman2018influence,leite2012long} and adults \cite{bodala2021teleoperated,Laban_blt_2023}.  Thus, robot-driven assessments might contribute to enhancing the wellbeing of the participants, potentially affecting follow-up assessments in a positive manner. In contrast, studies have also shown that repeated interactions with the robot may also cause diminished engagement \cite{serholt2016robots} and reduced interest to participate from the users \cite{leite2013social}. Previous research has addressed this issue by implementing variations in the interaction experience such as changes in session activities \cite{coninx2016towards} or adapting robot behaviour to align with user preferences to promote more social engagement \cite{Ahmad2017}.


In the context of our study, there were several key factors that influenced our methodology for using online-mediated interactions with social robots over other remote methods (such as chatbots, in-home robots or virtual animated characters of robots) that also aim to provide increased accessibility to the robots.
Firstly, while chatbots have become increasingly popular in the mental health domain, often being used in mental health applications, they require some initiative from the participants \cite{huberty2021evaluation}. This would be difficult to encourage in our study as the participants were children and could even lead to increased drop-out rates in the longitudinal study conducted here.

Secondly, in the case of in-home robots, there is a logistical disadvantage in terms of accessibility to every child \footnote{https://automationswitch.com/disadvantages-of-domestic-robots/}, present in different parts of the world in comparison with smart devices required for online-mediated robots that are more readily available in households \footnote{https://explodingtopics.com/blog/smartphone-stats}. As such, there would be monetary costs associated with purchasing either on the researchers' end or on the participants', creating inconvenience for each party. 

Thirdly, in the case of virtual animated robot characters, prior research\cite{li2015benefit} has suggested that they do not offer any notable advantage over online-mediated interaction with robots. Li et al \cite{li2015benefit}  have indicated that physical presence and not the physical embodiment of the robots have been shown to positively affect user responses. They demonstrated that co-present robots (present in the same room as the participants) have been shown to favourably influence users' behaviour and preference in comparison with virtual (animated character of robots) or telepresent/online (physical robots present on a camera feed).  In addition, Laban et al \cite{Laban_blt_2023, coping_ijsr, laban_icd_2022} have shown that online-mediated interactions with robots (Pepper robot in their case) using videoconferencing platforms like Zoom have been shown to improve self-disclosure and mood of the participants. This finding aligns well with the familiarity and comfort that children developed with online platforms like Zoom and Skype over the Covid-19 pandemic, thus facilitating a more convenient and accessible interaction environment\footnote{https://www.zoom.com/en/blog/how-zoom-supported-young-people-in-uk-during-pandemic/}. Moreover, we also wanted to make valid comparisons with our previous works conducted in \cite{abbasi2022can,abbasi2024analysing}. As such, we followed the experimental protocol conducted in \cite{abbasi2022can,abbasi2023humanoid, abbasi2024analysing} and only modified the mode of interaction from in-person to online. Using a virtual robot (or agent) would have changed the robot's form and appearance to an animated character making it difficult to ascertain if any changes in findings were due to the mode of the interaction (in-person vs. online) or the robot's form (virtual animated character or the actual physical robot). Spitale et al \cite{spitale2023robotic} have shown that in HRI for wellbeing studies changing robot form leads to significant changes in user perception and interaction. To this end, we employed robots that interacted with the participants through Zoom or Skype, to explore how they affect the assessment of mental wellbeing in children. In short, the aforementioned factors such as participants' engagement, logistical accessibility, familiarity with online platforms and most importantly, methodological consistency to make valid comparisons with previous works \cite{abbasi2022can, abbasi2023humanoid,abbasi2024analysing} influenced our approach to use online-mdeiated interactions with robots for the assessment of mental wellbeing of children.

Following the results of this study, and considering the future vision of integrating social robots into supporting mental health assessment, we propose a customised approach for a robot-led mental wellbeing assessments rather than a one-size-fits-all model. As such, the frequency of the evaluations should be decided in line with current guidelines \footnote{https://www.nhs.uk/mental-health/social-care-and-your-rights/mental-health-assessments/} that determine the severity of the conditions and the care provided. As seen in Figure \ref{pipeline} (b), robots' multimodal sensing capabilities can be utilised to sense and identify potential wellbeing concerns in children through verbal and non-verbal cues. Child mental health practitioners can then review the observations collected by robots and make informed decisions. This approach provides genuine data and insights that enhance and support clinical decision-making, rather than merely guiding it. In addition, robots can also be programmed with "memory" abilities to learn from prior assessments for a particular child and focus on areas that might warrant further investigation. Robotised assessments can harness the time efficiency of ecological momentary assessment and intervention programs like Artemis-A \footnote{https://www.artemis-a.org/} which utilise computerised adaptive testing to significantly reduce the duration of follow-up assessments. This can alleviate the workload and pressure on existing psychological resources that rely solely on the verbal and written responses provided by children \cite{robot_post}. As such, robotised assessments can be used as valuable tools to (I) help children become more candid with their responses,  (II) flag cases that should receive further attention from child mental health practitioners, (III) provide genuine data of verbal and non-verbal cues from authentic interactions that can support clinical decision making, 
and (IV) carry out time-efficient follow up assessments if needed, to develop a more comprehensive understanding of the children's mental wellbeing over time and in ecological settings (e.g., public spaces, local General Practitioner's surgery and schools). We aspire for robot-led assessments to serve as useful tools supporting the work of child mental health practitioners in diagnosing and creating treatment plans that monitor changes in the mental wellbeing of future generations in an ethical and responsible way.

Finally, the use of adaptive robot behaviour in the context of mental well-being and HRI presents significant potential for enhancing personalised interactions, particularly through the integration of large language models (LLMs)  \cite{zhang_llm_2023,Kim2024} and affective capabilities such as affect recognition from facial and vocal expressions, as well as behavioural cues \cite{spitale_acii22,Spitale2024}. By dynamically adjusting the robot’s responses based on users’ affective and psychological states, these adaptive systems can foster greater engagement, rapport, and deeper self-disclosure  \cite{Sharing2024}. However, to ensure accurate mental health assessments, these interactions should adhere to standard practices by following a structured, validated, and well-established assessment frameworks \cite{Abbasi2024}. Hence, certain aspects of interaction need to remain structured and consistent in order to adhere to established protocols of wellbeing assessments. For instance, the RCADS and the SMFQ need to be conducted in accordance with standardised guidelines to maintain the validity and reliability of the findings \cite{angold1995development, sharp2006short, chorpita2015revised}. The questionnaire statements could not be modified as doing so would compromise the validity of the psychological test under consideration \cite{Stewart2012}. As such, this highlights the need to balance flexibility with adherence to established protocols so as to ensure desired, valid and reliable outcomes \cite{Abbasi2024}. 

Consequently, the current adaptive approaches may not be particularly suitable for mental health assessments \cite[see][\textit{"(R4) Verbal adaptation should be limited to preserve well-being practice efficacy"}]{Axelsson2024}, especially with children. Moreover, despite the potential benefits of affective technologies to quality of interaction, there are critical ethical concerns related to privacy, data security, and the handling of sensitive emotional information. These adaptive capabilities rely on processing and sharing large amounts of personal data with third parties, raising concerns about the transparency of the algorithms and the potential for biased or inappropriate responses, especially in vulnerable populations like children  \cite{Kurian2024,Sharing2024}. Additionally, there are concerns regarding the potential for LLMs to inadvertently reinforce biases or generate inappropriate responses, which could be particularly harmful in sensitive mental health contexts \cite{Spitale2024}. For example, if the model produces responses misaligned with a child’s developmental needs or misinterprets the emotional nuances of a situation, it could negatively affect the child’s well-being \cite{Kurian2023}. The opaque nature of LLMs further complicates efforts to fully ensure the safety and appropriateness of these interactions, emphasising the need for real-time oversight and intervention  \cite{Liao2024,Harrer2023}. Therefore, while adaptive LLM-driven behaviour holds promise for improving mental well-being outcomes in HRI, it is imperative to rigorously test for biases, ensure transparent data handling, and implement safeguards to prevent unintended consequences \cite[e.g., see][]{Tam2024}. Future work should focus on refining these adaptive and affective capabilities to prioritise both the psychological benefits for users and the ethical management of their data, ensuring trust and safety in long-term HRI applications.

This calls for a further essential discussion addressing the ethical concerns surrounding the use of social robots in mental wellbeing assessments, particularly regarding trust, confidentiality, and managing children's expectations of privacy. While social robots can foster trust and encourage self-disclosure due to their non-judgmental presence, it is crucial to communicate the limits of confidentiality clearly. Children and their guardians must understand that disclosures related to safety concerns may need to be shared with trusted adults, in line with safeguarding protocols. To balance trust with safeguarding duties, a tiered privacy management strategy could be implemented, ensuring that disclosures involving harm are escalated appropriately \cite{Dietrich2023}. Furthermore, while robots may facilitate openness, they should serve as complementary tools, not substitutes for human caregivers, ensuring emotional support remains part of a broader care framework \cite{Sharing2024}. These ethical considerations deserve their own dedicated discussions, which we are currently exploring through parallel ongoing research \cite{levinson2024expert}. A key finding from our interviews with subject matter experts as mentioned in \cite{levinson2024expert} is that, while robots should not make judgement calls, they can be a valuable addition to the children’s safeguarding team. However, there is a clear need for interdisciplinary dialogue to thoughtfully inform policy and legal frameworks that define the appropriate roles of robots in children’s lives, particularly in intimate and personal settings \cite{levinson2024expert}.

\section{Limitations and Future Work}
Although we have addressed the research questions defined at the outset of our work, our study has several limitations that we hope to address in the future.


First, we have only considered verbal responses from two tasks of the study. Future work will also focus on verbal responses from other tasks (e.g., Task 1: recall of happy and sad memories, Task 3: picture-based task), children's drawings, and non-verbal responses (e.g., facial expressions, speech), enabling automatic multimodal robot-led assessment of children's mental wellbeing.
Second, our research is a feasibility study conducted on a non-clinical population to determine whether and how SARs can be used as novel tools to contribute to the holistic understanding of mental health status. In future, we do hope to conduct case-controlled designs in clinical settings with clinical populations through clinician-administered measures.
Third, we have not taken into consideration whether/how the varying ethnicities of the children might influence the robotised wellbeing assessment. In future, we aim to investigate how cultural differences might affect the robot-led wellbeing assessment of children. 

Fourth, we have not considered whether the children's past experiences with robots might affect their interaction. In future, we aim to investigate how their attitudes towards robots due to their prior experiences might influence their experience of the robotised wellbeing assessment.
Fifth, we have not taken into account wellbeing changes between online form-filling and the experiment sessions. In future, we aim to conduct self-report and parent-report questionnaires concurrently with the robot interaction session to avoid any potential fluctuations in mental wellbeing reporting. 

Sixth, the time frame of the longitudinal interactions has been kept relatively focused to 3-6 weeks. While this period allowed us to capture meaningful trends in children’s wellbeing assessments and perceptions of the robot, extending the study over a longer duration could offer additional insights into the sustained dynamics of CRI \cite{leite2013social}. A longer-term study would help to further explore how children’s engagement with the robot evolves over time, especially regarding the potential novelty effect \cite{smedegaard2019reframing}. Such a study could determine whether the robot continues to maintain children’s interest and provides sustained benefits, or if engagement declines after additional repeated interactions. Nevertheless, the findings of this study provide an important benchmark, demonstrating that over the 3–6 week period, children’s perceptions of the robot either remained stable or improved, suggesting that the novelty effect did not negatively influence engagement within this timeframe. Future research should seek to replicate and build upon these results by extending interactions to further assess the long-term potential of social robots in child wellbeing assessments. 

Additionally, while our study utilized the Nao robot, a widely recognized humanoid platform in HRI \cite{Amirova2021}, it is important to consider that different embodiments of robots may lead to variations in children's interactions and responses. Prior research has demonstrated that robot embodiment can significantly influence factors such as trust, engagement, and self-disclosure \cite[e.g.,][]{Laban2021,spitale2023robotic}, potentially affecting wellbeing assessments. While we expect that results would remain consistent with other humanoid robots that support verbal interaction \cite{afford2016,Cross2021}, future research should explore how other embodiments (e.g., more machine-like designs, or animated virtual humans) might evoke different emotional responses or alter children's perceptions of robots as confidants. Expanding this research to include robots with diverse embodiments will help clarify the generalizability of these findings across various robotic platforms and better inform the role of embodiment in robot-mediated mental health assessments.

Despite the aforementioned limitations, we hope that the findings from this work can provide a valuable stepping stone for future robot-led initiatives for mental wellbeing assessment of children, that are more accessible and take into account children's propensity and attitude.

\section{Conclusions}
Our work has unveiled compelling insights into longitudinally assessing children's mental wellbeing through online interactions with a robot. We hope that robots can form a valuable addition to the mental wellbeing assessment procedure of children, to aid clinicians and psychologists in identifying cases with potential mental wellbeing concerns so that necessary clinical intervention can be provided in a timely manner. Our study demonstrated that remote robotised wellbeing assessment platforms can effectively gauge children's wellbeing longitudinally. Furthermore, the user perception of these robot-mediated wellbeing assessment platforms has been sustained or in many instances improved over time. In addition, our exploratory analyses revealed gender-related variability, thus indicating important considerations for developing personalised interactions. As such, incorporating remote robotised platforms in mental wellbeing assessment frameworks offers valuable potential to improve accessibility and support the mental health of the future generation.


\section*{Acknowledgments}
N. I. Abbasi is supported by the W.D. Armstrong Trust PhD Studentship and the Cambridge Trusts. G. Laban and H. Gunes are supported by the EPSRC project ARoEQ under grant ref. EP/R030782/1. T. Ford and P. Jones acknowledge support from the NIHR ARC East of England and the Cambridge BRC. The equipment used in the study was supported by the University of Cambridge's Observatory for Human Machine Collaboration (OHMC) Small Equipment Funding. The views expressed are those of the authors and not necessarily those of the NIHR or the Department of Health and Social Care. The authors would also like to thank Miss Amina Itrat Abbasi for her help with the experimental setup.
\section*{Open Access} For the purpose of open access, the authors have applied a Creative Commons Attribution (CC BY 4) licence to any Author Accepted Manuscript version arising.
\section*{Data Access} Overall statistical analysis of research data underpinning this publication is available in the text of this publication. Additional raw data related to this publication cannot be openly released; the raw data contains transcripts of interviews, but none of the interviewees consented to data sharing.

\bibliographystyle{ACM-Reference-Format}
\bibliography{References}


\begin{thebibliography}{129}


\ifx \showCODEN    \undefined \def \showCODEN     #1{\unskip}     \fi
\ifx \showDOI      \undefined \def \showDOI       #1{#1}\fi
\ifx \showISBNx    \undefined \def \showISBNx     #1{\unskip}     \fi
\ifx \showISBNxiii \undefined \def \showISBNxiii  #1{\unskip}     \fi
\ifx \showISSN     \undefined \def \showISSN      #1{\unskip}     \fi
\ifx \showLCCN     \undefined \def \showLCCN      #1{\unskip}     \fi
\ifx \shownote     \undefined \def \shownote      #1{#1}          \fi
\ifx \showarticletitle \undefined \def \showarticletitle #1{#1}   \fi
\ifx \showURL      \undefined \def \showURL       {\relax}        \fi
\providecommand\bibfield[2]{#2}
\providecommand\bibinfo[2]{#2}
\providecommand\natexlab[1]{#1}
\providecommand\showeprint[2][]{arXiv:#2}

\bibitem[Abbasi et~al\mbox{.}(2023)]%
        {abbasi2023humanoid}
\bibfield{author}{\bibinfo{person}{Nida Abbasi}, \bibinfo{person}{Micol Spitale}, \bibinfo{person}{Joanna Anderson}, \bibinfo{person}{Tamsin Ford}, \bibinfo{person}{Peter Jones}, {and} \bibinfo{person}{Hatice Gunes}.} \bibinfo{year}{2023}\natexlab{}.
\newblock \showarticletitle{Humanoid Robots for Wellbeing Assessment in Children: How Does Anxiety towards the Robot Affect Perceptions of Robot Role, Behaviour and Capabilities?}
\newblock  (\bibinfo{year}{2023}).
\newblock


\bibitem[Abbasi et~al\mbox{.}(2024a)]%
        {Abbasi2024}
\bibfield{author}{\bibinfo{person}{Nida~Itrat Abbasi}, \bibinfo{person}{Guy Laban}, \bibinfo{person}{Tamsin Ford}, \bibinfo{person}{Peter~B. Jones}, {and} \bibinfo{person}{Hatice Gunes}.} \bibinfo{year}{2024}\natexlab{a}.
\newblock \showarticletitle{Robotising Psychometrics: Validating Wellbeing Assessment Tools in Child-Robot Interactions}. In \bibinfo{booktitle}{\emph{2024 33rd IEEE International Conference on Robot and Human Interactive Communication (ROMAN)}} (Pasadena, CA). \bibinfo{publisher}{IEEE}, \bibinfo{pages}{1651--1658}.
\newblock
\showISBNx{979-8-3503-7502-2}
\urldef\tempurl%
\url{https://doi.org/10.1109/RO-MAN60168.2024.10731253}
\showDOI{\tempurl}


\bibitem[Abbasi et~al\mbox{.}(2022)]%
        {abbasi2022can}
\bibfield{author}{\bibinfo{person}{Nida~Itrat Abbasi}, \bibinfo{person}{Micol Spitale}, \bibinfo{person}{Joanna Anderson}, \bibinfo{person}{Tamsin Ford}, \bibinfo{person}{Peter~B Jones}, {and} \bibinfo{person}{Hatice Gunes}.} \bibinfo{year}{2022}\natexlab{}.
\newblock \showarticletitle{Can robots help in the evaluation of mental wellbeing in children? an empirical study}. In \bibinfo{booktitle}{\emph{2022 31st IEEE International Conference on Robot and Human Interactive Communication (RO-MAN)}}. IEEE, \bibinfo{pages}{1459--1466}.
\newblock


\bibitem[Abbasi et~al\mbox{.}(2024b)]%
        {abbasi2024analysing}
\bibfield{author}{\bibinfo{person}{Nida~Itrat Abbasi}, \bibinfo{person}{Micol Spitale}, \bibinfo{person}{Joanna Anderson}, \bibinfo{person}{Tamsin Ford}, \bibinfo{person}{Peter~B Jones}, {and} \bibinfo{person}{Hatice Gunes}.} \bibinfo{year}{2024}\natexlab{b}.
\newblock \showarticletitle{Analysing Children’s Responses from Multiple Modalities During Robot-Assisted Assessment of Mental Wellbeing}.
\newblock \bibinfo{journal}{\emph{International Journal of Social Robotics}} (\bibinfo{year}{2024}), \bibinfo{pages}{1--48}.
\newblock


\bibitem[Abdelmohsen and Arafa(2021)]%
        {abdelmohsen2021virtual}
\bibfield{author}{\bibinfo{person}{Maha Abdelmohsen} {and} \bibinfo{person}{Yasmine Arafa}.} \bibinfo{year}{2021}\natexlab{}.
\newblock \showarticletitle{Virtual Social Robot Enhances the Social Skills of Children with HFA}. In \bibinfo{booktitle}{\emph{Social Robotics: 13th International Conference, ICSR 2021, Singapore, Singapore, November 10--13, 2021, Proceedings 13}}. Springer, \bibinfo{pages}{497--508}.
\newblock


\bibitem[Ahmad et~al\mbox{.}(2017)]%
        {Ahmad2017}
\bibfield{author}{\bibinfo{person}{Muneeb~Imtiaz Ahmad}, \bibinfo{person}{Omar Mubin}, {and} \bibinfo{person}{Joanne Orlando}.} \bibinfo{year}{2017}\natexlab{}.
\newblock \showarticletitle{Adaptive Social Robot for Sustaining Social Engagement during Long-Term Children–Robot Interaction}.
\newblock \bibinfo{journal}{\emph{International Journal of Human–Computer Interaction}}  \bibinfo{volume}{33} (\bibinfo{date}{12} \bibinfo{year}{2017}), \bibinfo{pages}{943--962}.
\newblock
Issue 12.
\showISSN{15327590}
\urldef\tempurl%
\url{https://doi.org/10.1080/10447318.2017.1300750}
\showDOI{\tempurl}


\bibitem[Aickin and Gensler(2011)]%
        {Aickin2011}
\bibfield{author}{\bibinfo{person}{Mikel Aickin} {and} \bibinfo{person}{Helen Gensler}.} \bibinfo{year}{2011}\natexlab{}.
\newblock \showarticletitle{Adjusting for multiple testing when reporting research results: the Bonferroni vs Holm methods.}
\newblock \bibinfo{journal}{\emph{American Journal of Public Health}}  \bibinfo{volume}{86} (\bibinfo{date}{10} \bibinfo{year}{2011}), \bibinfo{pages}{726--728}.
\newblock
Issue 5.
\showISSN{00900036}
\urldef\tempurl%
\url{https://doi.org/10.2105/AJPH.86.5.726}
\showDOI{\tempurl}


\bibitem[Amirova et~al\mbox{.}(2021)]%
        {Amirova2021}
\bibfield{author}{\bibinfo{person}{Aida Amirova}, \bibinfo{person}{Nazerke Rakhymbayeva}, \bibinfo{person}{Elmira Yadollahi}, \bibinfo{person}{Anara Sandygulova}, {and} \bibinfo{person}{Wafa Johal}.} \bibinfo{year}{2021}\natexlab{}.
\newblock \showarticletitle{10 Years of Human-NAO Interaction Research: A Scoping Review}.
\newblock \bibinfo{journal}{\emph{Frontiers in Robotics and AI}}  \bibinfo{volume}{8} (\bibinfo{date}{11} \bibinfo{year}{2021}), \bibinfo{pages}{744526}.
\newblock
\showISSN{22969144}
\urldef\tempurl%
\url{https://doi.org/10.3389/FROBT.2021.744526/BIBTEX}
\showDOI{\tempurl}


\bibitem[Angold et~al\mbox{.}(1995)]%
        {angold1995development}
\bibfield{author}{\bibinfo{person}{Adrian Angold}, \bibinfo{person}{Elizabeth~J Costello}, \bibinfo{person}{Stephen~C Messer}, {and} \bibinfo{person}{Andrew Pickles}.} \bibinfo{year}{1995}\natexlab{}.
\newblock \showarticletitle{Development of a short questionnaire for use in epidemiological studies of depression in children and adolescents.}
\newblock \bibinfo{journal}{\emph{International journal of methods in psychiatric research}} (\bibinfo{year}{1995}).
\newblock


\bibitem[Arsi{\'c} et~al\mbox{.}(2022)]%
        {arsic2022use}
\bibfield{author}{\bibinfo{person}{Bojana Arsi{\'c}}, \bibinfo{person}{Anja Gaji{\'c}}, \bibinfo{person}{Sara Vidojkovi{\'c}}, \bibinfo{person}{Dragana Ma{\'c}e{\v{s}}i{\'c}-Petrovi{\'c}}, \bibinfo{person}{Aleksandra Ba{\v{s}}i{\'c}}, {and} \bibinfo{person}{Ru{\v{z}}ica~Zdravkovi{\'c} Parezanovi{\'c}}.} \bibinfo{year}{2022}\natexlab{}.
\newblock \showarticletitle{The use of nao robots in teaching children with autism}.
\newblock \bibinfo{journal}{\emph{European Journal of Alternative Education Studies}} \bibinfo{volume}{7}, \bibinfo{number}{1} (\bibinfo{year}{2022}).
\newblock


\bibitem[Axelsson et~al\mbox{.}(2024)]%
        {Axelsson2024}
\bibfield{author}{\bibinfo{person}{Minja Axelsson}, \bibinfo{person}{Micol Spitale}, {and} \bibinfo{person}{Hatice Gunes}.} \bibinfo{year}{2024}\natexlab{}.
\newblock \showarticletitle{Robots as Mental Well-being Coaches: Design and Ethical Recommendations}.
\newblock \bibinfo{journal}{\emph{ACM Transactions on Human-Robot Interaction}}  \bibinfo{volume}{13} (\bibinfo{date}{6} \bibinfo{year}{2024}).
\newblock
Issue 2.
\showISSN{25739522}
\urldef\tempurl%
\url{https://doi.org/10.1145/3643457/ASSET/EBE92B19-CC4D-4F8C-88F8-8BABC7886953/ASSETS/GRAPHIC/THRI-2022-0075-PAGE50.JPG}
\showDOI{\tempurl}


\bibitem[Barnett et~al\mbox{.}(2005)]%
        {barnett2005regression}
\bibfield{author}{\bibinfo{person}{Adrian~G Barnett}, \bibinfo{person}{Jolieke~C Van Der~Pols}, {and} \bibinfo{person}{Annette~J Dobson}.} \bibinfo{year}{2005}\natexlab{}.
\newblock \showarticletitle{Regression to the mean: what it is and how to deal with it}.
\newblock \bibinfo{journal}{\emph{International journal of epidemiology}} \bibinfo{volume}{34}, \bibinfo{number}{1} (\bibinfo{year}{2005}), \bibinfo{pages}{215--220}.
\newblock


\bibitem[Bartneck et~al\mbox{.}(2009)]%
        {bartneck2009measurement}
\bibfield{author}{\bibinfo{person}{Christoph Bartneck}, \bibinfo{person}{Dana Kuli{\'c}}, \bibinfo{person}{Elizabeth Croft}, {and} \bibinfo{person}{Susana Zoghbi}.} \bibinfo{year}{2009}\natexlab{}.
\newblock \showarticletitle{Measurement instruments for the anthropomorphism, animacy, likeability, perceived intelligence, and perceived safety of robots}.
\newblock \bibinfo{journal}{\emph{International journal of social robotics}}  \bibinfo{volume}{1} (\bibinfo{year}{2009}), \bibinfo{pages}{71--81}.
\newblock


\bibitem[Bellak and Bellak(1949)]%
        {bellak1949children}
\bibfield{author}{\bibinfo{person}{Leopold Bellak} {and} \bibinfo{person}{Sonya~S Bellak}.} \bibinfo{year}{1949}\natexlab{}.
\newblock \showarticletitle{Children's Apperception Test.}
\newblock  (\bibinfo{year}{1949}).
\newblock


\bibitem[Bethel et~al\mbox{.}(2016)]%
        {bethel2016using}
\bibfield{author}{\bibinfo{person}{Cindy~L Bethel}, \bibinfo{person}{Zachary Henkel}, \bibinfo{person}{Kristen Stives}, \bibinfo{person}{David~C May}, \bibinfo{person}{Deborah~K Eakin}, \bibinfo{person}{Melinda Pilkinton}, \bibinfo{person}{Alexis Jones}, {and} \bibinfo{person}{Megan Stubbs-Richardson}.} \bibinfo{year}{2016}\natexlab{}.
\newblock \showarticletitle{Using robots to interview children about bullying: Lessons learned from an exploratory study}. In \bibinfo{booktitle}{\emph{2016 25th IEEE International Symposium on Robot and Human Interactive Communication (RO-MAN)}}. IEEE, \bibinfo{pages}{712--717}.
\newblock


\bibitem[Bodala et~al\mbox{.}(2021)]%
        {bodala2021teleoperated}
\bibfield{author}{\bibinfo{person}{Indu~P Bodala}, \bibinfo{person}{Nikhil Churamani}, {and} \bibinfo{person}{Hatice Gunes}.} \bibinfo{year}{2021}\natexlab{}.
\newblock \showarticletitle{Teleoperated robot coaching for mindfulness training: A longitudinal study}. In \bibinfo{booktitle}{\emph{2021 30th IEEE International Conference on Robot \& Human Interactive Communication (RO-MAN)}}. IEEE, \bibinfo{pages}{939--944}.
\newblock


\bibitem[Bremner et~al\mbox{.}(2016)]%
        {bremner2016personality}
\bibfield{author}{\bibinfo{person}{Paul Bremner}, \bibinfo{person}{Oya Celiktutan}, {and} \bibinfo{person}{Hatice Gunes}.} \bibinfo{year}{2016}\natexlab{}.
\newblock \showarticletitle{Personality perception of robot avatar tele-operators}. In \bibinfo{booktitle}{\emph{2016 11th ACM/IEEE International Conference on Human-Robot Interaction (HRI)}}. IEEE, \bibinfo{pages}{141--148}.
\newblock


\bibitem[Burn et~al\mbox{.}(2022)]%
        {burn2022developing}
\bibfield{author}{\bibinfo{person}{Anne-Marie Burn}, \bibinfo{person}{Tamsin~J Ford}, \bibinfo{person}{Jan Stochl}, \bibinfo{person}{Peter~B Jones}, \bibinfo{person}{Jesus Perez}, {and} \bibinfo{person}{Joanna~K Anderson}.} \bibinfo{year}{2022}\natexlab{}.
\newblock \showarticletitle{Developing a web-based app to assess mental health difficulties in secondary school pupils: qualitative user-centered design study}.
\newblock \bibinfo{journal}{\emph{JMIR Formative Research}} \bibinfo{volume}{6}, \bibinfo{number}{1} (\bibinfo{year}{2022}), \bibinfo{pages}{e30565}.
\newblock


\bibitem[Calvo-Barajas and Castellano(2022)]%
        {calvo2022understanding}
\bibfield{author}{\bibinfo{person}{Natalia Calvo-Barajas} {and} \bibinfo{person}{Ginevra Castellano}.} \bibinfo{year}{2022}\natexlab{}.
\newblock \showarticletitle{Understanding Children’s Trust Development through Repeated Interactions with a Virtual Social Robot}. In \bibinfo{booktitle}{\emph{2022 31st IEEE International Conference on Robot and Human Interactive Communication (RO-MAN)}}. IEEE, \bibinfo{pages}{1451--1458}.
\newblock


\bibitem[Campbell et~al\mbox{.}(2021)]%
        {campbell2021prevalence}
\bibfield{author}{\bibinfo{person}{Timothy~CH Campbell}, \bibinfo{person}{Andrea Reupert}, \bibinfo{person}{Keith Sutton}, \bibinfo{person}{Soumya Basu}, \bibinfo{person}{Gavin Davidson}, \bibinfo{person}{Christel~M Middeldorp}, \bibinfo{person}{Michael Naughton}, {and} \bibinfo{person}{Darryl Maybery}.} \bibinfo{year}{2021}\natexlab{}.
\newblock \showarticletitle{Prevalence of mental illness among parents of children receiving treatment within child and adolescent mental health services (CAMHS): a scoping review}.
\newblock \bibinfo{journal}{\emph{European Child \& Adolescent Psychiatry}}  \bibinfo{volume}{30} (\bibinfo{year}{2021}), \bibinfo{pages}{997--1012}.
\newblock


\bibitem[Cano et~al\mbox{.}(2021)]%
        {s21155166}
\bibfield{author}{\bibinfo{person}{Sandra Cano}, \bibinfo{person}{Carina~S. González}, \bibinfo{person}{Rosa~María Gil-Iranzo}, {and} \bibinfo{person}{Sergio Albiol-Pérez}.} \bibinfo{year}{2021}\natexlab{}.
\newblock \showarticletitle{Affective Communication for Socially Assistive Robots (SARs) for Children with Autism Spectrum Disorder: A Systematic Review}.
\newblock \bibinfo{journal}{\emph{Sensors}} \bibinfo{volume}{21}, \bibinfo{number}{15} (\bibinfo{year}{2021}).
\newblock
\showISSN{1424-8220}
\urldef\tempurl%
\url{https://doi.org/10.3390/s21155166}
\showDOI{\tempurl}


\bibitem[Cervin et~al\mbox{.}(2022)]%
        {cervin2022multi}
\bibfield{author}{\bibinfo{person}{Matti Cervin}, \bibinfo{person}{Alejandro Veas}, \bibinfo{person}{Jos{\'e}~A Piqueras}, {and} \bibinfo{person}{Agust{\'\i}n~E Mart{\'\i}nez-Gonz{\'a}lez}.} \bibinfo{year}{2022}\natexlab{}.
\newblock \showarticletitle{A multi-group confirmatory factor analysis of the revised children's anxiety and depression scale (RCADS) in Spain, Chile and Sweden}.
\newblock \bibinfo{journal}{\emph{Journal of Affective Disorders}}  \bibinfo{volume}{310} (\bibinfo{year}{2022}), \bibinfo{pages}{228--234}.
\newblock


\bibitem[Chandra and Minkovitz(2006)]%
        {chandra2006stigma}
\bibfield{author}{\bibinfo{person}{Anita Chandra} {and} \bibinfo{person}{Cynthia~S Minkovitz}.} \bibinfo{year}{2006}\natexlab{}.
\newblock \showarticletitle{Stigma starts early: Gender differences in teen willingness to use mental health services}.
\newblock \bibinfo{journal}{\emph{Journal of adolescent health}} \bibinfo{volume}{38}, \bibinfo{number}{6} (\bibinfo{year}{2006}), \bibinfo{pages}{754--e1}.
\newblock


\bibitem[Choi and Choung(2021)]%
        {Choi2021}
\bibfield{author}{\bibinfo{person}{Mina Choi} {and} \bibinfo{person}{Hyesun Choung}.} \bibinfo{year}{2021}\natexlab{}.
\newblock \showarticletitle{Mediated communication matters during the COVID-19 pandemic: The use of interpersonal and masspersonal media and psychological well-being:}.
\newblock \bibinfo{journal}{\emph{Journal of Social and Personal Relationships}}  \bibinfo{volume}{38} (\bibinfo{date}{7} \bibinfo{year}{2021}), \bibinfo{pages}{2397--2418}.
\newblock
Issue 8.
\urldef\tempurl%
\url{https://doi.org/10.1177/02654075211029378}
\showDOI{\tempurl}


\bibitem[Chorpita et~al\mbox{.}(2015)]%
        {chorpita2015revised}
\bibfield{author}{\bibinfo{person}{Bruce~F Chorpita}, \bibinfo{person}{C Ebesutani}, {and} \bibinfo{person}{SH Spence}.} \bibinfo{year}{2015}\natexlab{}.
\newblock \showarticletitle{Revised children’s anxiety and depression scale}.
\newblock \bibinfo{journal}{\emph{H{\"a}mtad fr{\aa}n}} (\bibinfo{year}{2015}).
\newblock


\bibitem[Chorpita and {\textit{et al.}}(2005)]%
        {chorpita2005psychometric}
\bibfield{author}{\bibinfo{person}{Bruce~F Chorpita} {and} \bibinfo{person}{{\textit{et al.}}}} \bibinfo{year}{2005}\natexlab{}.
\newblock \showarticletitle{Psychometric properties of the Revised Child Anxiety and Depression Scale in a clinical sample}.
\newblock \bibinfo{journal}{\emph{Behaviour research and therapy}} \bibinfo{volume}{43}, \bibinfo{number}{3} (\bibinfo{year}{2005}), \bibinfo{pages}{309--322}.
\newblock


\bibitem[Coninx et~al\mbox{.}(2016)]%
        {coninx2016towards}
\bibfield{author}{\bibinfo{person}{Alexandre Coninx}, \bibinfo{person}{Paul Baxter}, \bibinfo{person}{Elettra Oleari}, \bibinfo{person}{Sara Bellini}, \bibinfo{person}{Bert Bierman}, \bibinfo{person}{O Henkemans}, \bibinfo{person}{Lola Ca{\~n}amero}, \bibinfo{person}{Piero Cosi}, \bibinfo{person}{Valentin Enescu}, \bibinfo{person}{R Espinoza}, {et~al\mbox{.}}} \bibinfo{year}{2016}\natexlab{}.
\newblock \showarticletitle{Towards long-term social child-robot interaction: using multi-activity switching to engage young users}.
\newblock \bibinfo{journal}{\emph{Journal of Human-Robot Interaction}} (\bibinfo{year}{2016}).
\newblock


\bibitem[Conti et~al\mbox{.}(2019)]%
        {conti2019kindergarten}
\bibfield{author}{\bibinfo{person}{Daniela Conti}, \bibinfo{person}{Santo Di~Nuovo}, {and} \bibinfo{person}{Alessandro Di~Nuovo}.} \bibinfo{year}{2019}\natexlab{}.
\newblock \showarticletitle{Kindergarten children attitude towards humanoid robots: What is the effect of the first experience?}. In \bibinfo{booktitle}{\emph{2019 14th ACM/IEEE International Conference on Human-Robot Interaction (HRI)}}. IEEE, \bibinfo{pages}{630--631}.
\newblock


\bibitem[Corrigan et~al\mbox{.}(2014)]%
        {Corrigan2014}
\bibfield{author}{\bibinfo{person}{Patrick~W. Corrigan}, \bibinfo{person}{Benjamin~G. Druss}, {and} \bibinfo{person}{Deborah~A. Perlick}.} \bibinfo{year}{2014}\natexlab{}.
\newblock \showarticletitle{The Impact of Mental Illness Stigma on Seeking and Participating in Mental Health Care}.
\newblock \bibinfo{journal}{\emph{Psychological Science in the Public Interest}}  \bibinfo{volume}{15} (\bibinfo{date}{9} \bibinfo{year}{2014}), \bibinfo{pages}{37--70}.
\newblock
Issue 2.
\showISSN{15396053}
\urldef\tempurl%
\url{https://doi.org/10.1177/1529100614531398}
\showDOI{\tempurl}


\bibitem[Cross and Ramsey(2021)]%
        {Cross2021}
\bibfield{author}{\bibinfo{person}{Emily~S. Cross} {and} \bibinfo{person}{Richard Ramsey}.} \bibinfo{year}{2021}\natexlab{}.
\newblock \showarticletitle{Mind Meets Machine: Towards a Cognitive Science of Human–Machine Interactions}.
\newblock \bibinfo{journal}{\emph{Trends in Cognitive Sciences}}  \bibinfo{volume}{25} (\bibinfo{date}{3} \bibinfo{year}{2021}), \bibinfo{pages}{200--212}.
\newblock
Issue 3.
\showISSN{1364-6613}
\urldef\tempurl%
\url{https://doi.org/10.1016/J.TICS.2020.11.009}
\showDOI{\tempurl}


\bibitem[Crossman et~al\mbox{.}(2018)]%
        {crossman2018influence}
\bibfield{author}{\bibinfo{person}{Molly~K Crossman}, \bibinfo{person}{Alan~E Kazdin}, {and} \bibinfo{person}{Elizabeth~R Kitt}.} \bibinfo{year}{2018}\natexlab{}.
\newblock \showarticletitle{The influence of a socially assistive robot on mood, anxiety, and arousal in children.}
\newblock \bibinfo{journal}{\emph{Professional Psychology: Research and Practice}} \bibinfo{volume}{49}, \bibinfo{number}{1} (\bibinfo{year}{2018}), \bibinfo{pages}{48}.
\newblock


\bibitem[de~Greeff et~al\mbox{.}(2014)]%
        {de2014child}
\bibfield{author}{\bibinfo{person}{Joachim de Greeff}, \bibinfo{person}{Olivier Blanson~Henkemans}, \bibinfo{person}{Aafke Fraaije}, \bibinfo{person}{Lara Solms}, \bibinfo{person}{Noel Wigdor}, \bibinfo{person}{Bert Bierman}, \bibinfo{person}{Joris~B Janssen}, \bibinfo{person}{Rosemarijn Looije}, \bibinfo{person}{Paul Baxter}, \bibinfo{person}{Mark~A Neerincx}, {et~al\mbox{.}}} \bibinfo{year}{2014}\natexlab{}.
\newblock \showarticletitle{Child-robot interaction in the wild: Field testing activities of the ALIZ-E project}. In \bibinfo{booktitle}{\emph{Proceedings of the 2014 ACM/IEEE international conference on Human-robot interaction}}. \bibinfo{pages}{148--149}.
\newblock


\bibitem[De~Haas et~al\mbox{.}(2020)]%
        {de2020effects}
\bibfield{author}{\bibinfo{person}{Mirjam De~Haas}, \bibinfo{person}{Paul Vogt}, {and} \bibinfo{person}{Emiel Krahmer}.} \bibinfo{year}{2020}\natexlab{}.
\newblock \showarticletitle{The effects of feedback on children's engagement and learning outcomes in robot-assisted second language learning}.
\newblock \bibinfo{journal}{\emph{Frontiers in Robotics and AI}}  \bibinfo{volume}{7} (\bibinfo{year}{2020}), \bibinfo{pages}{101}.
\newblock


\bibitem[Di~Dio et~al\mbox{.}(2020)]%
        {di2020shall}
\bibfield{author}{\bibinfo{person}{Cinzia Di~Dio}, \bibinfo{person}{Federico Manzi}, \bibinfo{person}{Giulia Peretti}, \bibinfo{person}{Angelo Cangelosi}, \bibinfo{person}{Paul~L Harris}, \bibinfo{person}{Davide Massaro}, {and} \bibinfo{person}{Antonella Marchetti}.} \bibinfo{year}{2020}\natexlab{}.
\newblock \showarticletitle{Shall I trust you? From child--robot interaction to trusting relationships}.
\newblock \bibinfo{journal}{\emph{Frontiers in psychology}}  \bibinfo{volume}{11} (\bibinfo{year}{2020}), \bibinfo{pages}{469}.
\newblock


\bibitem[Di~Nuovo et~al\mbox{.}(2019)]%
        {di2019assessment}
\bibfield{author}{\bibinfo{person}{Alessandro Di~Nuovo}, \bibinfo{person}{Simone Varrasi}, \bibinfo{person}{Alexandr Lucas}, \bibinfo{person}{Daniela Conti}, \bibinfo{person}{John McNamara}, {and} \bibinfo{person}{Alessandro Soranzo}.} \bibinfo{year}{2019}\natexlab{}.
\newblock \showarticletitle{Assessment of cognitive skills via human-robot interaction and cloud computing}.
\newblock \bibinfo{journal}{\emph{Journal of bionic engineering}}  \bibinfo{volume}{16} (\bibinfo{year}{2019}), \bibinfo{pages}{526--539}.
\newblock


\bibitem[Dietrich et~al\mbox{.}(2023)]%
        {Dietrich2023}
\bibfield{author}{\bibinfo{person}{Manuel Dietrich}, \bibinfo{person}{Matti Krüger}, {and} \bibinfo{person}{Thomas~H. Weisswange}.} \bibinfo{year}{2023}\natexlab{}.
\newblock \showarticletitle{What should a robot disclose about me? A study about privacy-appropriate behaviors for social robots}.
\newblock \bibinfo{journal}{\emph{Frontiers in Robotics and AI}}  \bibinfo{volume}{10} (\bibinfo{date}{12} \bibinfo{year}{2023}), \bibinfo{pages}{1236733}.
\newblock
\showISSN{22969144}
\urldef\tempurl%
\url{https://doi.org/10.3389/FROBT.2023.1236733/BIBTEX}
\showDOI{\tempurl}


\bibitem[Faul et~al\mbox{.}(2009)]%
        {RefWorks:410}
\bibfield{author}{\bibinfo{person}{Franz Faul}, \bibinfo{person}{Edgar Erdfelder}, \bibinfo{person}{Axel Buchner}, {and} \bibinfo{person}{Albert-Georg Lang}.} \bibinfo{year}{2009}\natexlab{}.
\newblock \showarticletitle{Statistical power analyses using G*Power 3.1: Tests for correlation and regression analyses}.
\newblock \bibinfo{journal}{\emph{Behavior Research Methods}}  \bibinfo{volume}{41} (\bibinfo{year}{2009}), \bibinfo{pages}{1149--1160}.
\newblock
Issue 4.
\showISBNx{1554-3528}
\urldef\tempurl%
\url{https://doi.org/10.3758/BRM.41.4.1149}
\showDOI{\tempurl}
\newblock
\shownote{ID: Faul2009}.


\bibitem[Feil-Seifer et~al\mbox{.}(2020)]%
        {thri_covid_20}
\bibfield{author}{\bibinfo{person}{David Feil-Seifer}, \bibinfo{person}{Kerstin~S. Haring}, \bibinfo{person}{Silvia Rossi}, \bibinfo{person}{Alan~R. Wagner}, {and} \bibinfo{person}{Tom Williams}.} \bibinfo{year}{2020}\natexlab{}.
\newblock \showarticletitle{Where to Next? The Impact of COVID-19 on Human-Robot Interaction Research}.
\newblock \bibinfo{journal}{\emph{ACM Transactions on Human-Robot Interaction (THRI)}}  \bibinfo{volume}{10} (\bibinfo{date}{6} \bibinfo{year}{2020}).
\newblock
Issue 1.
\showISSN{25739522}
\urldef\tempurl%
\url{https://doi.org/10.1145/3405450}
\showDOI{\tempurl}


\bibitem[Fior et~al\mbox{.}(2010)]%
        {fior2010children}
\bibfield{author}{\bibinfo{person}{Meghann Fior}, \bibinfo{person}{Sarah Nugent}, \bibinfo{person}{Tanya~N Beran}, \bibinfo{person}{Alejandro Ramirez-Serrano}, {and} \bibinfo{person}{Roman Kuzyk}.} \bibinfo{year}{2010}\natexlab{}.
\newblock \showarticletitle{Children’s relationships with robots: robot is child’s new friend}.
\newblock  (\bibinfo{year}{2010}).
\newblock


\bibitem[Ford et~al\mbox{.}(2020)]%
        {ford2020data}
\bibfield{author}{\bibinfo{person}{Tamsin Ford}, \bibinfo{person}{Tim Vizard}, \bibinfo{person}{Katharine Sadler}, \bibinfo{person}{Sally McManus}, \bibinfo{person}{Anna Goodman}, \bibinfo{person}{Salah Merad}, \bibinfo{person}{Maria Tejerina-Arreal}, \bibinfo{person}{Dan Collinson}, {and} \bibinfo{person}{MHCYP Collaboration}.} \bibinfo{year}{2020}\natexlab{}.
\newblock \showarticletitle{Data resource profile: Mental health of children and young people (MHCYP) surveys}.
\newblock \bibinfo{journal}{\emph{International journal of epidemiology}} \bibinfo{volume}{49}, \bibinfo{number}{2} (\bibinfo{year}{2020}), \bibinfo{pages}{363--364g}.
\newblock


\bibitem[Gamborino et~al\mbox{.}(2019)]%
        {gamborino2019mood}
\bibfield{author}{\bibinfo{person}{Edwinn Gamborino}, \bibinfo{person}{Hsiu-Ping Yueh}, \bibinfo{person}{Weijane Lin}, \bibinfo{person}{Su-Ling Yeh}, {and} \bibinfo{person}{Li-Chen Fu}.} \bibinfo{year}{2019}\natexlab{}.
\newblock \showarticletitle{Mood estimation as a social profile predictor in an autonomous, multi-session, emotional support robot for children}. In \bibinfo{booktitle}{\emph{2019 28th IEEE International Conference on Robot and Human Interactive Communication (RO-MAN)}}. IEEE, \bibinfo{pages}{1--6}.
\newblock


\bibitem[Gittens(2021)]%
        {Gittens2021}
\bibfield{author}{\bibinfo{person}{Curtis~L. Gittens}.} \bibinfo{year}{2021}\natexlab{}.
\newblock \showarticletitle{Remote HRI: a Methodology for Maintaining COVID-19 Physical Distancing and Human Interaction Requirements in HRI Studies}.
\newblock \bibinfo{journal}{\emph{Information Systems Frontiers}}  \bibinfo{volume}{26} (\bibinfo{date}{2} \bibinfo{year}{2021}), \bibinfo{pages}{91--106}.
\newblock
Issue 1.
\showISSN{15729419}
\urldef\tempurl%
\url{https://doi.org/10.1007/S10796-021-10162-4/TABLES/5}
\showDOI{\tempurl}


\bibitem[Gittens and Garnes(2022a)]%
        {Gittens2022}
\bibfield{author}{\bibinfo{person}{Curtis~L. Gittens} {and} \bibinfo{person}{Damian Garnes}.} \bibinfo{year}{2022}\natexlab{a}.
\newblock \showarticletitle{Zenbo on Zoom: Evaluating the Human-Robot Interaction User Experience in a Video Conferencing Session}.
\newblock \bibinfo{journal}{\emph{Digest of Technical Papers - IEEE International Conference on Consumer Electronics}}  \bibinfo{volume}{2022-January} (\bibinfo{year}{2022}).
\newblock
\showISBNx{9781665441544}
\showISSN{0747668X}
\urldef\tempurl%
\url{https://doi.org/10.1109/ICCE53296.2022.9730259}
\showDOI{\tempurl}


\bibitem[Gittens and Garnes(2022b)]%
        {gittens2022zenbo}
\bibfield{author}{\bibinfo{person}{Curtis~L Gittens} {and} \bibinfo{person}{Damian Garnes}.} \bibinfo{year}{2022}\natexlab{b}.
\newblock \showarticletitle{Zenbo on zoom: Evaluating the human-robot interaction user experience in a video conferencing session}. In \bibinfo{booktitle}{\emph{2022 IEEE International Conference on Consumer Electronics (ICCE)}}. IEEE, \bibinfo{pages}{1--6}.
\newblock


\bibitem[Godoi et~al\mbox{.}(2020)]%
        {godoi2020proteger}
\bibfield{author}{\bibinfo{person}{Diogo Godoi}, \bibinfo{person}{Roseli~AF Romero}, \bibinfo{person}{Helio Azevedo}, \bibinfo{person}{Josu{\'e} Ramos}, \bibinfo{person}{Germano Beraldo~Filho}, {and} \bibinfo{person}{M{\'a}rcia Ap~Thome Garcia}.} \bibinfo{year}{2020}\natexlab{}.
\newblock \showarticletitle{Proteger: A social robotics system to support child psychological evaluation}. In \bibinfo{booktitle}{\emph{2020 Latin American Robotics Symposium (LARS), 2020 Brazilian Symposium on Robotics (SBR) and 2020 Workshop on Robotics in Education (WRE)}}. IEEE, \bibinfo{pages}{1--6}.
\newblock


\bibitem[Grist et~al\mbox{.}(2017)]%
        {Grist2017}
\bibfield{author}{\bibinfo{person}{Rebecca Grist}, \bibinfo{person}{Joanna Porter}, {and} \bibinfo{person}{Paul Stallard}.} \bibinfo{year}{2017}\natexlab{}.
\newblock \showarticletitle{Mental health mobile apps for preadolescents and adolescents: A systematic review}.
\newblock \bibinfo{journal}{\emph{Journal of Medical Internet Research}}  \bibinfo{volume}{19} (\bibinfo{date}{5} \bibinfo{year}{2017}), \bibinfo{pages}{e7332}.
\newblock
Issue 5.
\showISSN{14388871}
\urldef\tempurl%
\url{https://doi.org/10.2196/JMIR.7332}
\showDOI{\tempurl}


\bibitem[Gromada et~al\mbox{.}(2020)]%
        {gromada2020worlds}
\bibfield{author}{\bibinfo{person}{Anna Gromada}, \bibinfo{person}{Gwyther Rees}, {and} \bibinfo{person}{Yekaterina Chzhen}.} \bibinfo{year}{2020}\natexlab{}.
\newblock \bibinfo{booktitle}{\emph{Worlds of influence: Understanding what shapes child well-being in rich countries}}.
\newblock \bibinfo{publisher}{United Nations Children’s Fund}.
\newblock


\bibitem[Guracho et~al\mbox{.}(2023)]%
        {Guracho2023}
\bibfield{author}{\bibinfo{person}{Yonas~Deressa Guracho}, \bibinfo{person}{Susan~J. Thomas}, {and} \bibinfo{person}{Khin~Than Win}.} \bibinfo{year}{2023}\natexlab{}.
\newblock \showarticletitle{Smartphone application use patterns for mental health disorders: A systematic literature review and meta-analysis}.
\newblock \bibinfo{journal}{\emph{International Journal of Medical Informatics}}  \bibinfo{volume}{179} (\bibinfo{date}{11} \bibinfo{year}{2023}), \bibinfo{pages}{105217}.
\newblock
\showISSN{1386-5056}
\urldef\tempurl%
\url{https://doi.org/10.1016/J.IJMEDINF.2023.105217}
\showDOI{\tempurl}


\bibitem[Harrer(2023)]%
        {Harrer2023}
\bibfield{author}{\bibinfo{person}{Stefan Harrer}.} \bibinfo{year}{2023}\natexlab{}.
\newblock \showarticletitle{Attention is not all you need: the complicated case of ethically using large language models in healthcare and medicine}.
\newblock \bibinfo{journal}{\emph{eBioMedicine}}  \bibinfo{volume}{90} (\bibinfo{date}{4} \bibinfo{year}{2023}), \bibinfo{pages}{104512}.
\newblock
\showISSN{2352-3964}
\urldef\tempurl%
\url{https://doi.org/10.1016/J.EBIOM.2023.104512}
\showDOI{\tempurl}


\bibitem[Henkemans et~al\mbox{.}(2017)]%
        {henkemans2017design}
\bibfield{author}{\bibinfo{person}{Olivier A~Blanson Henkemans}, \bibinfo{person}{Bert~PB Bierman}, \bibinfo{person}{Joris Janssen}, \bibinfo{person}{Rosemarijn Looije}, \bibinfo{person}{Mark~A Neerincx}, \bibinfo{person}{Marierose~MM van Dooren}, \bibinfo{person}{Jitske~LE de Vries}, \bibinfo{person}{Gert~Jan van~der Burg}, {and} \bibinfo{person}{Sasja~D Huisman}.} \bibinfo{year}{2017}\natexlab{}.
\newblock \showarticletitle{Design and evaluation of a personal robot playing a self-management education game with children with diabetes type 1}.
\newblock \bibinfo{journal}{\emph{International Journal of Human-Computer Studies}}  \bibinfo{volume}{106} (\bibinfo{year}{2017}), \bibinfo{pages}{63--76}.
\newblock


\bibitem[Henschel et~al\mbox{.}(2021)]%
        {Henschel2021}
\bibfield{author}{\bibinfo{person}{Anna Henschel}, \bibinfo{person}{Guy Laban}, {and} \bibinfo{person}{Emily~S Cross}.} \bibinfo{year}{2021}\natexlab{}.
\newblock \showarticletitle{What Makes a Robot Social? A Review of Social Robots from Science Fiction to a Home or Hospital Near You}.
\newblock \bibinfo{journal}{\emph{Current Robotics Reports}} (\bibinfo{year}{2021}), \bibinfo{pages}{9--19}.
\newblock
Issue 2.
\showISSN{2662-4087}
\urldef\tempurl%
\url{https://doi.org/10.1007/s43154-020-00035-0}
\showDOI{\tempurl}


\bibitem[Honig and Oron-Gilad(2020)]%
        {Honig2020}
\bibfield{author}{\bibinfo{person}{Shanee Honig} {and} \bibinfo{person}{Tal Oron-Gilad}.} \bibinfo{year}{2020}\natexlab{}.
\newblock \showarticletitle{Comparing Laboratory User Studies and Video-Enhanced Web Surveys for Eliciting User Gestures in Human-Robot Interactions}.
\newblock \bibinfo{journal}{\emph{Companion of the 2020 ACM/IEEE International Conference on Human-Robot Interaction}}, \bibinfo{pages}{248--250}.
\newblock
\showISBNx{9781450370578}
\showISSN{21672148}
\urldef\tempurl%
\url{https://doi.org/10.1145/3371382.3378325}
\showDOI{\tempurl}


\bibitem[Horstmann and Kr{\"a}mer(2019)]%
        {horstmann2019great}
\bibfield{author}{\bibinfo{person}{Aike~C Horstmann} {and} \bibinfo{person}{Nicole~C Kr{\"a}mer}.} \bibinfo{year}{2019}\natexlab{}.
\newblock \showarticletitle{Great expectations? Relation of previous experiences with social robots in real life or in the media and expectancies based on qualitative and quantitative assessment}.
\newblock \bibinfo{journal}{\emph{Frontiers in psychology}}  \bibinfo{volume}{10} (\bibinfo{year}{2019}), \bibinfo{pages}{939}.
\newblock


\bibitem[Huberty et~al\mbox{.}(2021)]%
        {huberty2021evaluation}
\bibfield{author}{\bibinfo{person}{Jennifer Huberty}, \bibinfo{person}{Jeni Green}, \bibinfo{person}{Megan Puzia}, {and} \bibinfo{person}{Chad Stecher}.} \bibinfo{year}{2021}\natexlab{}.
\newblock \showarticletitle{Evaluation of mood check-in feature for participation in meditation mobile app users: retrospective longitudinal analysis}.
\newblock \bibinfo{journal}{\emph{JMIR mHealth and uHealth}} \bibinfo{volume}{9}, \bibinfo{number}{4} (\bibinfo{year}{2021}), \bibinfo{pages}{e27106}.
\newblock


\bibitem[Jachens and Houdmont(2019)]%
        {jachens2019effort}
\bibfield{author}{\bibinfo{person}{Liza Jachens} {and} \bibinfo{person}{Jonathan Houdmont}.} \bibinfo{year}{2019}\natexlab{}.
\newblock \showarticletitle{Effort-reward imbalance and job strain: a composite indicator approach}.
\newblock \bibinfo{journal}{\emph{International journal of environmental research and public health}} \bibinfo{volume}{16}, \bibinfo{number}{21} (\bibinfo{year}{2019}), \bibinfo{pages}{4169}.
\newblock


\bibitem[Jackson et~al\mbox{.}(2008)]%
        {jackson2008culture}
\bibfield{author}{\bibinfo{person}{Linda~A Jackson}, \bibinfo{person}{Yong Zhao}, \bibinfo{person}{Wei Qiu}, \bibinfo{person}{Anthony Kolenic~III}, \bibinfo{person}{Hiram~E Fitzgerald}, \bibinfo{person}{Rena Harold}, {and} \bibinfo{person}{Alexander von Eye}.} \bibinfo{year}{2008}\natexlab{}.
\newblock \showarticletitle{Culture, gender and information technology use: A comparison of Chinese and US children}.
\newblock \bibinfo{journal}{\emph{Computers in human behavior}} \bibinfo{volume}{24}, \bibinfo{number}{6} (\bibinfo{year}{2008}), \bibinfo{pages}{2817--2829}.
\newblock


\bibitem[Johnson et~al\mbox{.}(2021)]%
        {Johnson2021}
\bibfield{author}{\bibinfo{person}{Ashleigh~M. Johnson}, \bibinfo{person}{Carolyn~A. McCarty}, \bibinfo{person}{Lyscha~A. Marcynyszyn}, \bibinfo{person}{Douglas~F. Zatzick}, \bibinfo{person}{Sara~P.D. Chrisman}, {and} \bibinfo{person}{Frederick~P. Rivara}.} \bibinfo{year}{2021}\natexlab{}.
\newblock \showarticletitle{Child- compared with parent-report ratings on psychosocial measures following a mild traumatic brain injury among youth with persistent post-concussion symptoms}.
\newblock \bibinfo{journal}{\emph{Brain Injury}}  \bibinfo{volume}{35} (\bibinfo{date}{4} \bibinfo{year}{2021}), \bibinfo{pages}{574--586}.
\newblock
Issue 5.
\showISSN{1362301X}
\urldef\tempurl%
\url{https://doi.org/10.1080/02699052.2021.1889663}
\showDOI{\tempurl}


\bibitem[Kanda et~al\mbox{.}(2004)]%
        {kanda2004interactive}
\bibfield{author}{\bibinfo{person}{Takayuki Kanda}, \bibinfo{person}{Takayuki Hirano}, \bibinfo{person}{Daniel Eaton}, {and} \bibinfo{person}{Hiroshi Ishiguro}.} \bibinfo{year}{2004}\natexlab{}.
\newblock \showarticletitle{Interactive robots as social partners and peer tutors for children: A field trial}.
\newblock \bibinfo{journal}{\emph{Human--Computer Interaction}} \bibinfo{volume}{19}, \bibinfo{number}{1-2} (\bibinfo{year}{2004}), \bibinfo{pages}{61--84}.
\newblock


\bibitem[Kennedy et~al\mbox{.}(2015)]%
        {kennedy2015comparing}
\bibfield{author}{\bibinfo{person}{James Kennedy}, \bibinfo{person}{Paul Baxter}, {and} \bibinfo{person}{Tony Belpaeme}.} \bibinfo{year}{2015}\natexlab{}.
\newblock \showarticletitle{Comparing robot embodiments in a guided discovery learning interaction with children}.
\newblock \bibinfo{journal}{\emph{International Journal of Social Robotics}}  \bibinfo{volume}{7} (\bibinfo{year}{2015}), \bibinfo{pages}{293--308}.
\newblock


\bibitem[Kim et~al\mbox{.}(2024)]%
        {Kim2024}
\bibfield{author}{\bibinfo{person}{Callie~Y. Kim}, \bibinfo{person}{Christine~P. Lee}, {and} \bibinfo{person}{Bilge Mutlu}.} \bibinfo{year}{2024}\natexlab{}.
\newblock \showarticletitle{Understanding Large-Language Model (LLM)-powered Human-Robot Interaction}.
\newblock \bibinfo{journal}{\emph{ACM/IEEE International Conference on Human-Robot Interaction}} (\bibinfo{date}{3} \bibinfo{year}{2024}), \bibinfo{pages}{371--380}.
\newblock
\showISBNx{9798400703225}
\showISSN{21672148}
\urldef\tempurl%
\url{https://doi.org/10.1145/3610977.3634966}
\showDOI{\tempurl}


\bibitem[Kory-Westlund and Breazeal(2019)]%
        {kory2019long}
\bibfield{author}{\bibinfo{person}{Jacqueline~M Kory-Westlund} {and} \bibinfo{person}{Cynthia Breazeal}.} \bibinfo{year}{2019}\natexlab{}.
\newblock \showarticletitle{A long-term study of young children's rapport, social emulation, and language learning with a peer-like robot playmate in preschool}.
\newblock \bibinfo{journal}{\emph{Frontiers in Robotics and AI}}  \bibinfo{volume}{6} (\bibinfo{year}{2019}), \bibinfo{pages}{81}.
\newblock


\bibitem[Kozima and Nakagawa(2007)]%
        {kozima2007longitudinal}
\bibfield{author}{\bibinfo{person}{Hideki Kozima} {and} \bibinfo{person}{Cocoro Nakagawa}.} \bibinfo{year}{2007}\natexlab{}.
\newblock \showarticletitle{Longitudinal Child-Robot Interaction at Preschool.}. In \bibinfo{booktitle}{\emph{AAAI Spring Symposium: Multidisciplinary Collaboration for Socially Assistive Robotics}}. \bibinfo{pages}{27--32}.
\newblock


\bibitem[Kumar~Singh et~al\mbox{.}(2020)]%
        {kumar2020toy}
\bibfield{author}{\bibinfo{person}{Divyanshu Kumar~Singh}, \bibinfo{person}{Sumita Sharma}, \bibinfo{person}{Jainendra Shukla}, {and} \bibinfo{person}{Grace Eden}.} \bibinfo{year}{2020}\natexlab{}.
\newblock \showarticletitle{Toy, tutor, peer, or pet? preliminary findings from child-robot interactions in a community school}. In \bibinfo{booktitle}{\emph{Companion of the 2020 ACM/IEEE international conference on human-robot interaction}}. \bibinfo{pages}{325--327}.
\newblock


\bibitem[Kurian(2023)]%
        {Kurian2023}
\bibfield{author}{\bibinfo{person}{Nomisha Kurian}.} \bibinfo{year}{2023}\natexlab{}.
\newblock \showarticletitle{AI's empathy gap: The risks of conversational Artificial Intelligence for young children's well-being and key ethical considerations for early childhood education and care}.
\newblock \bibinfo{journal}{\emph{Contemporary Issues in Early Childhood}} (\bibinfo{date}{10} \bibinfo{year}{2023}).
\newblock
\showISSN{14639491}
\urldef\tempurl%
\url{https://doi.org/10.1177/14639491231206004}
\showDOI{\tempurl}


\bibitem[Kurian(2024)]%
        {Kurian2024}
\bibfield{author}{\bibinfo{person}{Nomisha Kurian}.} \bibinfo{year}{2024}\natexlab{}.
\newblock \showarticletitle{‘No, Alexa, no!’: designing child-safe AI and protecting children from the risks of the ‘empathy gap’ in large language models}.
\newblock \bibinfo{journal}{\emph{Learning, Media and Technology}} (\bibinfo{date}{6} \bibinfo{year}{2024}).
\newblock
\showISSN{17439892}
\urldef\tempurl%
\url{https://doi.org/10.1080/17439884.2024.2367052}
\showDOI{\tempurl}


\bibitem[Laban(2024)]%
        {dISC2024}
\bibfield{author}{\bibinfo{person}{Guy Laban}.} \bibinfo{year}{2024}\natexlab{}.
\newblock \showarticletitle{Studying and Eliciting Self-Disclosure: Interdisciplinary Review of Research Methodologies and Behavioural Paradigms}.
\newblock \bibinfo{journal}{\emph{PsyArxiv}} (\bibinfo{date}{5} \bibinfo{year}{2024}).
\newblock
\urldef\tempurl%
\url{https://doi.org/10.31234/OSF.IO/FHDEK}
\showDOI{\tempurl}


\bibitem[Laban et~al\mbox{.}(2022a)]%
        {robot_post}
\bibfield{author}{\bibinfo{person}{Guy Laban}, \bibinfo{person}{Ziv Ben-Zion}, {and} \bibinfo{person}{Emily~S. Cross}.} \bibinfo{year}{2022}\natexlab{a}.
\newblock \showarticletitle{Social Robots for Supporting Post-traumatic Stress Disorder Diagnosis and Treatment}.
\newblock \bibinfo{journal}{\emph{Frontiers in Psychiatry}}  \bibinfo{volume}{12} (\bibinfo{year}{2022}), \bibinfo{pages}{2610}.
\newblock
\showISSN{1664-0640}
\urldef\tempurl%
\url{https://doi.org/10.3389/FPSYT.2021.752874}
\showDOI{\tempurl}


\bibitem[Laban and Cross(2024)]%
        {Sharing2024}
\bibfield{author}{\bibinfo{person}{Guy Laban} {and} \bibinfo{person}{Emily~S. Cross}.} \bibinfo{year}{2024}\natexlab{}.
\newblock \showarticletitle{Sharing our Emotions with Robots: Why do we do it and how does it make us feel?}
\newblock \bibinfo{journal}{\emph{IEEE Transactions on Affective Computing}} (\bibinfo{year}{2024}), \bibinfo{pages}{1--18}.
\newblock
\showISSN{1949-3045}
\urldef\tempurl%
\url{https://doi.org/10.1109/TAFFC.2024.3470984}
\showDOI{\tempurl}


\bibitem[Laban et~al\mbox{.}(2021)]%
        {Laban2021}
\bibfield{author}{\bibinfo{person}{Guy Laban}, \bibinfo{person}{Jean-Noël George}, \bibinfo{person}{Val Morrison}, {and} \bibinfo{person}{Emily~S. Cross}.} \bibinfo{year}{2021}\natexlab{}.
\newblock \showarticletitle{Tell me more! Assessing interactions with social robots from speech}.
\newblock \bibinfo{journal}{\emph{Paladyn, Journal of Behavioral Robotics}}  \bibinfo{volume}{12} (\bibinfo{year}{2021}), \bibinfo{pages}{136--159}.
\newblock
Issue 1.
\showISSN{20814836}
\urldef\tempurl%
\url{https://doi.org/10.1515/pjbr-2021-0011}
\showDOI{\tempurl}


\bibitem[Laban et~al\mbox{.}(2023)]%
        {Laban2023}
\bibfield{author}{\bibinfo{person}{Guy Laban}, \bibinfo{person}{Arvid Kappas}, \bibinfo{person}{Val Morrison}, {and} \bibinfo{person}{Emily~S. Cross}.} \bibinfo{year}{2023}\natexlab{}.
\newblock \showarticletitle{Opening Up to Social Robots: How Emotions Drive Self-Disclosure Behavior}.
\newblock \bibinfo{journal}{\emph{2023 32nd IEEE International Conference on Robot and Human Interactive Communication (RO-MAN)}}, \bibinfo{pages}{1697--1704}.
\newblock
\showISBNx{979-8-3503-3670-2}
\urldef\tempurl%
\url{https://doi.org/10.1109/RO-MAN57019.2023.10309551}
\showDOI{\tempurl}


\bibitem[Laban et~al\mbox{.}(2024)]%
        {Laban_blt_2023}
\bibfield{author}{\bibinfo{person}{Guy Laban}, \bibinfo{person}{Arvid Kappas}, \bibinfo{person}{Val Morrison}, {and} \bibinfo{person}{Emily~S Cross}.} \bibinfo{year}{2024}\natexlab{}.
\newblock \showarticletitle{Building Long-Term Human–Robot Relationships: Examining Disclosure, Perception and Well-Being Across Time}.
\newblock \bibinfo{journal}{\emph{International Journal of Social Robotics}}  \bibinfo{volume}{16} (\bibinfo{year}{2024}), \bibinfo{pages}{1--27}.
\newblock
Issue 5.
\showISSN{1875-4805}
\urldef\tempurl%
\url{https://doi.org/10.1007/s12369-023-01076-z}
\showDOI{\tempurl}


\bibitem[Laban et~al\mbox{.}(2022b)]%
        {laban_icd_2022}
\bibfield{author}{\bibinfo{person}{Guy Laban}, \bibinfo{person}{Val Morrison}, \bibinfo{person}{Arvid Kappas}, {and} \bibinfo{person}{Emily~S Cross}.} \bibinfo{year}{2022}\natexlab{b}.
\newblock \showarticletitle{Informal Caregivers Disclose Increasingly More to a Social Robot Over Time}.
\newblock \bibinfo{journal}{\emph{CHI Conference on Human Factors in Computing Systems Extended Abstracts}}, \bibinfo{pages}{1--7}.
\newblock
\showISBNx{9781450391566}
\urldef\tempurl%
\url{https://doi.org/10.1145/3491101.3519666}
\showDOI{\tempurl}


\bibitem[Laban et~al\mbox{.}(2025)]%
        {coping_ijsr}
\bibfield{author}{\bibinfo{person}{Guy Laban}, \bibinfo{person}{Val Morrison}, \bibinfo{person}{Arvid Kappas}, {and} \bibinfo{person}{Emily~S. Cross}.} \bibinfo{year}{2025}\natexlab{}.
\newblock \showarticletitle{Coping with Emotional Distress via Self-Disclosure to Robots: Intervention with Caregivers}.
\newblock \bibinfo{journal}{\emph{International Journal of Social Robotics}} (\bibinfo{year}{2025}).
\newblock
\urldef\tempurl%
\url{https://doi.org/10.1007/s12369-024-01207-0}
\showDOI{\tempurl}


\bibitem[Leite et~al\mbox{.}(2012)]%
        {leite2012long}
\bibfield{author}{\bibinfo{person}{Iolanda Leite}, \bibinfo{person}{Ginevra Castellano}, \bibinfo{person}{Andr{\'e} Pereira}, \bibinfo{person}{Carlos Martinho}, {and} \bibinfo{person}{Ana Paiva}.} \bibinfo{year}{2012}\natexlab{}.
\newblock \showarticletitle{Long-term interactions with empathic robots: Evaluating perceived support in children}. In \bibinfo{booktitle}{\emph{Social Robotics: 4th International Conference, ICSR 2012, Chengdu, China, October 29-31, 2012. Proceedings 4}}. Springer, \bibinfo{pages}{298--307}.
\newblock


\bibitem[Leite et~al\mbox{.}(2013)]%
        {leite2013social}
\bibfield{author}{\bibinfo{person}{Iolanda Leite}, \bibinfo{person}{Carlos Martinho}, {and} \bibinfo{person}{Ana Paiva}.} \bibinfo{year}{2013}\natexlab{}.
\newblock \showarticletitle{Social robots for long-term interaction: a survey}.
\newblock \bibinfo{journal}{\emph{International Journal of Social Robotics}}  \bibinfo{volume}{5} (\bibinfo{year}{2013}), \bibinfo{pages}{291--308}.
\newblock


\bibitem[Levinson et~al\mbox{.}(2024)]%
        {levinson2024expert}
\bibfield{author}{\bibinfo{person}{Leigh~M Levinson}, \bibinfo{person}{Nida~Itrat Abbasi}, \bibinfo{person}{Selma Sabanovic}, {and} \bibinfo{person}{Hatice Gunes}.} \bibinfo{year}{2024}\natexlab{}.
\newblock \showarticletitle{Expert Insights on Robots for Safeguarding Children: How (not) and Why (not)?}. In \bibinfo{booktitle}{\emph{Proceedings of the 23rd Annual ACM Interaction Design and Children Conference}}. \bibinfo{pages}{600--611}.
\newblock


\bibitem[Li(2015)]%
        {li2015benefit}
\bibfield{author}{\bibinfo{person}{Jamy Li}.} \bibinfo{year}{2015}\natexlab{}.
\newblock \showarticletitle{The benefit of being physically present: A survey of experimental works comparing copresent robots, telepresent robots and virtual agents}.
\newblock \bibinfo{journal}{\emph{International Journal of Human-Computer Studies}}  \bibinfo{volume}{77} (\bibinfo{year}{2015}), \bibinfo{pages}{23--37}.
\newblock


\bibitem[Liao and Vaughan(2024)]%
        {Liao2024}
\bibfield{author}{\bibinfo{person}{Q.~Vera Liao} {and} \bibinfo{person}{Jennifer~Wortman Vaughan}.} \bibinfo{year}{2024}\natexlab{}.
\newblock \showarticletitle{AI Transparency in the Age of LLMs: A Human-Centered Research Roadmap}.
\newblock \bibinfo{journal}{\emph{Harvard Data Science Review}} (\bibinfo{date}{2} \bibinfo{year}{2024}).
\newblock
Issue Special Issue 5.
\showISSN{2644-2353}
\urldef\tempurl%
\url{https://doi.org/10.1162/99608F92.8036D03B}
\showDOI{\tempurl}


\bibitem[Lindsey et~al\mbox{.}(2010)]%
        {lindsey2010family}
\bibfield{author}{\bibinfo{person}{Michael~A Lindsey}, \bibinfo{person}{Sean Joe}, {and} \bibinfo{person}{Von Nebbitt}.} \bibinfo{year}{2010}\natexlab{}.
\newblock \showarticletitle{Family matters: The role of mental health stigma and social support on depressive symptoms and subsequent help seeking among African American boys}.
\newblock \bibinfo{journal}{\emph{Journal of Black Psychology}} \bibinfo{volume}{36}, \bibinfo{number}{4} (\bibinfo{year}{2010}), \bibinfo{pages}{458--482}.
\newblock


\bibitem[Lisøy et~al\mbox{.}(2022)]%
        {lisa2022}
\bibfield{author}{\bibinfo{person}{Carina Lisøy}, \bibinfo{person}{Simon~Peter Neumer}, \bibinfo{person}{Trine Waaktaar}, \bibinfo{person}{Jo~Magne Ingul}, \bibinfo{person}{Solveig Holen}, {and} \bibinfo{person}{Kristin Martinsen}.} \bibinfo{year}{2022}\natexlab{}.
\newblock \showarticletitle{Making high‐quality measures available in diverse contexts—The psychometric properties of the Revised Child Anxiety and Depression Scale in a Norwegian sample}.
\newblock \bibinfo{journal}{\emph{International Journal of Methods in Psychiatric Research}}  \bibinfo{volume}{31} (\bibinfo{date}{12} \bibinfo{year}{2022}), \bibinfo{pages}{e1935}.
\newblock
Issue 4.
\showISSN{15570657}
\urldef\tempurl%
\url{https://doi.org/10.1002/MPR.1935}
\showDOI{\tempurl}


\bibitem[Lytridis et~al\mbox{.}(2020)]%
        {lytridis2020distance}
\bibfield{author}{\bibinfo{person}{Chris Lytridis}, \bibinfo{person}{Christos Bazinas}, \bibinfo{person}{George Sidiropoulos}, \bibinfo{person}{George~A Papakostas}, \bibinfo{person}{Vassilis~G Kaburlasos}, \bibinfo{person}{Vasiliki-Aliki Nikopoulou}, \bibinfo{person}{Vasiliki Holeva}, {and} \bibinfo{person}{Athanasios Evangeliou}.} \bibinfo{year}{2020}\natexlab{}.
\newblock \showarticletitle{Distance special education delivery by social robots}.
\newblock \bibinfo{journal}{\emph{Electronics}} \bibinfo{volume}{9}, \bibinfo{number}{6} (\bibinfo{year}{2020}), \bibinfo{pages}{1034}.
\newblock


\bibitem[Mansfield et~al\mbox{.}(2020)]%
        {mansfield2020oxwell}
\bibfield{author}{\bibinfo{person}{Karen Mansfield}, \bibinfo{person}{Christoph Jindra}, {and} \bibinfo{person}{Mina Fazel}.} \bibinfo{year}{2020}\natexlab{}.
\newblock \showarticletitle{The OxWell School Survey 2020}.
\newblock \bibinfo{journal}{\emph{Report of Preliminary Findings}} (\bibinfo{year}{2020}).
\newblock


\bibitem[Marsh and Hau(2002)]%
        {marsh2002multilevel}
\bibfield{author}{\bibinfo{person}{Herbert~W Marsh} {and} \bibinfo{person}{Kit-Tai Hau}.} \bibinfo{year}{2002}\natexlab{}.
\newblock \showarticletitle{Multilevel modeling of longitudinal growth and change: Substantive effects or regression toward the mean artifacts?}
\newblock \bibinfo{journal}{\emph{Multivariate Behavioral Research}} \bibinfo{volume}{37}, \bibinfo{number}{2} (\bibinfo{year}{2002}), \bibinfo{pages}{245--282}.
\newblock


\bibitem[Meghdari et~al\mbox{.}(2018)]%
        {meghdari2018arash}
\bibfield{author}{\bibinfo{person}{Ali Meghdari}, \bibinfo{person}{Azadeh Shariati}, \bibinfo{person}{Minoo Alemi}, \bibinfo{person}{Gholamreza~R Vossoughi}, \bibinfo{person}{Abdollah Eydi}, \bibinfo{person}{Ehsan Ahmadi}, \bibinfo{person}{Behrad Mozafari}, \bibinfo{person}{Ali Amoozandeh~Nobaveh}, {and} \bibinfo{person}{Reza Tahami}.} \bibinfo{year}{2018}\natexlab{}.
\newblock \showarticletitle{Arash: A social robot buddy to support children with cancer in a hospital environment}.
\newblock \bibinfo{journal}{\emph{Proceedings of the Institution of Mechanical Engineers, Part H: Journal of Engineering in Medicine}} \bibinfo{volume}{232}, \bibinfo{number}{6} (\bibinfo{year}{2018}), \bibinfo{pages}{605--618}.
\newblock


\bibitem[Munir et~al\mbox{.}(2014)]%
        {munir2014occupational}
\bibfield{author}{\bibinfo{person}{Fehmidah Munir}, \bibinfo{person}{Jonathan Houdmont}, \bibinfo{person}{Stacy Clemes}, \bibinfo{person}{Kelly Wilson}, \bibinfo{person}{Robert Kerr}, {and} \bibinfo{person}{Ken Addley}.} \bibinfo{year}{2014}\natexlab{}.
\newblock \showarticletitle{Occupational sitting time and its association with work engagement and job demand-control}. In \bibinfo{booktitle}{\emph{European Academy of Occupational Health Psychology Conference}}. European Academy of Occupational Health Psychology.
\newblock


\bibitem[Murray et~al\mbox{.}(2022)]%
        {murray2022learning}
\bibfield{author}{\bibinfo{person}{Michael Murray}, \bibinfo{person}{Nick Walker}, \bibinfo{person}{Amal Nanavati}, \bibinfo{person}{Patricia Alves-Oliveira}, \bibinfo{person}{Nikita Filippov}, \bibinfo{person}{Allison Sauppe}, \bibinfo{person}{Bilge Mutlu}, {and} \bibinfo{person}{Maya Cakmak}.} \bibinfo{year}{2022}\natexlab{}.
\newblock \showarticletitle{Learning backchanneling behaviors for a social robot via data augmentation from human-human conversations}. In \bibinfo{booktitle}{\emph{Conference on robot learning}}. PMLR, \bibinfo{pages}{513--525}.
\newblock


\bibitem[Mutlu et~al\mbox{.}(2006)]%
        {mutlu2006task}
\bibfield{author}{\bibinfo{person}{Bilge Mutlu}, \bibinfo{person}{Steven Osman}, \bibinfo{person}{Jodi Forlizzi}, \bibinfo{person}{Jessica Hodgins}, {and} \bibinfo{person}{Sara Kiesler}.} \bibinfo{year}{2006}\natexlab{}.
\newblock \showarticletitle{Task structure and user attributes as elements of human-robot interaction design}. In \bibinfo{booktitle}{\emph{ROMAN 2006-The 15th IEEE International Symposium on Robot and Human Interactive Communication}}. IEEE, \bibinfo{pages}{74--79}.
\newblock


\bibitem[Nalin et~al\mbox{.}(2012a)]%
        {nalin2012children}
\bibfield{author}{\bibinfo{person}{Marco Nalin}, \bibinfo{person}{Ilaria Baroni}, \bibinfo{person}{Ivana Kruijff-Korbayov{\'a}}, \bibinfo{person}{Lola Canamero}, \bibinfo{person}{Matthew Lewis}, \bibinfo{person}{Aryel Beck}, \bibinfo{person}{Heriberto Cuay{\'a}huitl}, {and} \bibinfo{person}{Alberto Sanna}.} \bibinfo{year}{2012}\natexlab{a}.
\newblock \showarticletitle{Children's adaptation in multi-session interaction with a humanoid robot}. In \bibinfo{booktitle}{\emph{2012 IEEE RO-MAN: The 21st IEEE International Symposium on Robot and Human Interactive Communication}}. IEEE, \bibinfo{pages}{351--357}.
\newblock


\bibitem[Nalin et~al\mbox{.}(2012b)]%
        {nalin2012robotic}
\bibfield{author}{\bibinfo{person}{Marco Nalin}, \bibinfo{person}{Ilaria Baroni}, \bibinfo{person}{Alberto Sanna}, {and} \bibinfo{person}{Clara Pozzi}.} \bibinfo{year}{2012}\natexlab{b}.
\newblock \showarticletitle{Robotic companion for diabetic children: emotional and educational support to diabetic children, through an interactive robot}. In \bibinfo{booktitle}{\emph{Proceedings of the 11th international conference on interaction design and children}}. \bibinfo{pages}{260--263}.
\newblock


\bibitem[Nock and Kazdin(2001)]%
        {nock2001parent}
\bibfield{author}{\bibinfo{person}{Matthew~K Nock} {and} \bibinfo{person}{Alan~E Kazdin}.} \bibinfo{year}{2001}\natexlab{}.
\newblock \showarticletitle{Parent expectancies for child therapy: Assessment and relation to participation in treatment}.
\newblock \bibinfo{journal}{\emph{Journal of Child and Family Studies}}  \bibinfo{volume}{10} (\bibinfo{year}{2001}), \bibinfo{pages}{155--180}.
\newblock


\bibitem[Nomura et~al\mbox{.}(2006)]%
        {nomura2006measurement}
\bibfield{author}{\bibinfo{person}{Tatsuya Nomura}, \bibinfo{person}{Tomohiro Suzuki}, \bibinfo{person}{Takayuki Kanda}, {and} \bibinfo{person}{Kensuke Kato}.} \bibinfo{year}{2006}\natexlab{}.
\newblock \showarticletitle{Measurement of anxiety toward robots}. In \bibinfo{booktitle}{\emph{ROMAN 2006-The 15th IEEE International Symposium on Robot and Human Interactive Communication}}. IEEE, \bibinfo{pages}{372--377}.
\newblock


\bibitem[O'Reilly and \textit{et al.}(2015)]%
        {OReilly2015}
\bibfield{author}{\bibinfo{person}{Michelle O'Reilly} {and} \bibinfo{person}{\textit{et al.}}} \bibinfo{year}{2015}\natexlab{}.
\newblock \showarticletitle{Identifying the interactional processes in the first assessments in child mental health}.
\newblock \bibinfo{journal}{\emph{Child and Adolescent Mental Health}}  \bibinfo{volume}{20} (\bibinfo{date}{11} \bibinfo{year}{2015}), \bibinfo{pages}{195--201}.
\newblock
Issue 4.
\showISSN{1475-3588}
\urldef\tempurl%
\url{https://doi.org/10.1111/CAMH.12077}
\showDOI{\tempurl}


\bibitem[Pasupuleti et~al\mbox{.}(2022)]%
        {pasupuleti2022exploring}
\bibfield{author}{\bibinfo{person}{Devasena Pasupuleti}, \bibinfo{person}{Sreejith Sasidharan}, \bibinfo{person}{Rajesh Sharma}, {and} \bibinfo{person}{Gayathri Manikutty}.} \bibinfo{year}{2022}\natexlab{}.
\newblock \showarticletitle{Exploring collaborative game play with robots to encourage good hand hygiene practises among children}. In \bibinfo{booktitle}{\emph{2022 31st IEEE International Conference on Robot and Human Interactive Communication (RO-MAN)}}. IEEE, \bibinfo{pages}{308--315}.
\newblock


\bibitem[Pearson(2023)]%
        {pearson2023masculinity}
\bibfield{author}{\bibinfo{person}{Rebecca Pearson}.} \bibinfo{year}{2023}\natexlab{}.
\newblock \showarticletitle{Masculinity and emotionality in education: critical reflections on discourses of boys’ behaviour and mental health}.
\newblock \bibinfo{journal}{\emph{Educational Review}} \bibinfo{volume}{75}, \bibinfo{number}{6} (\bibinfo{year}{2023}), \bibinfo{pages}{1101--1130}.
\newblock


\bibitem[Punukollu and Marques(2019a)]%
        {punukollu2019use}
\bibfield{author}{\bibinfo{person}{Mallika Punukollu} {and} \bibinfo{person}{Mafalda Marques}.} \bibinfo{year}{2019}\natexlab{a}.
\newblock \showarticletitle{Use of mobile apps and technologies in child and adolescent mental health: a systematic review}.
\newblock \bibinfo{journal}{\emph{BMJ Ment Health}} \bibinfo{volume}{22}, \bibinfo{number}{4} (\bibinfo{year}{2019}), \bibinfo{pages}{161--166}.
\newblock


\bibitem[Punukollu and Marques(2019b)]%
        {Punukollu2019}
\bibfield{author}{\bibinfo{person}{Mallika Punukollu} {and} \bibinfo{person}{Mafalda Marques}.} \bibinfo{year}{2019}\natexlab{b}.
\newblock \showarticletitle{Use of mobile apps and technologies in child and adolescent mental health: a systematic review}.
\newblock \bibinfo{journal}{\emph{Evidence-Based Mental Health}}  \bibinfo{volume}{22} (\bibinfo{date}{11} \bibinfo{year}{2019}), \bibinfo{pages}{161}.
\newblock
Issue 4.
\showISSN{13620347}
\urldef\tempurl%
\url{https://doi.org/10.1136/EBMENTAL-2019-300093}
\showDOI{\tempurl}


\bibitem[Rapinda et~al\mbox{.}(2021)]%
        {rapinda2021examining}
\bibfield{author}{\bibinfo{person}{Karli~K Rapinda}, \bibinfo{person}{Tyler Kempe}, \bibinfo{person}{Richard~S Kruk}, \bibinfo{person}{Jason~D Edgerton}, \bibinfo{person}{Harold~R Wallbridge}, {and} \bibinfo{person}{Matthew~T Keough}.} \bibinfo{year}{2021}\natexlab{}.
\newblock \showarticletitle{Examining the temporal associations between depression and pathological gaming.}
\newblock \bibinfo{journal}{\emph{Canadian Journal of Behavioural Science/Revue canadienne des sciences du comportement}} \bibinfo{volume}{53}, \bibinfo{number}{3} (\bibinfo{year}{2021}), \bibinfo{pages}{274}.
\newblock


\bibitem[Revenson et~al\mbox{.}(2016)]%
        {Revenson_book1_2016}
\bibfield{author}{\bibinfo{person}{Tracey~A Revenson}, \bibinfo{person}{Konstadina Griva}, \bibinfo{person}{Aleksandra Luszczynska}, \bibinfo{person}{Val Morrison}, \bibinfo{person}{Efharis Panagopoulou}, \bibinfo{person}{Noa Vilchinsky}, {and} \bibinfo{person}{Mariët Hagedoorn}.} \bibinfo{year}{2016}\natexlab{}.
\newblock \bibinfo{booktitle}{\emph{What Is Caregiving and How Should We Study It?}}
\newblock \bibinfo{publisher}{Palgrave Macmillan UK}, \bibinfo{pages}{1--14}.
\newblock
\showISBNx{978-1-137-55898-5}
\urldef\tempurl%
\url{https://doi.org/10.1057/9781137558985_1}
\showDOI{\tempurl}


\bibitem[Rikkers et~al\mbox{.}(2016)]%
        {rikkers2016internet}
\bibfield{author}{\bibinfo{person}{Wavne Rikkers}, \bibinfo{person}{David Lawrence}, \bibinfo{person}{Jennifer Hafekost}, {and} \bibinfo{person}{Stephen~R Zubrick}.} \bibinfo{year}{2016}\natexlab{}.
\newblock \showarticletitle{Internet use and electronic gaming by children and adolescents with emotional and behavioural problems in Australia--results from the second Child and Adolescent Survey of Mental Health and Wellbeing}.
\newblock \bibinfo{journal}{\emph{BMC public health}}  \bibinfo{volume}{16} (\bibinfo{year}{2016}), \bibinfo{pages}{1--16}.
\newblock


\bibitem[Robinette et~al\mbox{.}(2016)]%
        {robinette2016assessment}
\bibfield{author}{\bibinfo{person}{Paul Robinette}, \bibinfo{person}{Alan~R Wagner}, {and} \bibinfo{person}{Ayanna~M Howard}.} \bibinfo{year}{2016}\natexlab{}.
\newblock \showarticletitle{Assessment of robot to human instruction conveyance modalities across virtual, remote and physical robot presence}. In \bibinfo{booktitle}{\emph{2016 25th IEEE International Symposium on Robot and Human Interactive Communication (RO-MAN)}}. IEEE, \bibinfo{pages}{1044--1050}.
\newblock


\bibitem[Rossi et~al\mbox{.}(2022)]%
        {rossi2022using}
\bibfield{author}{\bibinfo{person}{Silvia Rossi}, \bibinfo{person}{Silvano~Junior Santini}, \bibinfo{person}{Daniela Di~Genova}, \bibinfo{person}{Gianpaolo Maggi}, \bibinfo{person}{Alberto Verrotti}, \bibinfo{person}{Giovanni Farello}, \bibinfo{person}{Roberta Romualdi}, \bibinfo{person}{Anna Alisi}, \bibinfo{person}{Alberto~Eugenio Tozzi}, {and} \bibinfo{person}{Clara Balsano}.} \bibinfo{year}{2022}\natexlab{}.
\newblock \showarticletitle{Using the social robot NAO for emotional support to children at a pediatric emergency department: Randomized clinical trial}.
\newblock \bibinfo{journal}{\emph{Journal of Medical Internet Research}} \bibinfo{volume}{24}, \bibinfo{number}{1} (\bibinfo{year}{2022}), \bibinfo{pages}{e29656}.
\newblock


\bibitem[Rudenko et~al\mbox{.}(2024)]%
        {Rudenko2024}
\bibfield{author}{\bibinfo{person}{Irina Rudenko}, \bibinfo{person}{Andrey Rudenko}, \bibinfo{person}{Achim~J. Lilienthal}, \bibinfo{person}{Kai~O. Arras}, {and} \bibinfo{person}{Barbara Bruno}.} \bibinfo{year}{2024}\natexlab{}.
\newblock \showarticletitle{The Child Factor in Child–Robot Interaction: Discovering the Impact of Developmental Stage and Individual Characteristics}.
\newblock \bibinfo{journal}{\emph{International Journal of Social Robotics}}  \bibinfo{volume}{16} (\bibinfo{date}{8} \bibinfo{year}{2024}), \bibinfo{pages}{1879--1900}.
\newblock
Issue 8.
\showISSN{18754805}
\urldef\tempurl%
\url{https://doi.org/10.1007/S12369-024-01121-5/TABLES/2}
\showDOI{\tempurl}


\bibitem[Russell et~al\mbox{.}(2021)]%
        {russell2021use}
\bibfield{author}{\bibinfo{person}{Jeremy~K Russell}, \bibinfo{person}{Esben Strodl}, {and} \bibinfo{person}{David Kavanagh}.} \bibinfo{year}{2021}\natexlab{}.
\newblock \showarticletitle{Use of a social robot in the implementation of a narrative intervention for young people with cystic fibrosis: a feasibility study}.
\newblock \bibinfo{journal}{\emph{International Journal of Social Robotics}} (\bibinfo{year}{2021}), \bibinfo{pages}{1--15}.
\newblock


\bibitem[Salter et~al\mbox{.}(2004)]%
        {salter2004robots}
\bibfield{author}{\bibinfo{person}{Tamie Salter}, \bibinfo{person}{Kerstin Dautenhahn}, {and} \bibinfo{person}{R Bockhorst}.} \bibinfo{year}{2004}\natexlab{}.
\newblock \showarticletitle{Robots moving out of the laboratory-detecting interaction levels and human contact in noisy school environments}. In \bibinfo{booktitle}{\emph{RO-MAN 2004. 13th IEEE International Workshop on Robot and Human Interactive Communication (IEEE Catalog No. 04TH8759)}}. IEEE, \bibinfo{pages}{563--568}.
\newblock


\bibitem[Sandoval et~al\mbox{.}(2014)]%
        {sandoval2014human}
\bibfield{author}{\bibinfo{person}{Eduardo~Benitez Sandoval}, \bibinfo{person}{Omar Mubin}, {and} \bibinfo{person}{Mohammad Obaid}.} \bibinfo{year}{2014}\natexlab{}.
\newblock \showarticletitle{Human robot interaction and fiction: A contradiction}. In \bibinfo{booktitle}{\emph{Social Robotics: 6th International Conference, ICSR 2014, Sydney, NSW, Australia, October 27-29, 2014. Proceedings 6}}. Springer, \bibinfo{pages}{54--63}.
\newblock


\bibitem[Scassellati et~al\mbox{.}(2018)]%
        {scassellati2018improving}
\bibfield{author}{\bibinfo{person}{Brian Scassellati}, \bibinfo{person}{Laura Boccanfuso}, \bibinfo{person}{Chien-Ming Huang}, \bibinfo{person}{Marilena Mademtzi}, \bibinfo{person}{Meiying Qin}, \bibinfo{person}{Nicole Salomons}, \bibinfo{person}{Pamela Ventola}, {and} \bibinfo{person}{Frederick Shic}.} \bibinfo{year}{2018}\natexlab{}.
\newblock \showarticletitle{Improving social skills in children with ASD using a long-term, in-home social robot}.
\newblock \bibinfo{journal}{\emph{Science Robotics}} \bibinfo{volume}{3}, \bibinfo{number}{21} (\bibinfo{year}{2018}), \bibinfo{pages}{eaat7544}.
\newblock


\bibitem[Serholt and Barendregt(2016)]%
        {serholt2016robots}
\bibfield{author}{\bibinfo{person}{Sofia Serholt} {and} \bibinfo{person}{Wolmet Barendregt}.} \bibinfo{year}{2016}\natexlab{}.
\newblock \showarticletitle{Robots tutoring children: Longitudinal evaluation of social engagement in child-robot interaction}. In \bibinfo{booktitle}{\emph{Proceedings of the 9th nordic conference on human-computer interaction}}. \bibinfo{pages}{1--10}.
\newblock


\bibitem[Sharp et~al\mbox{.}(2006)]%
        {sharp2006short}
\bibfield{author}{\bibinfo{person}{Carla Sharp}, \bibinfo{person}{Ian~M Goodyer}, {and} \bibinfo{person}{Tim~J Croudace}.} \bibinfo{year}{2006}\natexlab{}.
\newblock \showarticletitle{The Short Mood and Feelings Questionnaire (SMFQ): a unidimensional item response theory and categorical data factor analysis of self-report ratings from a community sample of 7-through 11-year-old children}.
\newblock \bibinfo{journal}{\emph{Journal of abnormal child psychology}}  \bibinfo{volume}{34} (\bibinfo{year}{2006}), \bibinfo{pages}{365--377}.
\newblock


\bibitem[Shields et~al\mbox{.}(2023)]%
        {Shields2023}
\bibfield{author}{\bibinfo{person}{Alan~L. Shields}, \bibinfo{person}{Fiona Taylor}, \bibinfo{person}{Roger~E. Lamoureux}, \bibinfo{person}{Brad Padilla}, \bibinfo{person}{Kas Severson}, \bibinfo{person}{Tanya Green}, \bibinfo{person}{Anthony~L. Boral}, \bibinfo{person}{Cem Akin}, \bibinfo{person}{Frank Siebenhaar}, {and} \bibinfo{person}{Brenton Mar}.} \bibinfo{year}{2023}\natexlab{}.
\newblock \showarticletitle{Psychometric evaluation of the Indolent Systemic Mastocytosis Symptom Assessment Form (ISM-SAF©) and determination of a threshold score for moderate symptoms}.
\newblock \bibinfo{journal}{\emph{Orphanet Journal of Rare Diseases}}  \bibinfo{volume}{18} (\bibinfo{date}{12} \bibinfo{year}{2023}), \bibinfo{pages}{69}.
\newblock
Issue 1.
\showISSN{17501172}
\urldef\tempurl%
\url{https://doi.org/10.1186/S13023-023-02661-1}
\showDOI{\tempurl}


\bibitem[Simmons et~al\mbox{.}(2011)]%
        {Simmons2011}
\bibfield{author}{\bibinfo{person}{Joseph~P. Simmons}, \bibinfo{person}{Leif~D. Nelson}, {and} \bibinfo{person}{Uri Simonsohn}.} \bibinfo{year}{2011}\natexlab{}.
\newblock \showarticletitle{False-positive psychology: Undisclosed flexibility in data collection and analysis allows presenting anything as significant}.
\newblock \bibinfo{journal}{\emph{Psychological Science}}  \bibinfo{volume}{22} (\bibinfo{date}{10} \bibinfo{year}{2011}), \bibinfo{pages}{1359--1366}.
\newblock
Issue 11.
\showISSN{14679280}
\urldef\tempurl%
\url{https://doi.org/10.1177/0956797611417632/ASSET/IMAGES/LARGE/10.1177_0956797611417632-FIG2.JPEG}
\showDOI{\tempurl}


\bibitem[Smart and Wegner(2000)]%
        {smart_wegner2000}
\bibfield{author}{\bibinfo{person}{Laura Smart} {and} \bibinfo{person}{Daniel~M Wegner}.} \bibinfo{year}{2000}\natexlab{}.
\newblock \bibinfo{booktitle}{\emph{The hidden costs of hidden stigma.}}
\newblock \bibinfo{publisher}{The Guilford Press}, \bibinfo{pages}{220--242}.
\newblock
\showISBNx{1-57230-573-8 (Hardcover)}


\bibitem[Smedegaard(2019)]%
        {smedegaard2019reframing}
\bibfield{author}{\bibinfo{person}{Catharina~Vesterager Smedegaard}.} \bibinfo{year}{2019}\natexlab{}.
\newblock \showarticletitle{Reframing the role of novelty within social HRI: from noise to information}. In \bibinfo{booktitle}{\emph{2019 14th acm/ieee international conference on human-robot interaction (hri)}}. IEEE, \bibinfo{pages}{411--420}.
\newblock


\bibitem[Sorter et~al\mbox{.}(2024)]%
        {sorter2024addressing}
\bibfield{author}{\bibinfo{person}{Michael Sorter}, \bibinfo{person}{Lori~J Stark}, \bibinfo{person}{Tracy Glauser}, \bibinfo{person}{Jessica McClure}, \bibinfo{person}{John Pestian}, \bibinfo{person}{Katherine Junger}, {and} \bibinfo{person}{Tina~L Cheng}.} \bibinfo{year}{2024}\natexlab{}.
\newblock \showarticletitle{Addressing the pediatric mental health crisis: Moving from a reactive to a proactive system of care}.
\newblock \bibinfo{journal}{\emph{The Journal of Pediatrics}}  \bibinfo{volume}{265} (\bibinfo{year}{2024}).
\newblock


\bibitem[Spitale et~al\mbox{.}(2023)]%
        {spitale2023robotic}
\bibfield{author}{\bibinfo{person}{Micol Spitale}, \bibinfo{person}{Minja Axelsson}, {and} \bibinfo{person}{Hatice Gunes}.} \bibinfo{year}{2023}\natexlab{}.
\newblock \showarticletitle{Robotic mental well-being coaches for the workplace: An in-the-wild study on form}. In \bibinfo{booktitle}{\emph{Proceedings of the 2023 ACM/IEEE International Conference on Human-Robot Interaction}}. \bibinfo{pages}{301--310}.
\newblock


\bibitem[Spitale et~al\mbox{.}(2024)]%
        {Spitale2024}
\bibfield{author}{\bibinfo{person}{Micol Spitale}, \bibinfo{person}{Minja Axelsson}, \bibinfo{person}{Sooyeon Jeong}, \bibinfo{person}{Paige Tuttosı}, \bibinfo{person}{Caitlin~A. Stamatis}, \bibinfo{person}{Guy Laban}, \bibinfo{person}{Angelica Lim}, {and} \bibinfo{person}{Hatice Gune}.} \bibinfo{year}{2024}\natexlab{}.
\newblock \showarticletitle{Past, Present, and Future: A Survey of The Evolution of Affective Robotics For Well-being}.
\newblock  (\bibinfo{date}{7} \bibinfo{year}{2024}).
\newblock
\urldef\tempurl%
\url{https://doi.org/10.48550/arXiv.2407.02957}
\showDOI{\tempurl}


\bibitem[Spitale and Gunes(2022)]%
        {spitale_acii22}
\bibfield{author}{\bibinfo{person}{M Spitale} {and} \bibinfo{person}{H Gunes}.} \bibinfo{year}{2022}\natexlab{}.
\newblock \showarticletitle{Affective Robotics For Wellbeing: A Scoping Review}. In \bibinfo{booktitle}{\emph{2022 10th International Conference on Affective Computing and Intelligent Interaction Workshops and Demos (ACIIW)}}. \bibinfo{pages}{1--8}.
\newblock
\urldef\tempurl%
\url{https://doi.org/10.1109/ACIIW57231.2022.10085995}
\showDOI{\tempurl}


\bibitem[Stenseng et~al\mbox{.}(2020)]%
        {stenseng2020time}
\bibfield{author}{\bibinfo{person}{Frode Stenseng}, \bibinfo{person}{Beate~Wold Hygen}, {and} \bibinfo{person}{Lars Wichstr{\o}m}.} \bibinfo{year}{2020}\natexlab{}.
\newblock \showarticletitle{Time spent gaming and psychiatric symptoms in childhood: cross-sectional associations and longitudinal effects}.
\newblock \bibinfo{journal}{\emph{European child \& adolescent psychiatry}} \bibinfo{volume}{29}, \bibinfo{number}{6} (\bibinfo{year}{2020}), \bibinfo{pages}{839--847}.
\newblock


\bibitem[Stewart et~al\mbox{.}(2012)]%
        {Stewart2012}
\bibfield{author}{\bibinfo{person}{Anita~L. Stewart}, \bibinfo{person}{Angela~D. Thrasher}, \bibinfo{person}{Jack Goldberg}, {and} \bibinfo{person}{Judy~A. Shea}.} \bibinfo{year}{2012}\natexlab{}.
\newblock \showarticletitle{A Framework for Understanding Modifications to Measures for Diverse Populations}.
\newblock \bibinfo{journal}{\emph{Journal of aging and health}}  \bibinfo{volume}{24} (\bibinfo{date}{9} \bibinfo{year}{2012}), \bibinfo{pages}{992}.
\newblock
Issue 6.
\showISSN{08982643}
\urldef\tempurl%
\url{https://doi.org/10.1177/0898264312440321}
\showDOI{\tempurl}


\bibitem[Strait et~al\mbox{.}(2015)]%
        {strait2015gender}
\bibfield{author}{\bibinfo{person}{Megan Strait}, \bibinfo{person}{Priscilla Briggs}, {and} \bibinfo{person}{Matthias Scheutz}.} \bibinfo{year}{2015}\natexlab{}.
\newblock \showarticletitle{Gender, more so than age, modulates positive perceptions of language-based human-robot interactions}. In \bibinfo{booktitle}{\emph{4th international symposium on new frontiers in human robot interaction}}. \bibinfo{pages}{21--22}.
\newblock


\bibitem[Tam et~al\mbox{.}(2024)]%
        {Tam2024}
\bibfield{author}{\bibinfo{person}{Thomas Yu~Chow Tam}, \bibinfo{person}{Sonish Sivarajkumar}, \bibinfo{person}{Sumit Kapoor}, \bibinfo{person}{Alisa~V. Stolyar}, \bibinfo{person}{Katelyn Polanska}, \bibinfo{person}{Karleigh~R. McCarthy}, \bibinfo{person}{Hunter Osterhoudt}, \bibinfo{person}{Xizhi Wu}, \bibinfo{person}{Shyam Visweswaran}, \bibinfo{person}{Sunyang Fu}, \bibinfo{person}{Piyush Mathur}, \bibinfo{person}{Giovanni~E. Cacciamani}, \bibinfo{person}{Cong Sun}, \bibinfo{person}{Yifan Peng}, {and} \bibinfo{person}{Yanshan Wang}.} \bibinfo{year}{2024}\natexlab{}.
\newblock \showarticletitle{A framework for human evaluation of large language models in healthcare derived from literature review}.
\newblock \bibinfo{journal}{\emph{npj Digital Medicine 2024 7:1}}  \bibinfo{volume}{7} (\bibinfo{date}{9} \bibinfo{year}{2024}), \bibinfo{pages}{1--20}.
\newblock
Issue 1.
\showISSN{2398-6352}
\urldef\tempurl%
\url{https://doi.org/10.1038/s41746-024-01258-7}
\showDOI{\tempurl}


\bibitem[Tanaka et~al\mbox{.}(2015)]%
        {tanaka2015pepper}
\bibfield{author}{\bibinfo{person}{Fumihide Tanaka}, \bibinfo{person}{Kyosuke Isshiki}, \bibinfo{person}{Fumiki Takahashi}, \bibinfo{person}{Manabu Uekusa}, \bibinfo{person}{Rumiko Sei}, {and} \bibinfo{person}{Kaname Hayashi}.} \bibinfo{year}{2015}\natexlab{}.
\newblock \showarticletitle{Pepper learns together with children: Development of an educational application}. In \bibinfo{booktitle}{\emph{2015 IEEE-RAS 15th International Conference on Humanoid Robots (Humanoids)}}. IEEE, \bibinfo{pages}{270--275}.
\newblock


\bibitem[Thepthien et~al\mbox{.}(2019)]%
        {thepthien2019self}
\bibfield{author}{\bibinfo{person}{Bang-on Thepthien}, \bibinfo{person}{Pakaporn Busprachong}, {and} \bibinfo{person}{Nate Hongkeilert}.} \bibinfo{year}{2019}\natexlab{}.
\newblock \showarticletitle{Self-Disclosure Among Youth with Problematic Methamphetamine Use Who Received Treatment in Public Health Centers of the Bangkok Metropolitan Administration: A Qualitative Analysis}.
\newblock \bibinfo{journal}{\emph{Journal of Child \& Adolescent Substance Abuse}} \bibinfo{volume}{28}, \bibinfo{number}{5} (\bibinfo{year}{2019}), \bibinfo{pages}{363--375}.
\newblock


\bibitem[Tourangeau et~al\mbox{.}(2000)]%
        {tourangeau2000psychology}
\bibfield{author}{\bibinfo{person}{Roger Tourangeau}, \bibinfo{person}{Lance~J Rips}, {and} \bibinfo{person}{Kenneth Rasinski}.} \bibinfo{year}{2000}\natexlab{}.
\newblock \showarticletitle{The psychology of survey response}.
\newblock  (\bibinfo{year}{2000}).
\newblock


\bibitem[Vallverdú and Trovato(2016)]%
        {afford2016}
\bibfield{author}{\bibinfo{person}{Jordi Vallverdú} {and} \bibinfo{person}{Gabriele Trovato}.} \bibinfo{year}{2016}\natexlab{}.
\newblock \showarticletitle{Emotional affordances for human–robot interaction}.
\newblock \bibinfo{journal}{\emph{Adaptive Behavior}}  \bibinfo{volume}{24} (\bibinfo{date}{10} \bibinfo{year}{2016}), \bibinfo{pages}{320--334}.
\newblock
Issue 5.
\showISSN{17412633}
\urldef\tempurl%
\url{https://doi.org/10.1177/1059712316668238}
\showDOI{\tempurl}


\bibitem[Van Der~Drift et~al\mbox{.}(2014)]%
        {van2014remote}
\bibfield{author}{\bibinfo{person}{Esther~JG Van Der~Drift}, \bibinfo{person}{Robbert-Jan Beun}, \bibinfo{person}{Rosemarijn Looije}, \bibinfo{person}{Olivier~A Blanson~Henkemans}, {and} \bibinfo{person}{Mark~A Neerincx}.} \bibinfo{year}{2014}\natexlab{}.
\newblock \showarticletitle{A remote social robot to motivate and support diabetic children in keeping a diary}. In \bibinfo{booktitle}{\emph{Proceedings of the 2014 ACM/IEEE international conference on Human-robot interaction}}. \bibinfo{pages}{463--470}.
\newblock


\bibitem[Villatoro et~al\mbox{.}(2018)]%
        {Villatoro2018}
\bibfield{author}{\bibinfo{person}{Alice~P. Villatoro}, \bibinfo{person}{Melissa~J. DuPont-Reyes}, \bibinfo{person}{Jo~C. Phelan}, \bibinfo{person}{Kirstin Painter}, {and} \bibinfo{person}{Bruce~G. Link}.} \bibinfo{year}{2018}\natexlab{}.
\newblock \showarticletitle{Parental recognition of preadolescent mental health problems: Does stigma matter?}
\newblock \bibinfo{journal}{\emph{Social Science \& Medicine}}  \bibinfo{volume}{216} (\bibinfo{date}{11} \bibinfo{year}{2018}), \bibinfo{pages}{88--96}.
\newblock
\showISSN{0277-9536}
\urldef\tempurl%
\url{https://doi.org/10.1016/J.SOCSCIMED.2018.09.040}
\showDOI{\tempurl}


\bibitem[Yamazaki et~al\mbox{.}(2023)]%
        {yamazaki2023long}
\bibfield{author}{\bibinfo{person}{Ryuji Yamazaki}, \bibinfo{person}{Shuichi Nishio}, \bibinfo{person}{Yuma Nagata}, \bibinfo{person}{Yuto Satake}, \bibinfo{person}{Maki Suzuki}, \bibinfo{person}{Hideki Kanemoto}, \bibinfo{person}{Miyae Yamakawa}, \bibinfo{person}{David Figueroa}, \bibinfo{person}{Hiroshi Ishiguro}, {and} \bibinfo{person}{Manabu Ikeda}.} \bibinfo{year}{2023}\natexlab{}.
\newblock \showarticletitle{Long-term effect of the absence of a companion robot on older adults: A preliminary pilot study}.
\newblock \bibinfo{journal}{\emph{Frontiers in Computer Science}}  \bibinfo{volume}{5} (\bibinfo{year}{2023}), \bibinfo{pages}{1129506}.
\newblock


\bibitem[Yang et~al\mbox{.}(2019)]%
        {yang2019channel}
\bibfield{author}{\bibinfo{person}{Diyi Yang}, \bibinfo{person}{Zheng Yao}, \bibinfo{person}{Joseph Seering}, {and} \bibinfo{person}{Robert Kraut}.} \bibinfo{year}{2019}\natexlab{}.
\newblock \showarticletitle{The channel matters: Self-disclosure, reciprocity and social support in online cancer support groups}. In \bibinfo{booktitle}{\emph{Proceedings of the 2019 chi conference on human factors in computing systems}}. \bibinfo{pages}{1--15}.
\newblock


\bibitem[Zhang et~al\mbox{.}(2023)]%
        {zhang_llm_2023}
\bibfield{author}{\bibinfo{person}{Ceng Zhang}, \bibinfo{person}{Junxin Chen}, \bibinfo{person}{Jiatong Li}, \bibinfo{person}{Yanhong Peng}, {and} \bibinfo{person}{Zebing Mao}.} \bibinfo{year}{2023}\natexlab{}.
\newblock \showarticletitle{Large language models for human–robot interaction: A review}.
\newblock \bibinfo{journal}{\emph{Biomimetic Intelligence and Robotics}}  \bibinfo{volume}{3} (\bibinfo{date}{12} \bibinfo{year}{2023}), \bibinfo{pages}{100131}.
\newblock
Issue 4.
\showISSN{2667-3797}
\urldef\tempurl%
\url{https://doi.org/10.1016/J.BIROB.2023.100131}
\showDOI{\tempurl}


\end{thebibliography}

\end{document}